\documentclass{atlasnote_modified}
\graphicspath{{figures/}}

\usepackage{atlasphysics} 

\usepackage{bm}
\usepackage{booktabs}
\usepackage{xspace}
\usepackage{multirow}
\usepackage{bigstrut}
\usepackage{subfigure}
\usepackage{hyperref}
\usepackage{placeins}

\usepackage{mathrsfs}
\usepackage{slashed}
\usepackage{array}
\usepackage{multirow}
\usepackage{xspace}
\usepackage{amssymb,amsmath}

\usepackage{authblk}
\usepackage{tabularx}
\usepackage{dcolumn}

\newcommand{\mjjj}{\mbox{$m_{\rm jjj}$}}

\newcommand{\AKT}{anti-$k_{t}$}
\newcommand{\Zg}{\Zboson/\gamma^*}

\newcommand{\muplus} {\ensuremath{\mu}+jets\xspace}
\newcommand{\eplus} {\ensuremath{e}+jets\xspace}

\newcommand{\nl} {N^{\mathrm{loose}}}
\newcommand{\nlr} {N^{\mathrm{loose}}_{\mathrm{real}}}
\newcommand{\nlf} {N^{\mathrm{loose}}_{\mathrm{fake}}}

\newcommand{\nt} {N^{\mathrm{std}}}

\newcommand{\mT} {\ensuremath{m_{\mathrm{T}}}}

\newcommand{\epsr} {\ensuremath{r}}

\newcommand{\epsf} {\ensuremath{f}}

\newcommand{\pmasym}[2]{^{+#1}_{-#2}}
\newcommand{\sigmattbar}{\ensuremath{\sigma_{\ttbar}}}

\newcommand{\lumitot}{\mbox{2.9\,pb$^{-1}$}}

\newcommand{\ncandlj}{37}
\newcommand{\ncandll}{9}

\newcommand{\nbglj}{\ensuremath{12.2 \pm 3.9}}
\newcommand{\nbgll}{\ensuremath{2.5 \pm 0.6}}

\newcommand{\sigcomb}{4.8}

\newcommand{\xsecctot}{145}
\newcommand{\xseccstat}{\pm 31}
\newcommand{\xseccsyst}{~{}^{+42}_{-27}}

\newcommand{\singlefigure}[4]{
\begin{figure}[htbp]
\centering
\includegraphics[width=#2]{#1}
\caption{\label{#3}#4}
\end{figure}
}

\skipbeforetitle{100pt}

\title{Measurement of the top quark-pair production cross section\\
       with ATLAS in pp collisions at $\sqrt{s}=7\TeV$}

\author{The ATLAS Collaboration}
\atlasnote{CERN-PH-EP-2010-064 (Submitted to EPJC)}
\usepackage{authblk}
\renewcommand\Authands{, } 
\renewcommand\Affilfont{\itshape\small} 

\abstracttext{ 

A measurement of the production cross-section for top quark pairs
($\ttbar$) in $pp$ collisions at $\sqrt{s}=7~\TeV$ is presented using
data recorded with the ATLAS detector at the Large Hadron
Collider. Events are selected in two different topologies: single
lepton (electron $e$ or muon $\mu$) with large missing transverse
energy and at least four jets, and dilepton ($ee$, $\mu\mu$ or
$e\mu$) with large missing transverse energy and at least two jets. In
a data sample of $\lumitot$, $\ncandlj$ candidate events are observed
in the single-lepton topology and $\ncandll$ events in the dilepton
topology. The corresponding expected backgrounds from non-$\ttbar$
Standard Model processes are estimated using data-driven methods and
determined to be $\nbglj$ events and $\nbgll$ events,
respectively. The kinematic properties of the selected events are
consistent with SM $\ttbar$ production. The inclusive top quark pair
production cross-section is measured to be
\[
\sigmattbar=\xsecctot \xseccstat \xseccsyst\,~\rm pb
\]
\noindent where the first uncertainty is statistical and the second systematic.
The measurement agrees with perturbative QCD calculations.  
}

\begin{document}

\section{Introduction}\label{s:intro}

The observation of top quark pair ($\ttbar$) production is one of the
milestones for the early LHC physics programme. The measurement of the
top quark pair production cross-section (\sigmattbar) in the various
decay channels is interesting for several reasons.  Uncertainties on
the theoretical predictions are now at the level of $10\%$ and a
comparison with experimental measurements performed in different
channels will ultimately allow a precision test of the predictions of
perturbative QCD.  In addition, the abundant $\ttbar$ sample which is
expected to be produced in the first years of data-taking can be
exploited for improving many aspects of detector performance. Finally,
$\ttbar$ production is an important background in various searches for
physics beyond the Standard Model, and new physics may also give rise
to additional $\ttbar$ production mechanisms or modification of the
top quark decay channels.

In the Standard Model (SM)~\cite{SM} the $\ttbar$ production
cross-section in $pp$ collisions is calculated to be
164.6~$\pmasym{11.4}{15.7}$~pb~\cite{topxs} at a centre of mass energy
$\sqrt{s}=7~\TeV$ assuming a top mass of 172.5 GeV, and top quarks are
predicted to decay to a $W$ boson and a $b$-quark ($t\rightarrow Wb$)
nearly 100\% of the time. Events with a \ttbar\ pair can be classified
as `single-lepton', `dilepton', or `all hadronic' by the decays of the
two $W$ bosons: a pair of quarks ($W\rightarrow\qqbar$) or a
lepton-neutrino pair ($W\rightarrow\ell\nu$), where $\ell$ refers to a
lepton. At the Tevatron the dominant production mechanism is
$q\bar{q}$ annihilation, and the \ttbar\ cross section at
$\sqrt{s}=1.8$ GeV and at $\sqrt{s}=1.96$ GeV have been measured by D0
and CDF~\cite{cdfd0} in all channels.  The production of \ttbar\ at
the LHC is dominated by $gg$ fusion. Recently, the CMS collaboration
has presented a cross-section measurement, $\sigmattbar=194 \pm
72\;{\rm(stat.)} \pm 24\;{\rm(syst.)} \pm 21\;{\rm(lumi.)}$~pb in the
dilepton channel using 3.1~\ipb\ of data~\cite{cmstopdileptons}.

The results described in this paper are based on reconstructed electrons and
muons and include small contributions from leptonically decaying tau
leptons.  The single-lepton mode, with a branching ratio\footnote{The
quoted branching ratios also include small contributions from
leptonically decaying taus.} of 37.9\% (combining $e$ and $\mu$
channels), and the dilepton mode, with a branching ratio of 6.5\%
(combining $ee$, $\mu\mu$ and $e\mu$ channels), both give rise to final
states with at least one lepton, missing transverse energy and jets,
some with $b$ flavour. The cross-section measurements in both modes
are based on a straightforward counting method. The number of signal
events is obtained in a signal enriched sample after background
subtraction. The main background contributions are determined using
data-driven methods, since the theoretical uncertainties on the
normalisation of these backgrounds are relatively large.  For both
single-lepton and dilepton channels, alternative methods of signal
extraction and/or background estimation are explored.  In particular,
two template shape fitting methods, which use additional signal regions to exploit
the kinematic information in the events, are developed for the single-lepton mode. 
In this paper these two fitting methods serve as
important cross-checks of the counting method. The methods also
provide alternative data-driven estimates of backgrounds and are
expected to become more powerful when more data become available.

\section{Detector and data sample}\label{s:detdata}

The ATLAS detector~\cite{atlasdet} at the LHC covers nearly the entire
solid angle\footnote{In the right-handed ATLAS coordinate system, the
pseudorapidity $\eta$ is defined as $\eta=-\ln[\tan(\theta/2)]$, where
the polar angle $\theta$ is measured with respect to the LHC
beamline. The azimuthal angle $\phi$ is measured with respect to the
$x$-axis, which points towards the centre of the LHC ring.  The
$z$-axis is parallel to the anti-clockwise beam viewed from above.
Transverse momentum and energy are defined as $\pT = p\sin\theta$ and
$\ET = E\sin\theta$, respectively.}  around the collision point. It
consists of an inner tracking detector surrounded by a thin
superconducting solenoid, electromagnetic and hadronic calorimeters,
and an external muon spectrometer incorporating three large
superconducting toroid magnet assemblies. 

The inner-detector system is immersed in a 2\,T axial magnetic field
and provides charged particle tracking in the range $|\eta|<2.5$.  The
high-granularity silicon pixel detector covers the vertex region and
provides typically three measurements per track, followed by the
silicon microstrip tracker (SCT) which provides four measurements from
eight strip layers. These silicon detectors are complemented by the
transition radiation tracker (TRT), which enables extended track
reconstruction up to $|\eta|=2.0$. In giving typically more than 30
straw-tube measurements per track, the TRT is essential to the inner
detector momentum resolution, and also provides electron
identification information.

The calorimeter system covers the pseudorapidity range $|\eta|<4.9$.
Within the region $|\eta|<3.2$, electromagnetic calorimetry is
provided by barrel and endcap lead-liquid argon (LAr) electromagnetic
calorimeters, with an additional thin LAr presampler covering
$|\eta|<1.8$ to correct for energy loss in material upstream of the
calorimeters. Hadronic calorimetry is provided by the
steel/scintillating-tile calorimeter, segmented into three barrel
structures within $|\eta|<1.7$, and two copper/LAr hadronic endcap
calorimeters. The solid angle coverage is completed with forward
copper/LAr and tungsten/LAr calorimeter modules optimised for
electromagnetic and hadronic measurements respectively.

The muon spectrometer comprises separate trigger and high-precision
tracking chambers measuring the deflection of muons in a magnetic
field with a bending integral from 2 to 8 Tm in the central region,
generated by three superconducting air-core toroids. The precision
chamber system covers the region $|\eta|<2.7$ with three layers of
monitored drift tubes, complemented by cathode strip chambers in the
forward region, where the background is highest. The muon trigger
system covers the range $|\eta|<2.4$ with resistive plate chambers in
the barrel, and thin gap chambers in the endcap regions.

A three-level trigger system is used to select interesting events. The
\mbox{level-1} trigger is implemented in hardware and uses a subset of
detector information to reduce the event rate to a design value of at most
75\,kHz. This is followed by two software-based trigger levels,
\mbox{level-2} and the event filter, which together reduce the event rate to
about 200\,Hz.

Only data where all subsystems described above are fully operational
are used. Applying these requirements to $\sqrt{s}=7\TeV$ $pp$
collision data taken in stable beam conditions and recorded until
30$\rm ^{th}$ August 2010 results in a data sample of \lumitot.
This luminosity value has a relative uncertainty of 11\%~\cite{lumi}.

\section{Simulated event samples}\label{s:mc}
\label{mc.section}

Monte-Carlo simulation samples are used to develop and validate the
analysis procedures, to calculate the acceptance for \ttbar\ events
and to evaluate the contributions from some background processes.  For
the \ttbar\ signal the next-to-leading order (NLO) generator {\sc
MC@NLO} v3.41~\cite{mcatnlo1}, is used with an assumed top-quark mass of
$172.5\GeV$ and with the NLO parton density function (PDF) set
CTEQ66~\cite{cteq66}.

For the main backgrounds, consisting of QCD multi-jet events and $W/Z$ boson
production in association with multiple jets, {\sc Alpgen}
v2.13~\cite{alpgen} is used, which implements the exact LO matrix
elements for final states with up to 6 partons. Using the LO PDF set
CTEQ6L1~\cite{cteq6l}, the following backgrounds are generated:
$W$+jets events with up to 5 partons, $\Zg$+jets events with up to 5
partons and with the dilepton invariant mass $m_{\ell\ell}>40\GeV$;
QCD multi-jet events with up to 6 partons, and diboson $WW$+jets, $WZ$+jets and
$ZZ$+jets events. A separate sample of $Z$ boson production generated with {\sc Pythia} is used to cover the
region $10\GeV<m_{\ell\ell}<40\GeV$.
\noindent The `MLM' matching scheme of the {\sc Alpgen} generator is
used to remove overlaps between the $n$ and $n+1$ parton samples with
parameters {\tt RCLUS}=0.7 and {\tt ETCLUS}=$20\GeV$.  For all but
the diboson processes, separate samples are generated that include
$b\bar{b}$ and $c\bar{c}$ quark pair production at the matrix element
level. In addition, for the $W$+jets process, a separate sample
containing $W$+$c$+jets events is produced.  For the small background
of single-top production {\sc MC@NLO} is used, invoking the `diagram
removal scheme'~\cite{diagrem} to remove overlaps between the single-top
and the $\ttbar$ final states.

In simulation, the cross-section of \ttbar\ production is normalized to 164.6 pb
obtained from approximate NNLO calculations~\cite{topxs}. The cross-sections
for $W/Z$+jets and diboson with jets have been rescaled by a
factor 1.22 to match NNLO calculations of their inclusive cross-sections,
as is done in~\cite{CSCbook}. The QCD multi-jet sample
has not been rescaled as it is only used for validation studies.

Unless otherwise noted, all events are hadronised with {\sc Herwig},
using {\sc Jimmy} for the underlying event model. The same
underlying-event tune has been used for all samples. After event
generation, all samples are processed by the standard ATLAS detector
and trigger simulation~\cite{atlsim} and subject to the same
reconstruction algorithms as the data.

\subsection{Systematic uncertainties on the simulated samples}\label{s:mcsyst}

The use of simulated \ttbar\ samples to calculate the signal
acceptance gives rise to systematic uncertainties from the choice of
generator, the amount of initial and final state radiation (ISR/FSR)
and uncertainties on the PDF. The uncertainty due
to the choice of generator is evaluated by comparing the predictions
of {\sc MC@NLO} with those of {\sc Powheg}~\cite{powheg} interfaced to
both {\sc Herwig} or {\sc Pythia}. The uncertainty due to ISR/FSR is
evaluated by studies using the {\sc AcerMC} generator~\cite{Acer}
interfaced to {\sc Pythia}, and by varying the parameters controlling ISR
and FSR in a range consistent with experimental data~\cite{CSCbook}.
Finally, the uncertainty in the PDFs used to
generate \ttbar\ and single-top events is evaluated using a range of
current PDF sets with the procedure described in~\cite{CSCbook}.
In addition, the impact of the assumed top-quark mass is tested with
a set of samples generated with different masses.

Simulation-based predictions of $W/Z$+jets background events have
uncertainties on their total cross-section, on the contribution of
events with jets from heavy-flavour ($b,c$) quarks, and on the shape
of kinematic distributions. The predictions of the total cross-section
have uncertainties of up to $O(50\%)$~\cite{wjetstheo} increasing with
jet multiplicity. Total $W/Z$ cross-section predictions are not used
in the cross-section analysis, but are used in simulation predictions
shown in selected Figures. The heavy-flavor fractions in the
$W/Z$+jets samples are always taken from simulation, as the present
data sample is too small to measure them. Here a fully correlated
100\% uncertainty on the predicted fractions of $b\bar{b}$ and
$c\bar{c}$ quark pairs is assumed, as well as a separate 100\% uncertainty on
the fraction of events with a single $c$ quark.  
The uncertainty on the shape of kinematic distributions, used in fit-based
cross-checks of the single-lepton analysis, is assessed by varying
internal generator parameters, and by comparing {\sc Alpgen} with {\sc
Sherpa}~\cite{sherpa}.

For the small backgrounds from single-top and diboson production, only
overall normalisation uncertainties are considered and these are taken
to be 10\% and 5\%, respectively.

\section{Object and event selection} \label{s:obj}

For both the single lepton and the dilepton analysis, events are triggered by
a single lepton trigger (electron or muon)~\cite{wobspaper}. 
The detailed trigger
requirements vary through the data-taking period due to the rapidly
increasing LHC luminosity and the commissioning of the trigger system,
but the thresholds are always low enough to ensure that leptons with $\pT>20\GeV$
lie in the efficiency plateau.

The electron selection requires a \mbox{level-1} electromagnetic cluster with
 $\pT>10\GeV$. A more refined electromagnetic cluster
selection is required in the \mbox{level-2} trigger. Subsequently, a match
between the selected calorimeter electromagnetic cluster and an inner
detector track is required in the event filter.
Muons are selected requiring a $\pT>10\GeV$ momentum threshold muon trigger chamber
track at \mbox{level-1}, matched by a muon reconstructed in the precision
chambers at the event filter.

After the trigger selections, events must have at least one
offline-reconstructed primary vertex with at least five tracks, and
are discarded if any jet with $\pT>10\GeV$ at the EM scale is
identified as out-of-time activity or calorimeter
noise~\cite{jetqual}. 

The reconstruction of $\ttbar$ events makes use of
electrons, muons and jets, and of missing transverse
energy $\MET$ which is a measure of the energy imbalance in the
transverse plane and is used as an indicator of undetected neutrinos. 

Electron candidates are required to pass the electron selection as
defined in Ref.~\cite{wobspaper}, with $\pT>20\GeV$ and
$|\eta_{\mathrm{cluster}}|<2.47$, 
where $\eta_{\mathrm{cluster}}$ is the pseudorapidity of the calorimeter
cluster associated to the candidate. 
Candidates in the calorimeter transition
region at $1.37<|\eta_{\mathrm{cluster}}|<1.52$ are excluded.
In addition, the ratio $E/p$ of
electron cluster energy measured in the calorimeter to momentum in the
tracker must be consistent with that expected for an electron. Also,
in order to suppress the background from photon conversions, the track
must have an associated hit in the innermost pixel layer, except when
the track passes through one of the 2\% of pixel modules known to be
dead.  Muon candidates are reconstructed from track segments in the
different layers of the muon chambers~\cite{muonperf}. These segments
are then combined starting from the outermost layer, with a procedure
that takes material effects into account, and matched with tracks
found in the inner detector. The final candidates are refitted using
the complete track information from both detector systems, and
required to satisfy $\pT>20\GeV$ and $|\eta|<2.5$.

To reduce the background due to leptons from decays of hadrons
(including heavy flavours) produced in jets, the leptons in each event
are required to be isolated. For electrons, the $\ET$ deposited in the
calorimeter towers in a cone in $\eta$-$\phi$ space of radius $\Delta
R=0.2$ around the electron position\footnote{The radius $\Delta R$
between the object axis and the edge of the object cone is defined as
$\Delta R = \sqrt{{\Delta \phi}^2+{\Delta \eta}^2}$.} is summed, and
the $\ET$ due to the electron ($\ET^{e}$) is subtracted. The remaining
$\ET$ is required to be less than $4~\GeV+0.023 \cdot \ET^{e}$. For
muons, the corresponding calorimeter isolation energy in a cone of
$\Delta R=0.3$ is required to be less than $4\GeV$, and the scalar sum
of track transverse momenta in a cone of $\Delta R=0.3$ is also
required to be less than $4\GeV$ after subtraction of the muon $\pT$.
Additionally, muons are required to have a separation $\Delta R > 0.4$
from any jet with $\pT>20\GeV$, to further suppress muons from heavy
flavour decays inside jets.

Jets are reconstructed with the \AKT\ algorithm~\cite{antikt} ($\Delta
R=0.4$) from topological clusters~\cite{dijets} of energy deposits in
the calorimeters, calibrated at the electromagnetic (EM) scale
appropriate for the energy deposited by electrons or photons.  These
jets are then calibrated to the hadronic energy scale, using a
correction factor obtained from simulation~\cite{dijets} which depends
upon $\pT$ and $\eta$.  If the closest object to an electron candidate
is a jet with a separation $\Delta R<0.2$ the jet is removed in order
to avoid double-counting of electrons as jets.

Jets originating from b-quarks are selected by exploiting the long
lifetime of b-hadrons (about 1.5 ps) which leads to typical flight
paths of a few millimeters which are observable in the detector. The
SV0 b-tagging algorithm\cite{btagefi} used in this analysis explicitly
reconstructs a displaced vertex from the decay products of the
long-lived b-hadron. As input, the SV0 tagging algorithm is given a
list of tracks associated to the calorimeter jet. Only tracks
fulfilling certain quality criteria are used in the secondary vertex
fit. Secondary vertices are reconstructed in an inclusive way starting
from two- track vertices which are merged into a common vertex. Tracks
giving large $\chi^2$ contributions are then iteratively removed until
the reconstructed vertex fulfills certain quality criteria. Two-track
vertices at a radius consistent with the radius of one of the three
pixel detector layers are removed, as these vertices likely originate
from material interactions. A jet is considered b-tagged if it
contains a secondary vertex, reconstructed with the SV0 tagging
algorithm, with $L/\sigma(L) > 5.72$, where $L$ is the decay length
and $\sigma(L)$ its uncertainty. This operating point yields a 50\%
b-tagging efficiency in simulated \ttbar\ events. The sign of
$L/\sigma(L)$ is given by the sign of the projection of the decay
length vector on the jet axis.

The missing transverse energy is constructed from the vector sum of
all calorimeter cells contained in topological clusters. 
Calorimeter cells are associated with a parent physics object in a
chosen order: electrons, jets and muons, such that a cell is uniquely
associated to a single physics object~\cite{metpaper}. Cells belonging
to electrons are calibrated at the electron energy scale, but omitting
the out-of-cluster correction to avoid double cell-energy counting,
while cells belonging to jets are taken at the corrected energy scale
used for jets.  Finally, the contributions from muons passing
selection requirements are included, and the contributions from any
calorimeter cells associated to the muons are subtracted. The
remaining clustered energies not associated to electrons or jets are
included at the EM scale.

The modelled acceptances and efficiencies are verified by comparing
Monte-Carlo simulations with data in control regions which are
depleted of \ttbar{} events. Lepton efficiencies are derived from data
in the $Z$ boson mass window, and are validated by using them to estimate
inclusive $W$ and $Z$ boson cross-sections. The acceptances for the jet
multiplicity and \met{} cuts are validated using a number of control
regions surrounding the \ttbar{} signal region in phase-space.

\subsection{Systematic uncertainties for reconstructed objects}
\label{s:objsyst}

The uncertainties due to Monte-Carlo simulation modelling of the lepton
trigger, reconstruction and selection efficiencies are assessed using
leptons from $Z\rightarrow ee$ and $Z\rightarrow\mu\mu$ events
selected from the same data sample used for the $\ttbar$ analyses.
Scale factors are applied to Monte-Carlo samples when calculating
acceptances. 
The statistical and systematic uncertainties on the scale
factors are included in the uncertainties on the acceptance values.
The modelling of the lepton energy scale and resolution are studied using
reconstructed $Z$ boson mass distributions, and used to adjust the
simulation accordingly.

The jet energy scale (JES) and its uncertainty are derived by
combining information from test-beam data, LHC collision data and
simulation~\cite{dijets}. The JES uncertainty varies in the range
6--10\% as a function of jet $\pT$ and $\eta$. The jet energy
resolution (JER) and jet finding efficiency measured in data and in
simulation are in agreement. The limited statistical precision of the
comparisons for the energy resolution (14\%) and the efficiency (1\%)
are taken as the systematic uncertainties in each case.

The $b$-tagging efficiency and mistag fraction of the SV0 
$b$-tagging algorithm have been measured on
data~\cite{btagefi}. The efficiency measurement is based on a sample
of jets containing muons and makes use of the transverse momentum of a
muon relative to the jet axis. The measurement of the mistag fraction
is performed on an inclusive jet sample and includes two methods, one
which uses the invariant mass spectrum of tracks associated to
reconstructed secondary vertices to separate light- and heavy-flavour
jets and one which is based on the rate at which secondary vertices
with negative decay-length significance are present in the data. Both
the $b$-tagging efficiency and mistag fraction measured in data depend
strongly on the jet kinematics. In the range $25 < \pT({\rm jet}) < 85$~GeV, the
$b$-tagging efficiency rises from 40\% to 60\%, while the mistag fraction
increases from 0.2\% to 1\% between 20 and 150~GeV.  The measurements
of the $b$-tagging efficiencies and mistag fractions are provided in the
form of $\pT$-dependent scale factors correcting the $b$-tagging
performance in simulation to that observed in data. The relative
statistical (systematic) uncertainties for the $b$-tagging efficiency
range from 3\% to 10\% (10\% to 12\%). For the $b$-tagging efficiency, the
scale factor is close to one for all values of jet $\pT$. For
light-flavour jets, the simulation underestimates the tagging
efficiency by factors of $1.27 \pm 0.26$ for jets with $\pT < 40 \gev$
and $1.07 \pm 0.25$ for jets with $\pT > 40 \gev$.

The LHC instantaneous luminosity varied by several orders of magnitude
during the data-taking period considered for this measurement, reaching a peak of about $1\times
10^{31}$ cm$^{-2}$s$^{-1}$. At this luminosity, an average of about
two extra $pp$ interactions were superimposed on each hard
proton-proton interaction. This `pileup' background produces
additional activity in the detector, affecting variables like jet
reconstruction and isolation energy. No attempts to correct the event
reconstruction for these effects are made, since the data-driven
determination of object identification and trigger efficiencies and
backgrounds naturally include them.  The residual effects on the
$\ttbar$ event acceptance are assessed by using $\ttbar$ simulation
samples with additional pileup interactions, simulated with {\sc
Pythia}, that were overlayed during event digitisation and
reconstruction. In a scenario where on average two pileup interactions
are added to each event, corresponding to conditions that exceed
those observed during the data taking period, the largest change of
acceptance observed in any of the channels is 3.6$\%$. As the effect
of pileup is small even in this pessimistic scenario,
it is neglected in the acceptance systematics evaluation.

\section{Single lepton analysis}
\label{ljets.section}

\subsection{Event selection}
\label{s:ljets_selection}
The single lepton $\ttbar$ final state is characterized by an isolated
lepton with relatively high $p_T$ and missing transverse energy corresponding to the neutrino from the $W$ leptonic decay, two $b$ quark jets and two light jets from the hadronic $W$ decay.

The selection of events for the single-lepton analysis consists of a series of
requirements on the reconstructed objects defined in Section~\ref{s:obj}, designed to 
select events with the above topology. For each lepton flavour, the following event 
selections are first applied:

\begin{itemize}
 \item the appropriate single-electron or single-muon trigger has
 fired;
\item the event contains one and only reconstructed lepton (electron or muon)
 with $p_T>20\GeV$, matching the corresponding high-level trigger object;

 \item $\met>20\GeV$ and $\met+m{_T}(W)>60\GeV$\footnote{Here $m_T(W)$ is the $W$-boson transverse mass, defined as
$\sqrt{ 2 p_T^{\ell} p_T^{\nu} (1-\cos ( \phi^{\ell} - \phi^{\nu} )) }$ where the measured missing \et\ vector provides the neutrino information.
}. 
The cut on $\met$ rejects a significant fraction of the
QCD  multi-jet background. Further rejection can be achieved by applying a cut in
the ($\met$, $m{_T}(W)$) plane; true $W \rightarrow \ell \nu$ decays with large $\met$ have also large $m{_T}(W)$, while mis-measured jets in QCD multi-jet events may result in large $\met$ but small $m{_T}(W)$. The requirement on the sum of $\met$ and $m{_T}(W)$ discriminates between the two cases;

\item finally, the event is required to have $\ge 1$jet with $p_T>25\GeV$ and $|\eta|<2.5$.
 The requirement on the $p_T$ and the pseudorapidity of the jets
 is a compromise between the efficiency of the $\ttbar$ events selection, and the rejection
 of $W$+jets and QCD multi-jet background.

\end{itemize}

\noindent Events are then classified by the number of jets with $p_T>25\GeV$ and $|\eta|<2.5$, being either 1, 2, 3 or at least 4.
These samples are labeled `1-jet pre-tag' through `$\geq$4-jet
pre-tag', where the number corresponds to the jet multiplicity as
defined above and pre-tag refers to the fact that no $b$-tagging
information has been used.  Subsets of these samples are then defined
with the additional requirement that at least one of the jets with
$p_T>25\GeV$ is tagged as a $b$-jet. They are referred to as the
`1-jet tagged' through `$\geq$4-jet tagged' samples.

\begin{figure}[tbp]
\begin{center}
\begin{tabular}{ccc}
\includegraphics[width=0.315\textwidth]{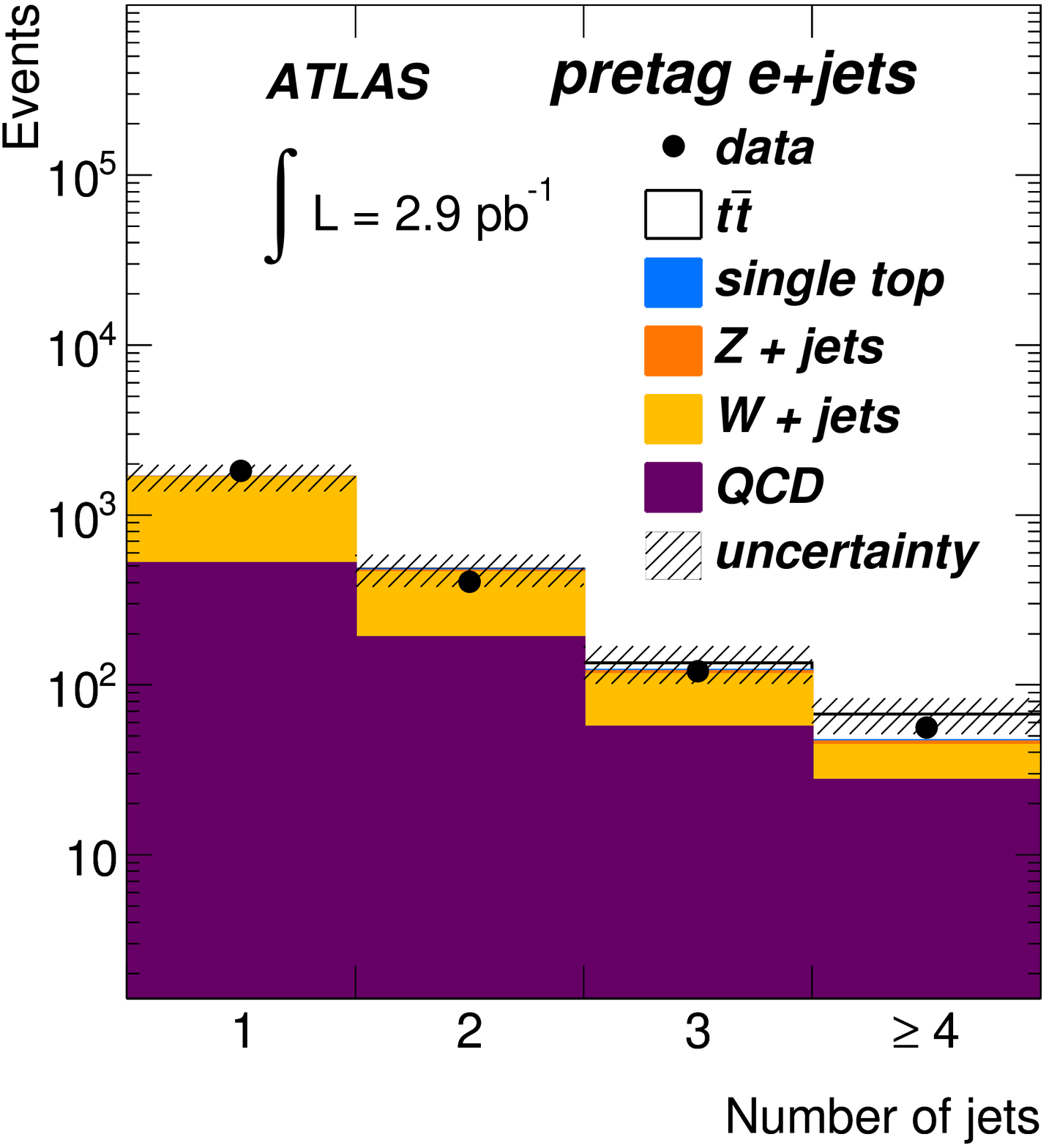} &
\includegraphics[width=0.315\textwidth]{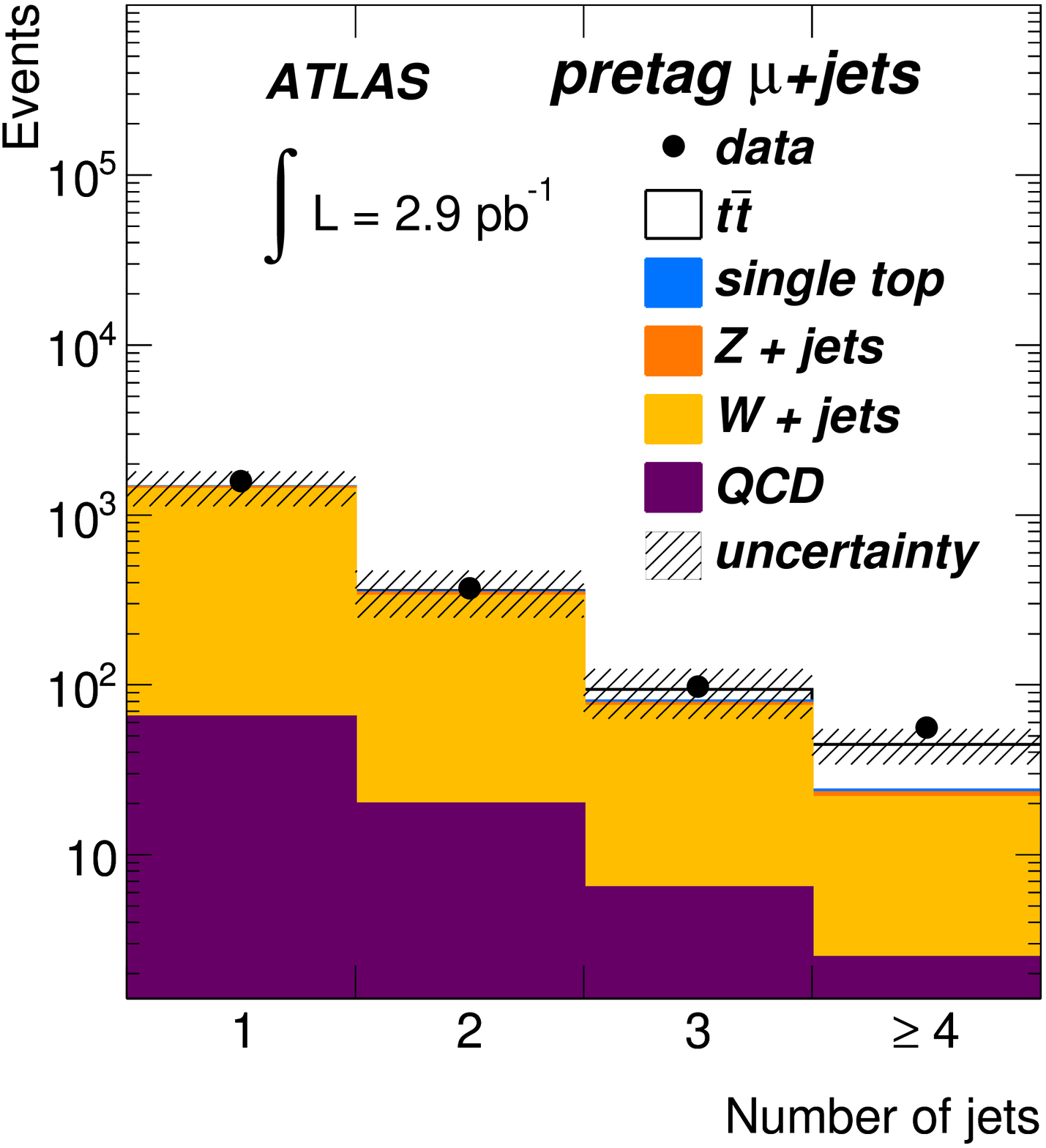} &
\includegraphics[width=0.315\textwidth]{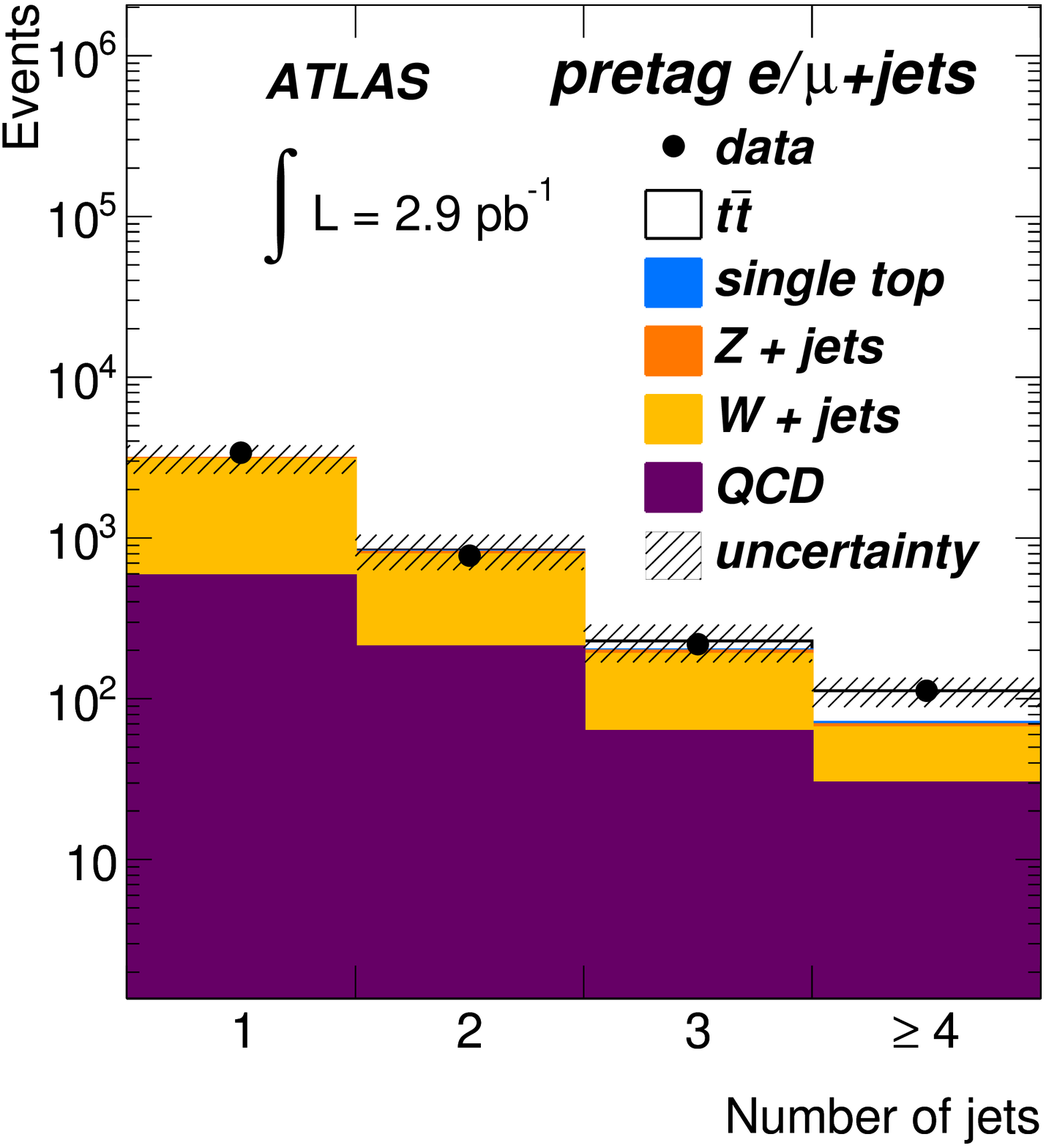} \\
(a) & (b) & (c) \\
\includegraphics[width=0.315\textwidth]{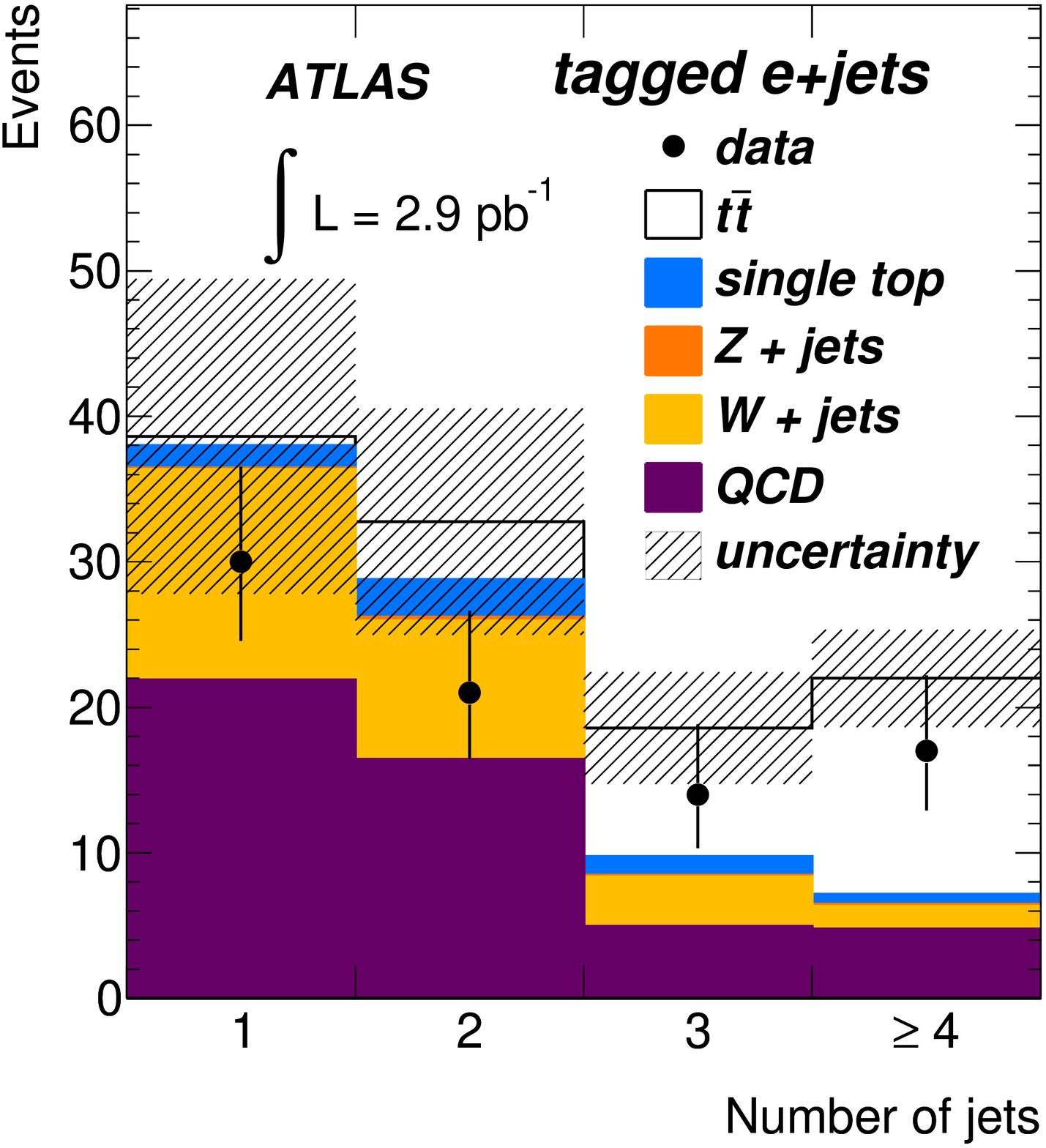} &
\includegraphics[width=0.315\textwidth]{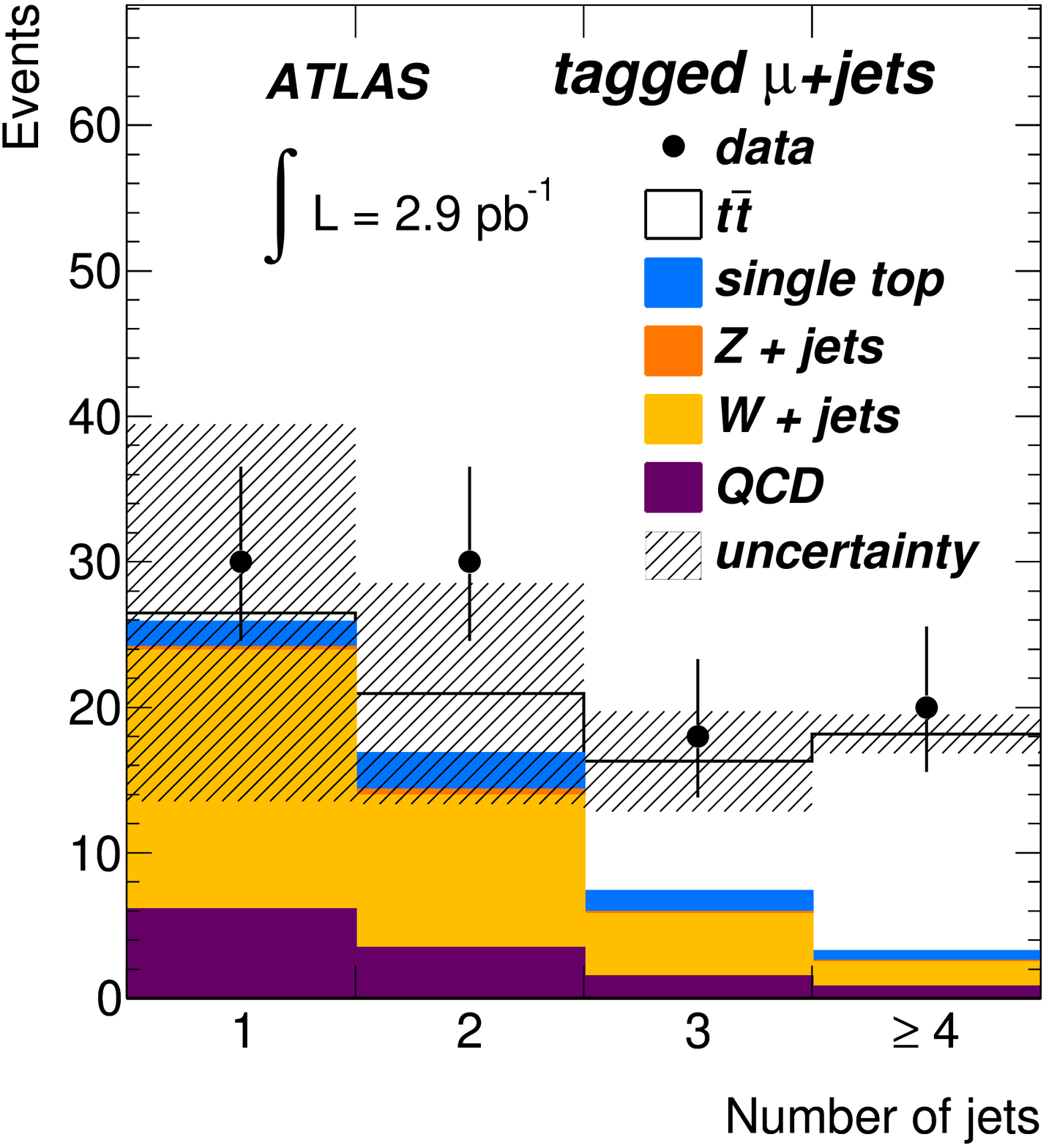} &
\includegraphics[width=0.315\textwidth]{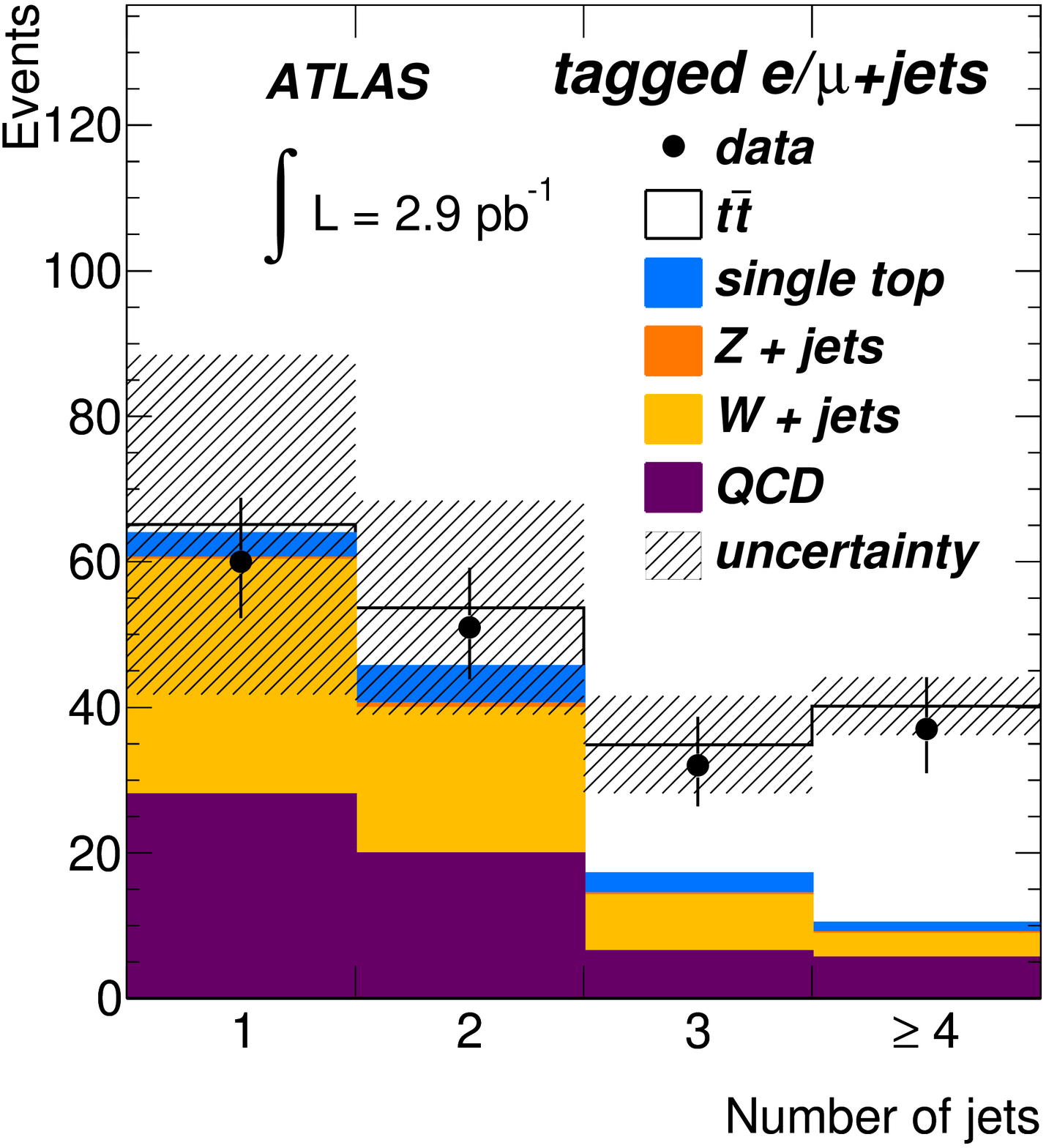} \\
(d) & (e) & (f) \\
\end{tabular}
\end{center}
\caption{\label{f:jetmult}
Jet multiplicity distributions (\textit{i.e.} number of jets with
$p_T>25$ GeV).  Top row - pre-tag samples: (a) electron channel, (b)
muon channel and (c) electron/muon combined.  Bottom row - tagged
samples: (d) electron channel, (e) muon channel and (f) electron/muon
combined.  The data are compared to the sum of all expected
contributions. For the totals shown, simulation estimates are
used for all contributions except QCD multi-jet, where a data-driven
technique is used. The background uncertainty on the
total expectation is represented by the hatched area. The
$\ge$4-jet bin in the tagged sample represents the signal region.}
\end{figure}

Figure~\ref{f:jetmult} shows the observed jet multiplicity for events
in the pre-tag and tagged samples,
together with the sum of all expected contributions as expected from simulation, except for QCD multi-jet, which is taken from a data-driven technique discussed in Section~\ref{s:StrategyBck}.
The largest fraction of $\ttbar$ events is concentrated in $\geq$4-jets bin of the tagged sample, which is defined as the signal region and used for the \ttbar\ signal extraction in the primary method described in Section~\ref{s:counting}.
One of the cross-check methods, discussed in Section~\ref{sec:fits}, uses in addition the 3-jet tagged sample for signal extraction.  Other regions are used as control samples for the determination of backgrounds.

Table~\ref{tab:num_el_mu} lists the numbers of events in
the four tagged samples, as well as the number of events in the 3-jet
and $\geq$4-jet zero-tag samples, which comprise the events not
containing $b$-tagged jets. These events are used for background normalisation 
in the second cross-check method described in Section~\ref{sec:fits}.
For all samples, Table~\ref{tab:num_el_mu} also lists the contributions
estimated from Monte Carlo simulation for \ttbar, $W$+jets, $Z$+jets
and single-top events. The quoted uncertainties are from object
reconstruction and identification. 
For the data-driven estimates of $W$+jets and QCD multi-jet, the results of the procedures that will be detailed in
Sections~\ref{s:QCD} and~\ref{s:wjets} are quoted. The uncertainty on the background prediction is mostly systematic and largely correlated between bins, and is also different 
in the electron and muon channels due to different sample composition in terms of QCD 
and $W$+jets fractions. QCD is larger than $W$+jets in the electron channel, while it 
is smaller for muons.

The estimated product of acceptance and branching fraction for \ttbar\
events in the $\ge$4-jet tagged signal region, measured from Monte
Carlo samples, are $(3.1 \pm 0.7)\%$ and $(3.2 \pm 0.7)\%$ for
\eplus\ and \muplus{}, respectively. About $90\%$ of the selected \ttbar\ events come from the
corresponding $t \rightarrow W \rightarrow e$ or $\mu$ decay including
leptonic $\tau$ decays, and the acceptance for those events is $15 \pm
3\%$. The remaining 10\% comes from dilepton events where one of the
leptons was not reconstructed as electron or muon. The contribution
from fully hadronic \ttbar\ events is negligible. The uncertainties on
the acceptance originate from physics process modelling and object
selection uncertainties detailed in Sections
\ref{s:mcsyst} and
\ref{s:objsyst}.  

\begin{table}[htp]
\begin{center}
\begin{tabular}{|c|c|c|c|c|c|c|}
\hline
\multicolumn{7}{|c|}{$e$+jets channel}\\
\hline
      	 & 1-jet & 2-jet & 3-jet & $\geq$4-jet & 3-jet & $\geq$4-jet \\
      	 & tagged	 & tagged	 & tagged	 & tagged	 & zero-tag	 & zero-tag \\
\hline
QCD (DD)	 & 21.9 $\pm$ 3.4	 & 16.4 $\pm$ 4.0     & 4.9 $\pm$ 2.7	 & 4.8 $\pm$ 3.1       & 52.0 $\pm$ 19	 & 23.0 $\pm$ 11 \\
$W$+jets (MC)	 & 14.5 $\pm$ 10	 & 9.5  $\pm$ 6.6     & 3.4 $\pm$ 2.7	 & 1.5 $\pm$ 1.4       & 55.1 $\pm$ 26	 & 15.1 $\pm$ 10 \\
$W$+jets (DD)	 & -			 & - 		      & -	         & 1.9 $\pm$ 1.1       & -	                 & 9.3  $\pm$ 4.0 \\
$Z$+jets (MC)	 & 0.1 $\pm$ 0.1	 & 0.3 $\pm$ 0.1      & 0.1 $\pm$ 0.1	 & 0.2 $\pm$ 0.1       & 4.6 $\pm$ 2.2	         & 1.7 $\pm$ 1.3 \\
Single top (MC)	 & 1.6 $\pm$ 0.3	 & 2.6 $\pm$ 0.6      & 1.3 $\pm$ 0.3	 & 0.7 $\pm$ 0.2       & 0.9 $\pm$ 0.2	         & 0.4 $\pm$ 0.1 \\
\hline
Total  (non \ttbar\ )& 38.1 $\pm$ 11	 & 28.8 $\pm$ 7.7     & 9.7 $\pm$ 3.8	 & 7.2 $\pm$ 3.4       & 112.6 $\pm$ 32	 & 40.2 $\pm$ 15  \\
\hline
\ttbar\ (MC)     & 0.6 $\pm$ 0.2       & 4.0 $\pm$ 1.0         & 8.8 $\pm$ 1.8    &14.9 $\pm$ 3.5          & 4.5 $\pm$ 0.8           & 5.4 $\pm$ 1.2 \\
\hline
Total expected   & 39 $\pm$ 11   & 33 $\pm$ 8    & 19 $\pm$ 4    & 22 $\pm$ 5     & 117 $\pm$ 32  & 46 $\pm$ 15 \\
\hline\hline
Observed	 & 30	 & 21	 & 14	 & 17	 & 106 & 39 \\
\hline
\end{tabular}

(a)
\vspace{1cm}

\begin{tabular}{|c|c|c|c|c|c|c|}
\hline
\multicolumn{7}{|c|}{$\mu$+jets channel}\\
\hline
      	 & 1-jet & 2-jet & 3-jet & $\geq$4-jet & 3-jet & $\geq$4-jet \\
      	 & tagged	 & tagged	 & tagged	 & tagged	 & zero-tag	 & zero-tag \\
\hline
QCD (DD)	 & 6.1 $\pm$ 2.9	 & 3.4 $\pm$ 1.8	 & 1.5 $\pm$ 0.8	 & 0.8 $\pm$ 0.5	 & 4.9 $\pm$ 2.3	 & 1.7 $\pm$ 1.1 \\
$W$+jets (MC)	 & 17.8 $\pm$ 12	 & 10.5 $\pm$ 7.4   	 & 4.3 $\pm$ 3.3	 & 1.7 $\pm$ 1.6	 & 63.6 $\pm$ 28	 & 17.6 $\pm$ 12 \\
$W$+jets (DD)	 & -			 & -	 		 & -			 & 3.2 $\pm$ 1.7	 & -                     & 15.7 $\pm$ 4.5 \\
$Z$+jets (MC)	 & 0.3 $\pm$ 0.1	 & 0.4 $\pm$ 0.2	 & 0.1 $\pm$ 0.1	 & 0.1 $\pm$ 0.1	 & 3.3 $\pm$ 1.6	 & 1.3 $\pm$ 0.8 \\
Single top (MC)	 & 1.7 $\pm$ 0.4	 & 2.5 $\pm$ 0.5	 & 1.5 $\pm$ 0.3	 & 0.7 $\pm$ 0.2	 & 1.1 $\pm$ 0.2	 & 0.3 $\pm$ 0.1 \\
\hline
Total  (non \ttbar\ )	& 25.9 $\pm$ 13 & 16.8 $\pm$ 7.6	 & 7.4 $\pm$ 3.4	 & 3.3 $\pm$ 1.7	 & 72.9 $\pm$ 29	 & 20.9 $\pm$ 13 \\
\hline
\ttbar (MC)      & 0.7 $\pm$ 0.2       & 4.1 $\pm$ 1.1         & 9.0 $\pm$ 1.8         & 15.0 $\pm$ 3.4  & 4.6 $\pm$ 0.7         & 5.5 $\pm$ 1.2 \\
\hline
Total expected   & 27 $\pm$ 13   & 21 $\pm$ 8    & 16 $\pm$ 4    & 18 $\pm$4        & 78 $\pm$ 29   & 26 $\pm$ 13 \\
\hline\hline
Observed	 & 30	 & 30	 & 18	 & 20	 & 80 & 36 \\
\hline
\end{tabular}

(b)
\caption{\label{tab:num_el_mu} Number of tagged and zero-tag events
with different jet multiplicities in (a) the single-electron and (b)
the single-muon channel.  The observed number of events are shown,
together with the Monte-Carlo simulation estimates (MC) for \ttbar,
$W$+jets, $Z$+jets and single-top events, normalised to the data
integrated luminosity of \lumitot. The data-driven estimates (DD) for
QCD multi-jet (see Section~\ref{s:QCD}) and $W$+jets (see
Section~\ref{s:wjets}) backgrounds are also shown. The `Total (non
{\ttbar})' row uses the simulation estimate for $W$+jets for all
samples.  The uncertainties on all data-driven background estimates
include the statistical uncertainty and all systematic
uncertainties. The numbers in the `Total expected' rows are rounded to
a precision commensurate with the uncertainty.}
\end{center}
\end{table}

\subsection{Background determination strategy}
\label{s:StrategyBck}

The expected dominant backgrounds in the single-lepton channel are
$W$+jets, which can give rise to the same final state as \ttbar\
signal, and QCD multi-jet events. QCD multi-jet events only contribute to
the signal selection if the reconstructed \met\ is sufficiently large
and a fake lepton is reconstructed. Fake leptons originate in
misidentified jets or are non-prompt leptons, e.g. from semileptonic
decays of heavy quarks.

In the pre-tag samples both $W$+jets and QCD multi-jet are dominated
by events with light quarks and gluons. In the $b$-tagged samples,
light-quark and gluon final states are strongly suppressed and their
contributions become comparable to those with $b\bar{b}$ pairs,
$c\bar{c}$ pairs and single $c$ quarks, which are all of a similar
magnitude.

The contribution of $W$+jet events and QCD multi-jet events to the
$\geq$4-jet bin are both measured with data-driven methods, as
detector simulation and/or theoretical predictions are insufficiently
precise. The remaining smaller backgrounds, notably single-top
production and $Z$+jets production, are estimated from simulation.

\subsection{Background with fake and non-prompt leptons}
\label{s:QCD}

\subsubsection{Background estimate in the $\mu$+jets channel}
\label{s:mujets_fakes}

In the $\mu$+jets channel, 
the background to `real' (prompt) muons coming from `fake' muons in QCD multi-jet events, 
is predominantly due to final states with a non-prompt muon.
As all other processes (\ttbar , $W$+jets, $Z$+jets and single-top) in this channel
feature a prompt muon from a $W$ or $Z$ boson decay, it is sufficient to estimate the
number of events with a non-prompt muon to quantify the QCD multi-jet background.

The number of events in the sample with a non-prompt muon can be
extracted from the data by considering the event count in the signal
region with two sets of muon identification criteria. The `standard'
and `loose' criteria comprise the standard muon definition
described in Section \ref{s:obj}, with and without, respectively, the
requirements on the lepton isolation.

The procedure followed at this point is the so-called `matrix method':
the number of events selected by the loose and by the standard cuts,
$\nl$ and $\nt$ respectively, can be expressed as linear combinations
of the number of events with a `real' (prompt) or a `fake' muon:
\begin{eqnarray}
\nl & = & \nlr + \nlf, \nonumber \\
\nt & = & \epsr  \nlr +  \epsf \nlf,
\label{eq:MM}
\end{eqnarray}

\noindent where $\epsr$ is the fraction of `real' (prompt)
muons in the loose selection that also pass the standard selection
and $\epsf$ is the fraction of `fake' (non-prompt) muons in the
loose selection that also pass the standard selection. If $\epsr$ and
$\epsf$ are known, the number of events with non-prompt muons can be
calculated from Equation~\ref{eq:MM} given a measured $\nl$ and $\nt$. The
relative efficiencies $\epsr$ and $\epsf$ are measured in data in
control samples enriched in either prompt or non-prompt muons. The key
issue in selecting these control regions is that they should be
kinematically representative of the signal region so that the measured
control-region efficiency can be applied in the signal region.

An inclusive $Z\to \mu^+\mu^-$ control sample is used to measure
the prompt muon efficiency $\epsr = 0.990\pm 0.003$.
No statistically significant dependence on
the jet multiplicity is observed. For the measurement of the
non-prompt muon efficiency two control regions are used: a
Sample~A with low missing transverse energy ($\met<10\GeV$) and at
least one jet with $p_T>25$ GeV, and a Sample~B with the nominal
missing transverse energy requirement ($\met>20\GeV$), at least one
jet with $p_T>25$ GeV, and a high muon impact parameter
significance. Sample A is dominated by QCD multi-jet events as most
QCD multi-jet events have little true \met\ and the cross-section is
comparatively large. The contribution from events with
prompt muons from $W$/$Z$+jets which remains in the $\met<10\GeV$
region has to be subtracted.
Since the contribution of these processes is not accurately known,
it is evaluated in an iterative procedure: the initial value
obtained for $\epsf$ is used to predict the number of leptons
in the full $\met$ range. The excess of candidate lepton events in
data is attributed to prompt muons from $W$/$Z$+jets, whose
contribution to the $\met<10\GeV$ region is then subtracted,
obtaining a new value for $\epsf$. The procedure converges in few 
iterations and it results in $\epsf^{A} = 0.382 \pm
0.007$, where the quoted uncertainty is statistical only. Sample B is
kinematically close to the signal region, but the large impact
parameter significance requirement selects muons that are
incompatible with originating from the primary vertex and the sample 
is thus enriched in non-prompt muons. Here a value
$\epsf^{B} = 0.295 \pm 0.025$ is measured, where the uncertainty is again
statistical only.

Since both samples A and B are reasonable, but imperfect,
approximations of the signal region in terms of event kinematics, the
unweighted average $\epsf = 0.339 \pm 0.013\;{\rm (stat.)} \pm 0.061\;{\rm (syst.)}$
is taken as the central value.
The systematic uncertainty is determined by half the difference
between the control regions, multiplied by $\sqrt{2}$ to obtain an
unbiased estimate of the underlying uncertainty,
assuming that the two control regions have similar kinematics
as the signal region.
A single value of $\epsf$ is used to estimate the background in each
of the four pre-tag $\mu$+jets samples using Equation~\ref{eq:MM}. The
validity of this approach has been verified on samples of simulated
events.

For the tagged samples, the estimated background in each pre-tag
sample is multiplied by the measured probability for a similar QCD multi-jet
event to
have at least one $b$-tagged jet.  This results in a more precise
measurement of the tagged event rate than a measurement of $\epsf$ in
a tagged control sample, which has a large statistical uncertainty
due to the relatively small number of tagged events.
The $b$-tagging probabilities for QCD multi-jet events are
 0.09$\pm0.02$, 0.17$\pm0.03$, 0.23$\pm0.06$ and 0.31$\pm0.10$ for
1 through $\ge$4-jet, respectively. These per-event
$b$-tag probabilities have been measured in a sample defined by the
pre-tag criteria, but without the \met\ cut, and by relaxing the muon
selection to the loose criteria.  The systematic uncertainty on this
per-event tagging probability is evaluated by varying the selection
criteria of the sample used for the measurement.

\begin{figure}[htbp]
\begin{center}
\begin{tabular}{ccc}
\includegraphics[width=0.315\textwidth]{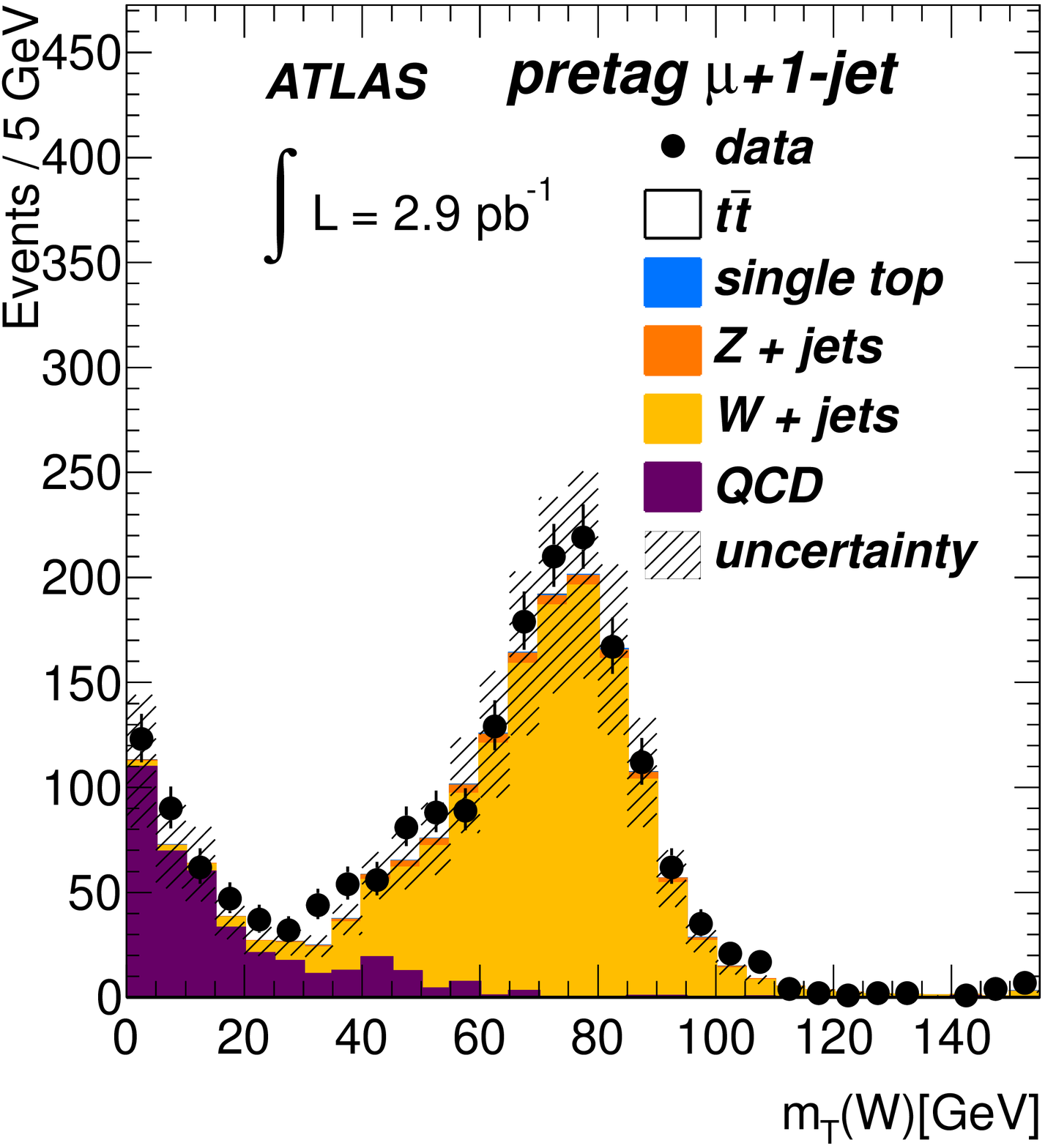} &
\includegraphics[width=0.315\textwidth]{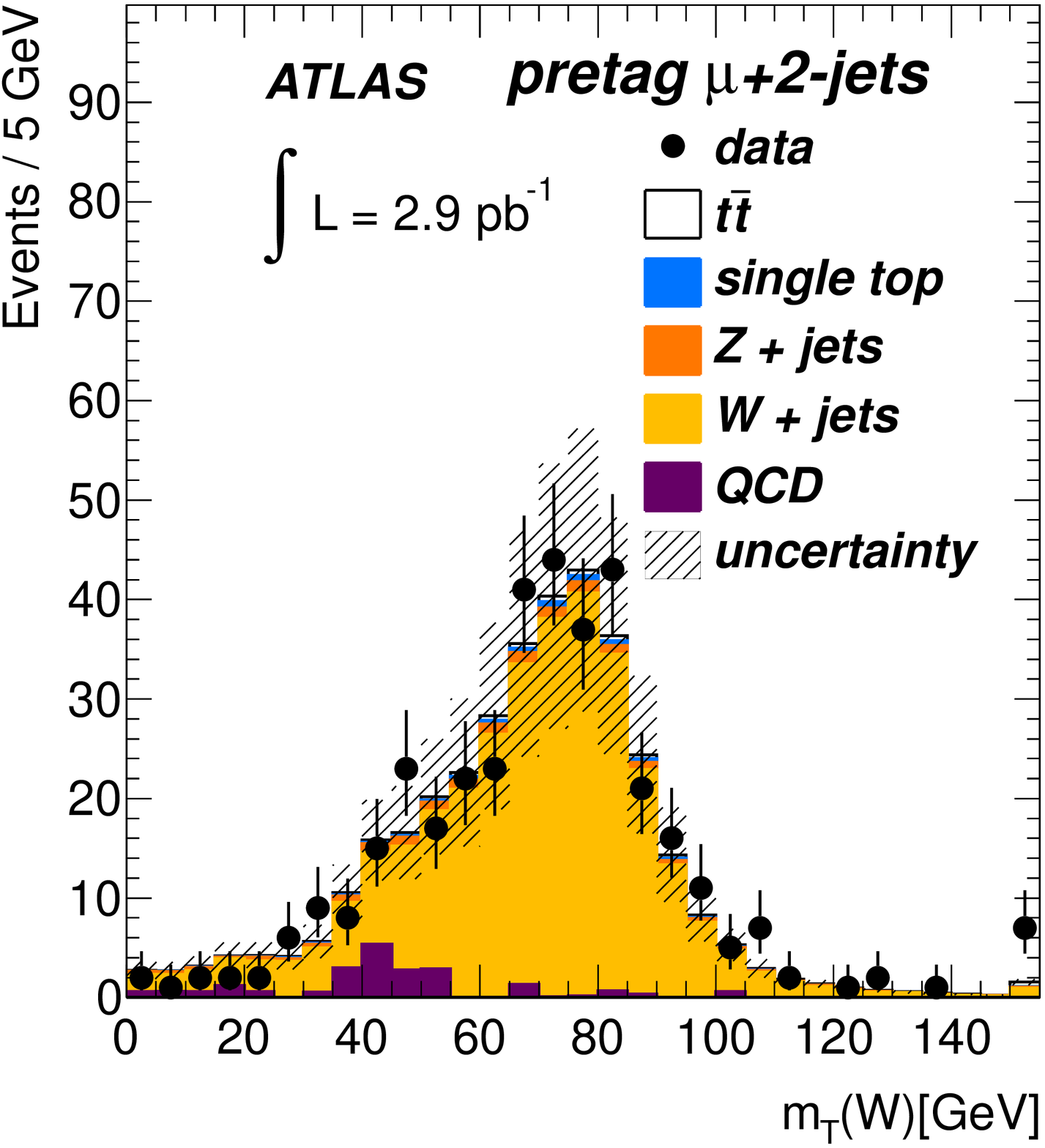} &
\includegraphics[width=0.315\textwidth]{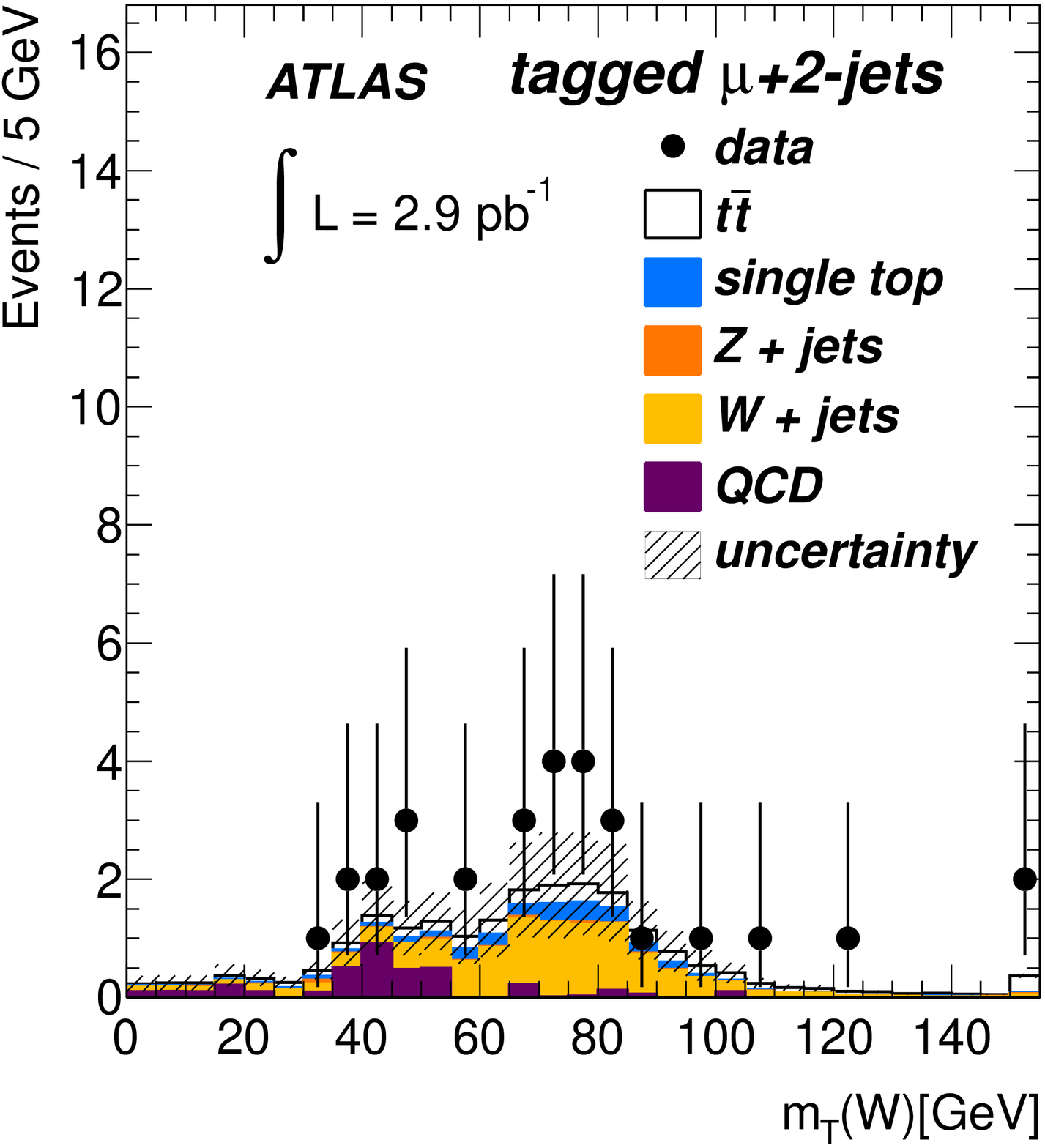} \\
(a) & (b) & (c) \\
\includegraphics[width=0.315\textwidth]{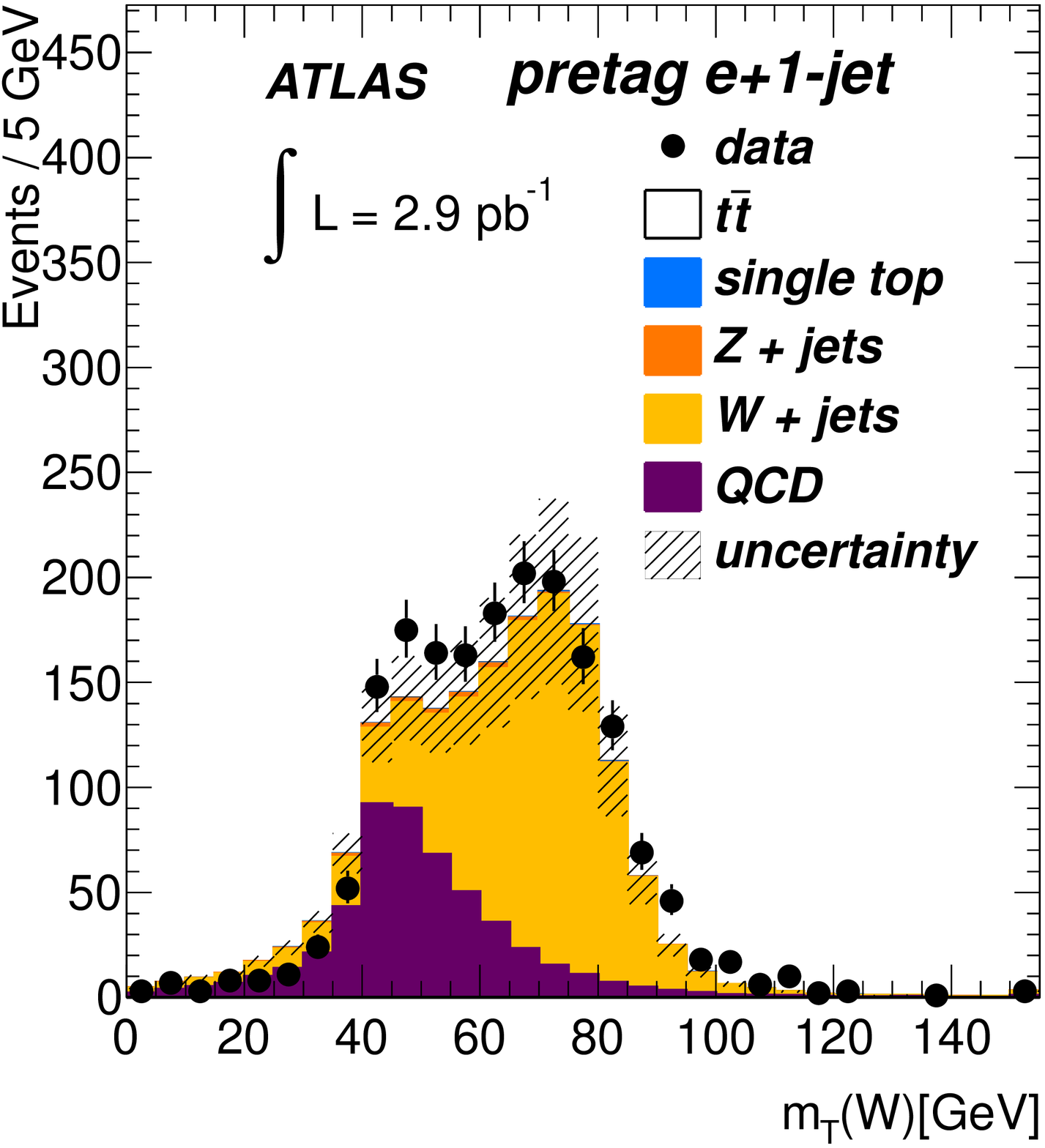} &
\includegraphics[width=0.315\textwidth]{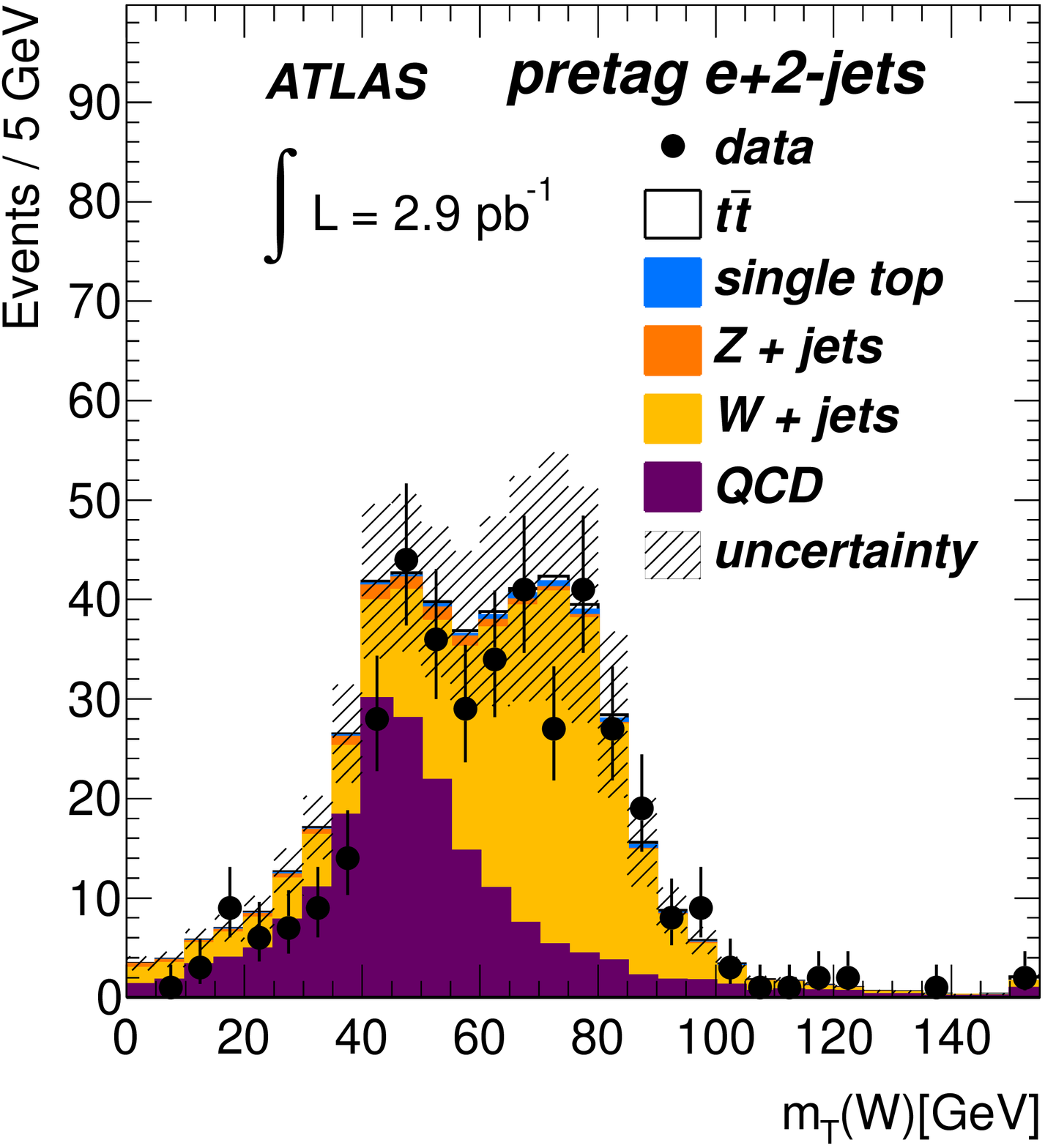} &
\includegraphics[width=0.315\textwidth]{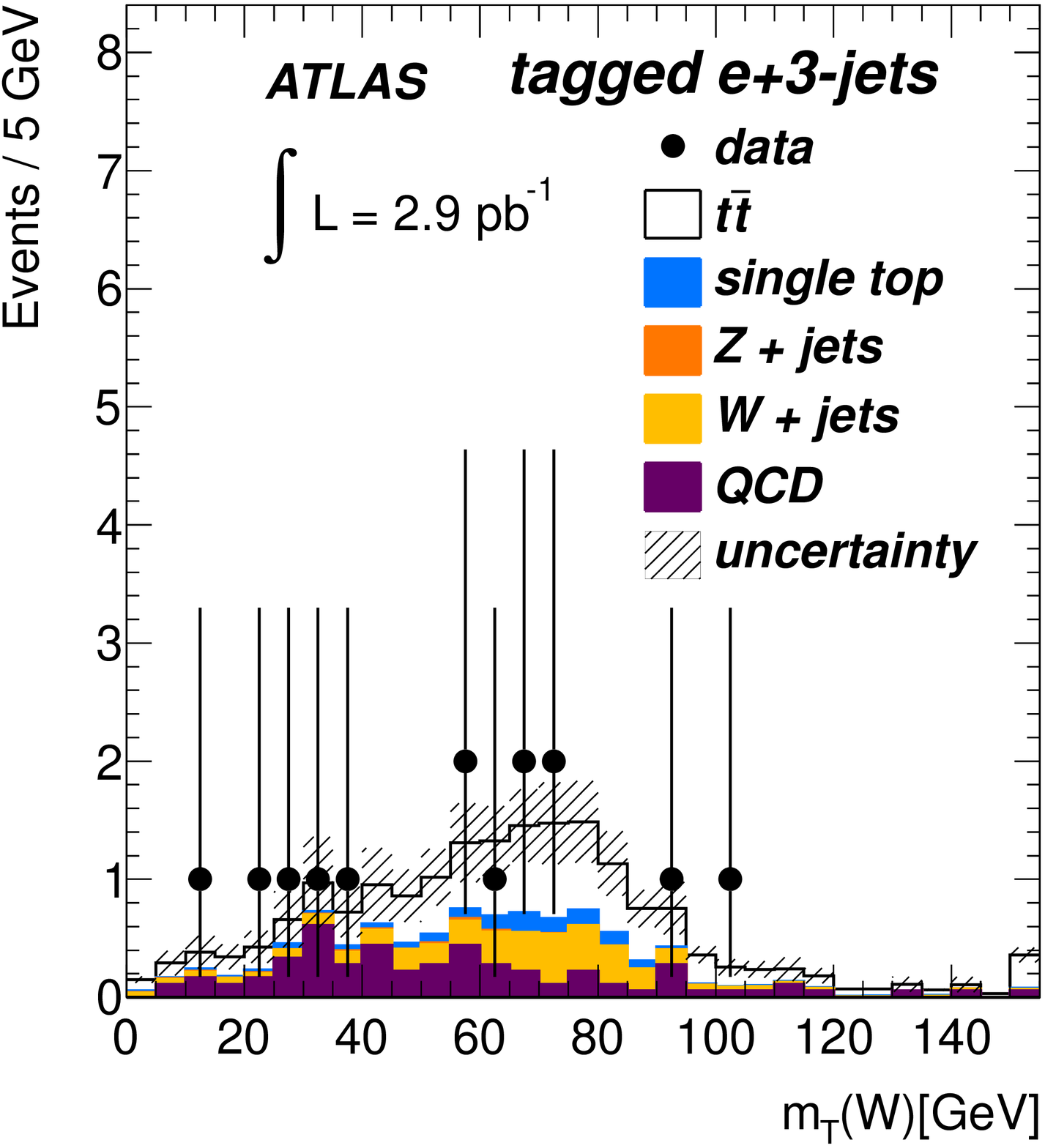} \\
(d) & (e) & (f) \\
\end{tabular}
\end{center}
\vspace{-0.5cm}
\caption{\label{f:mt_control} Distributions of $\mT(W)$.  Top row -
\muplus\ channel : (a) the 1-jet pre-tag sample (where the $\met+ \mT(W)$ requirement is not applied), 
(b) the 2-jet pre-tag sample and (c) the 2-jet
tagged sample. Bottom row - \eplus\ channel: (d) the 1-jet pre-tag
sample, (e) the 2-jet pre-tag sample and (f) the 3-jet tagged sample.  In
each plot data are compared to the sum of the
data-driven QCD estimate plus the contributions from $W$/$Z$+jets and
top from simulation. The background uncertainty on the total expectation is represented by the hatched area. }
\end{figure}

The estimated yields of QCD multi-jet events in the tagged $\mu$ + (1, 2, 3 and $\ge$4-jet),
zero-tag $\mu$ + (3 and $\ge$4-jet) and the pre-tag $\mu$ + (1 and 2-jet)
are summarised in Table~\ref{tab:num_el_mu} (b) and also shown in Table~\ref{tab:WCR}. 
Figure~\ref{f:mt_control} (a) shows the
distribution of $\mT(W)$ for the 1-jet pre-tag sample without
the $\met+\mT(W)$ requirement, while Figures~\ref{f:mt_control} (b) and (c) show
 $\mT(W)$ for the 2-jet pre-tag and for the 2-jet tagged samples respectively
after the $\met+\mT(W)$ requirement. Good agreement is observed comparing the data 
to the estimated rate of QCD multi-jet events summed with the other (non-QCD) 
simulation predictions.

The full QCD multi-jet background estimation procedure has been validated
by applying the procedure on a sample of simulated events and
comparing the result with the known amount of QCD multi-jet background
in the sample. The systematic uncertainty on the \muplus\ multi-jet
background estimate is due to the control region uncertainty
described above, and up to a relative 30\% uncertainty originating from the
method validation studies on the simulation and, for the tagged samples,
the uncertainty originating from the per-event $b$-tagging
probabilities.

\subsubsection{Background estimate in the \eplus\ channel}
\label{s:ejets_fakes}

In the \eplus\ channel, the background consists of both non-prompt electrons
and fake electrons where the latter include both electrons from photon conversion
and misidentified jets with high EM fractions.
The relative magnitude of the non-prompt and fake components is not
well known, as it depends on the details of electron misreconstruction effects
that are not perfectly modelled in the simulation as well as on the fraction of
QCD multi-jet events with non-prompt electrons in the final state. As the ratio also varies with the event kinematics, the
method of Equation~\ref{eq:MM}, which relies on a representative control
region to measure the input values of $\epsf$, is not well suited for the
electron channel.

A method, based on a binned likelihood template fit of the \met\
distribution, is used for the background estimate.  For each
previously defined pre-tag and tagged sample, the data are fitted to a
sum of four templates describing the \met\ distribution of the QCD
multi-jet, \ttbar\ , $W$+jets and $Z$+jets components
respectively. The fit is performed in the region with $\met<20$ GeV
which is complementary to the signal region.  To improve the
statistical precision the requirement on $\met+\mT(W)$ is not applied.
The QCD multi-jet template is extracted from the data as described in
the next paragraph, while the templates for the other processes are
taken from the simulation.  The fraction of QCD multi-jet events in
the signal region is then calculated by extrapolating the expected
fraction of events for each component to the signal region using the
template shape and accounting for the efficiency of the $\met+\mT(W)$
cut for each template.  The output of the fit is $\rho_\textrm{QCD}$,
the predicted fraction of QCD multi-jet events in the signal region,
which is then multiplied by the observed event count.

The templates for the QCD multi-jet \met\ distributions are obtained from two
data control regions. In the first region called `jet-electrons', events are
selected which have, instead of the standard electron, an additional
jet which passes the standard electron kinematic cuts and has at least
4 tracks and an EM fraction of 80-95\%. In the second region
called `non-electrons', the standard event selection is applied, except that
the electron candidate must fail the track quality cut in the innermost layers of
the tracking detector.
Since both control samples are
approximations of the signal region in terms of event kinematics, the
unweighted average of $\rho_\textrm{QCD}$ predicted by the template fits using
the jet-electron and non-electron templates, respectively, is taken for
the QCD multi-jet component. The uncertainty on $\rho_\textrm{QCD}$ has a component from
the template fit uncertainty, a component that quantifies the
uncertainty related to the choice of control region, evaluated as
the difference in $\rho_\textrm{QCD}$ between the two regions divided by $\sqrt{2}$,
and a component related to the
method calibration performed on simulation samples. The latter varies
between 2\% and 36\% depending on the sample.

The results for the QCD multi-jet background contribution to the \eplus{}
channel are summarised in Table~\ref{tab:num_el_mu} (a), and are also shown in
Table~\ref{tab:WCR}.
The estimates for the tagged \eplus\ samples are performed
directly in tagged control samples which have a sufficiently large
number of events, and no per-event $b$-tagging probabilities are used.

Figure \ref{f:mt_control} (bottom row)
shows the distributions of $\mT(W)$ for (d) the $e$ + 1-jet pre-tag,
(e) the $e$ + 2-jet pre-tag, and (f) the $e$ + 3-jet tagged samples.
Acceptable agreement is observed between data
and the sum of the QCD multi-jet background estimated with the
fitting method and the other backgrounds estimated from simulation.

\subsection{$W$+jets background}
\label{s:wjets}

The data-driven estimate for the $W$+jets background
in both electron and muon channels is constructed 
by multiplying the corresponding background contribution in the pre-tag sample by the per-event $b$-tagging probability:

\begin{equation}
\label{eq:Wdd1}
 W^{\ge4\textrm{-jet}}_{\textrm{tagged}} = W^{\ge4\textrm{-jet}}_{\textrm{pre-tag}}
\cdot f_\textrm{tagged}^{\ge4\textrm{-jet}}.
\end{equation}

\noindent Here $W^{\ge4\textrm{-jet}}_{\textrm{pre-tag}}$ is an estimate of the
$W$+jets event count in the pre-tag $\ge$4 jet sample and $f_\textrm{
tagged}^{\ge4\textrm{-jet}}$ is the fraction of these events that are
tagged, calculated as

\begin{equation}
\label{eq:Wdd2}
  f_{\mathrm{tagged}}^{\ge4\textrm{-jet}} = f_{\mathrm{tagged}}^{\textrm{2-jet}} \cdot f^{\mathrm{corr}}_{2 \rightarrow \ge4},
\end{equation}

\noindent where $f_\textrm{tagged}^\textrm{2-jet}$ is a measurement
of the $W$+jets tag fraction in the 2-jet sample and $f^{\mathrm
{corr}}_{2 \rightarrow \ge4}$ accounts for the difference in flavour
composition between the 2-jet and $\ge$4-jet samples as well as
differences in the per-flavour event tagging probabilities, which may
lead to different event rates after $b$-tagging.

For the first ingredient, $W^{\ge4\textrm{-jet}}_{\textrm{pre-tag }}$,
the fact that the ratio of $W$+$n$+1 jets to $W$+$n$ jets is expected
to be approximately constant as a function of $n$ is
exploited~\cite{Ber91, Ellis85}.  This is supported by the good
agreement with the Standard Model expectation as shown in
Figure~\ref{f:jetmult}. The number of $W$ events in the $\ge$4-jet
pre-tag sample can thus be estimated as

\begin{equation}
\label{eq:V}
W^{\ge4\textrm{-jet}}_{\textrm{pre-tag}} = W^{\textrm{2-jet}}_{\textrm{pre-tag}}
  \cdot \sum_{n=2}^{\infty}
  ( W^{\textrm{2-jet}}_{\textrm{pre-tag}}/W^{\textrm{1-jet}}_{\textrm{pre-tag}})^{n},
\end{equation}

\noindent where the sum is used to
extrapolate to a sample with four or more jets. These rates
are obtained by subtracting the estimated non-$W$ boson contributions from
the event count in the pre-tag 1-jet and 2-jet bins.  The QCD multi-jet
contribution is estimated from data as described in Section~\ref{s:QCD}
and simulation-based estimates are used for the other
backgrounds. The scaling behaviour of Equation~\ref{eq:V} does not apply to
$W \to \tau \nu$ events as their selection efficiency depends
significantly on the jet multiplicity. This contribution
is subtracted from the observed event count
in the $W^{\textrm{1-jet}}_{\textrm{pre-tag}}$ and $W^{\textrm{2-jet}}_{\textrm{pre-tag}}$
control samples and  is estimated separately in the electron and the muon channel
using the simulation to predict the ratio of ($W \to \tau \nu$ / $W \to \ell \nu$).
The data-driven technique is used for the estimation of
the $W \to e \nu$ background in the electron channel and the $W \to \mu \nu$
background in the muon channel. 
Table~\ref{tab:WCR} compares the
observed event yields in both the 1-jet and 2-jet samples with the
estimated pre-tag backgrounds for both the electron and muon channels.
Figures~\ref{f:mt_control} (b) and \ref{f:mt_control} (e) show the $m_T(W)$
distribution for the 2-jet pre-tag samples in the muon and electron channels, respectively.

\begin{table}[!htbp]
\begin{center}
\tabcolsep=3mm
\begin{tabular}{|c|c|c|c|c|}
\hline
                    & 1-jet pre-tag $e$                & 1-jet pre-tag $\mu$          &  2-jet pre-tag $e$ & 2-jet pre-tag $\mu$ \\
\hline \hline						  												
 Observed                & 1815                           & 1593                        & 404                 & 370 \\
\hline							  											
QCD multijet (DD)        & $517 \pm 89 $                  & $ 65 \pm 28$                & $190 \pm 43 $       & $20.0 \pm 9.7$ \\
$W$($\tau\nu$)+jets (MC) & $39  \pm 10$                   & $ 43 \pm 11$                & $11.7 \pm 4.4$      & $13.6 \pm 5.1 $  \\
$Z$+jets (MC)            & $19.0  \pm 9.1$                  & $ 48  \pm 12$               & $11.6 \pm 5.2$      & $14.0 \pm 4.8$ \\
$t\bar{t}$ (MC)          & $1.7  \pm 0.8$                 & $ 1.7 \pm 0.8 $             & $7.0 \pm 3.0$       & $7.7  \pm 3.3 $ \\
single-$t$ (MC)          & $4.4  \pm 0.7$                 & $ 5.0 \pm 0.8 $             & $5.2 \pm 0.8$       & $5.1  \pm 0.8 $ \\
diboson (MC)             & $4.8  \pm 4.8$                 & $ 5.7 \pm 5.7 $             & $3.8 \pm 3.8$       & $4.4 \pm 4.4 $ \\
\hline
Total (non $ W(l \nu)$+jets) & $585  \pm 90$               & $168    \pm 33$            & $229 \pm 44 $   & $65  \pm 13$ \\ \hline
Estimated $ W(l \nu)$+jets   & $1230 \pm 100$              & $1425   \pm 52$            & $175 \pm 49 $   & $305 \pm 23$ \\  \hline		
\end{tabular}
\caption{\label{tab:WCR}
Observed event yields in the pre-tag 1-jet and 2-jet samples and estimated contributions from non-$W$ processes and $W \to \tau \nu$.
The estimation for QCD multi-jet events is data-driven~(DD), all other estimates are based on simulation~(MC).
The last row  gives the number of $W(l \nu)$+jet events, estimated
as the observed event count minus all other contributions.
 }
\end{center}
\end{table}

The ratio between the 2-jet and 1-jet rates is measured with
significantly poorer precision in the electron channel, because of the
larger QCD multi-jet contamination.  Since the ratio between the 2-jet and
1-jet rates is expected to be independent of the $W$ boson decay mode, the muon
channel estimation is used also for the electron channel, giving

\begin{equation*}
 W^{\ge4\textrm{-jet}}_{\textrm{pre-tag }}   = 11.2 \pm 2.2 (\mathrm{stat.}) \pm 4.0 (\mathrm{syst.}), \quad {e {\ \rm  channel}},
\end{equation*}
\begin{equation*}
 W^{\ge4\textrm{-jet}}_{\textrm{pre-tag }} = 18.9 \pm 4.1(\mathrm{stat.}) \pm 5.0 (\mathrm{syst.}),\quad {\mu {\ \rm channel}}.
\end{equation*}

\noindent The leading systematic uncertainties are the uncertainty on the
purity of the low jet multiplicity control samples and the uncertainty
associated with the assumption that the ($W$ + $n$ + 1 jets)/($W$ + $n$ jets) ratio is
constant. The latter relative uncertainty has been evaluated to be 24\% from the results
reported in~\cite{alwall}. \\

For the second ingredient, $f_{\mathrm{tagged}}^{\textrm{2-jet}}$, the
pre-tag yield is taken from Table \ref{tab:WCR} and the pre-tag non-$W$ boson
backgrounds (also from Table \ref{tab:WCR}) are subtracted from this
yield. This gives an estimate of the $W$+jets contribution in the
2-jet pre-tag sample. The same is done in the tagged sample: the
estimated non-$W$ boson backgrounds, as shown in Table \ref{tab:num_el_mu},
are subtracted from the measured yield after applying the tagging
criteria resulting in an estimate of the $W$+jets contribution in the
2-jet sample after tagging. The ratio of the tagged to the pre-tag
contributions represents the estimate of the fraction of tagged events
in the 2-jet sample

\begin{equation*}
  f_{\mathrm{tagged}}^{\textrm{2-jet}} = 0.060 \pm 0.018 (\mathrm{stat.}) \pm 0.007 (\mathrm{syst.}).
\end{equation*}

\noindent This quantity is computed from the muon channel only, due to
the large uncertainty originating from the QCD multi-jet contamination in the electron
channel. Figures \ref{f:mt_control} (b) and \ref{f:mt_control} (c) show the distribution of the
transverse mass $m_T(W)$ for the \muplus\ 2-jet pre-tag and tagged samples respectively.
Clear $W$ signals are evident in both samples.

The final ingredient, the correction factor $f^{\mathrm{corr}}_{2
  \rightarrow \ge4}$, is defined as $f^{\mathrm{corr}}_{2 \rightarrow
  \ge4} =
f_{\mathrm{tagged}}^{\ge4\textrm{-jet}}/f_{\mathrm{tagged}}^{\textrm{2-jet}}$. It
is obtained from simulation studies on {\sc Alpgen} $W$+jets events and is
determined to be:

\begin{equation}
  f^{\mathrm{corr}}_{2 \rightarrow \ge4} = 2.8 \pm 0.8 (\mathrm{syst.}).
\end{equation}

\noindent The quoted uncertainty on $f^{\mathrm{corr}}_{2 \rightarrow \ge4}$ reflects
uncertainties on the assumed flavour composition of the pre-tag 2-jet
sample, the uncertainty on the scaling factors for the $b$-tagging efficiency 
for $b$, $c$ and light-quark jets,
and the uncertainty on the ratio of fractions in the
2-jet bin and the $\ge$4-jet bin for $W$+$b\bar{b}$+jets,
$W$+$c\bar{c}$+jets and
$W$+$c$+jets.  The leading uncertainty on $f^{\mathrm{corr}}_{2
\rightarrow \ge4}$ is due to the uncertainty on the
predicted ratios of flavour fractions in the 2-jet and $\ge$4-jet bin.
This is estimated by the variation of several {\sc Alpgen }generator
parameters that are known to influence these ratios~\cite{alpgen}, 
and adds up to a relative 40\%-60\%
per ratio. The uncertainty on the flavour composition in the 2-jet
bin, while large in itself, has a small effect on $f^{\mathrm{corr}}_{2
\rightarrow \ge4}$ due to effective cancellations in the ratio.

Applying Equation~(\ref{eq:Wdd1}) and Equation~(\ref{eq:Wdd2}) the estimated
yields for $W$+jets in the $\ge$4-jet tagged samples are

\begin{equation*}
 W^{\ge4\textrm{-jet}}_{\textrm{tagged}} = 1.9 \pm 0.7 (\mathrm{stat.}) \pm 0.9 (\mathrm{syst.}), \quad {e {\ \rm channel}},
\end{equation*}
\begin{equation*}
 W^{\ge4\textrm{-jet}}_{\textrm{tagged}} = 3.2 \pm 1.2 (\mathrm{stat.}) \pm 1.2 (\mathrm{syst.}),\quad {\mu {\ \rm channel}}.
\end{equation*}

\noindent as reported in Table \ref{tab:num_el_mu}.

\subsection{Cross-section measurement}\label{s:xs}
\subsubsection{Counting-based measurement of the cross-section in the $\ge$4-jet bin}
\label{s:counting}

In the $\ge$4-jet tagged sample the
\ttbar\ signal yield is obtained by subtracting the estimated rate of
all backgrounds from the observed event yield. This method depends
crucially on the understanding of the background, but makes minimal
assumptions on \ttbar\ signal properties for the yield calculation. For
the QCD multi-jet and $W$+jets backgrounds, the data-driven estimates
described in detail in Sections \ref{s:QCD} and \ref{s:wjets} are
used, while for the expected background from $Z$+jets and single-top
production, simulation estimates are used.  Table~\ref{tab:num_el_mu}
shows the complete overview of background contributions that are used
in this calculation. The observed yields, the total expected
background yields and the resulting \ttbar\ signal yields for the
{\eplus} , \muplus\ and combined channels are shown in
Table~\ref{t:sigyield}.  

\begin{table}[hbt]
\centering
\begin{tabular}{|l|c|c|c|}\hline
                          &  \eplus\                      & \muplus\                    &  combined            \\ \hline \hline
Observed                  &  17                       &  20                      &   37 \\ \hline
Total est. background     & $7.5  \pm 3.1$            &  $4.7 \pm 1.7$           &  $12.2 \pm 3.9$  \\ \hline \hline
 \ttbar                   & $9.5  \pm 4.1 \pm 3.1$    &  $15.3 \pm 4.4 \pm 1.7$  &  $24.8 \pm 6.1 \pm 3.9$  \\ \hline
\end{tabular}
\caption{\label{t:sigyield} Observed event yield, estimated total
background and \ttbar\ signal using the counting method in the
$b$-tagged $\ge$4-jet bin, for electrons and muons separately and
combined. The total background consists of the sum of individual
backgrounds listed in Table \ref{tab:num_el_mu}, choosing the
data-driven estimate for $W$+jets (instead of the simulation-based
$W$+jets estimate used in the `total (non-\ttbar\ )' row of Table
\ref{tab:num_el_mu}). The uncertainty on the total background includes
statistical uncertainties in control regions and systematic
uncertainties. The first quoted uncertainty on the \ttbar\ signal
yield is statistical, while the second is from the systematics on the
background estimation.}
\end{table}

\noindent 
The product of acceptance and branching fraction of \ttbar\ events in
the $\ge$4-jet tagged signal region, measured from Monte-Carlo samples
and quoted in Section \ref{s:ljets_selection}, is used together with
the value of the integrated luminosity to extract the cross-section
($\sigma_{\ttbar}$) from the observed event yield. The resulting
cross-sections are shown in Table~\ref{t:sigxs}. 

Table~\ref{tab:xssyst} provides a detailed breakdown of the total
systematic uncertainties on the cross-section for this method. The
components listed under `Object selection' relate to sources discussed
in Section \ref{s:objsyst}. The components listed under `Background
rates' relate to the uncertainties on background estimates detailed in
Sections \ref{s:QCD} and \ref{s:wjets}. The components listed under
`Signal simulation' relate to sources discussed in
Section~\ref{s:mcsyst}.  The largest systematic uncertainty is due to
the normalisation of the QCD multi-jet background in the \eplus{}
channel, followed by the uncertainties which affect mainly the \ttbar\
acceptance, like jet energy reconstruction, $b$-tagging and
ISR/FSR. The dependence of the measured cross-section on the assumed
top-quark mass is small. A change of $\pm 1$\GeV in the assumed
top-quark mass results in a change of $\mp 1$\% in the cross-section.

\begin{table}[hbt]
\begin{center}
\small
\begin{tabular}{| l | c | c |}
\hline
      & \multicolumn{2}{c|}{Relative cross-section uncertainty [\%]}   \\
\hline
   Source  & \hspace{0.6cm} \eplus\ \hspace{0.6cm}  & \hspace{0.6cm} \muplus\ \hspace{0.6cm} \\ \hline\hline
Statistical uncertainty       &  $\pm43$      &   $\pm29$      \\
\hline
{\em Object selection}  & &  \\
Lepton reconstruction, identification, trigger  &  $\pm3$     &   $\pm2$          \\

Jet energy reconstruction         &  $\pm13$    &   $\pm11$          \\
$b$-tagging             &  -10 / +15    & -10 / +14    \\
\hline
{\em Background rates }  & &   \\
 QCD normalisation               &  $\pm30$    &  $\pm2$       \\
 $W$+jets normalisation            &  $\pm11$      &  $\pm11$          \\
 Other backgrounds normalisation         &  $\pm1$     &  $\pm1$       \\
\hline
{\em Signal simulation}      & & \\
Initial/final state radiation  & -6 / +13  & $\pm8$    \\
  Parton distribution functions                       &   $\pm2$     &   $\pm2$       \\
Parton shower and hadronisation &   $\pm1$     &   $\pm3$          \\
Next-to-leading-order generator                &   $\pm4$     &   $\pm6$            \\
\hline
Integrated luminosity       & -11 / +14   & -10 / +13   \\ \hline
\hline
Total systematic uncertainty  & -38 / +43         & -23 / +27        \\
\hline
Statistical + systematic uncertainty   & -58  / +61   &  -37 / +40 \\ 
\hline
\end{tabular}
\end{center}
\caption{\label{tab:xssyst} Summary of individual systematic
uncertainty contributions to the single-lepton cross-section
determination using the counting method. The combined uncertainties
listed in the bottom two rows include the luminosity uncertainty.
}
\end{table}

\begin{figure}[htbp]
\begin{center}
\begin{tabular}{cc}
\includegraphics[width=0.33\textwidth]{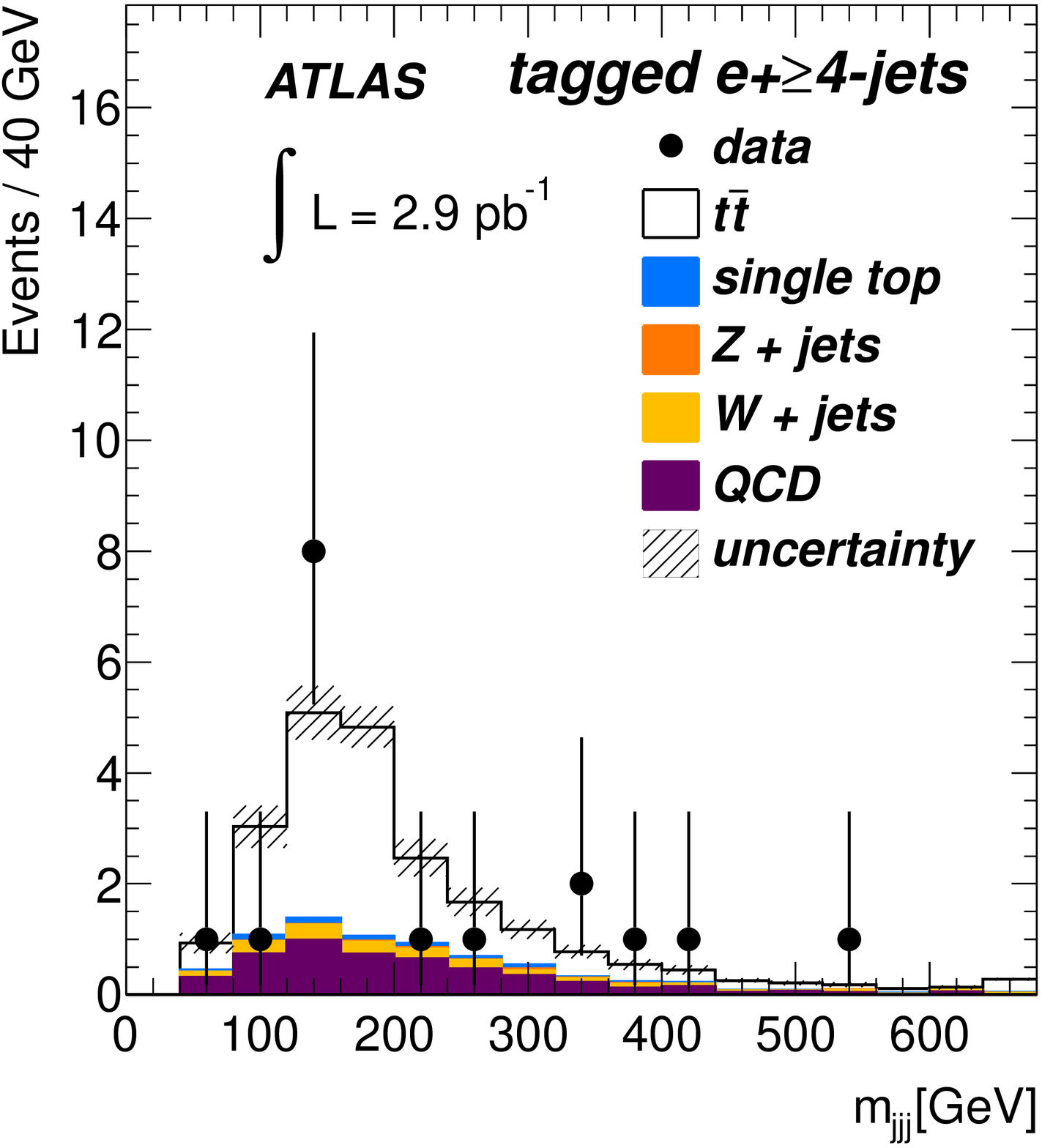} &
\includegraphics[width=0.33\textwidth]{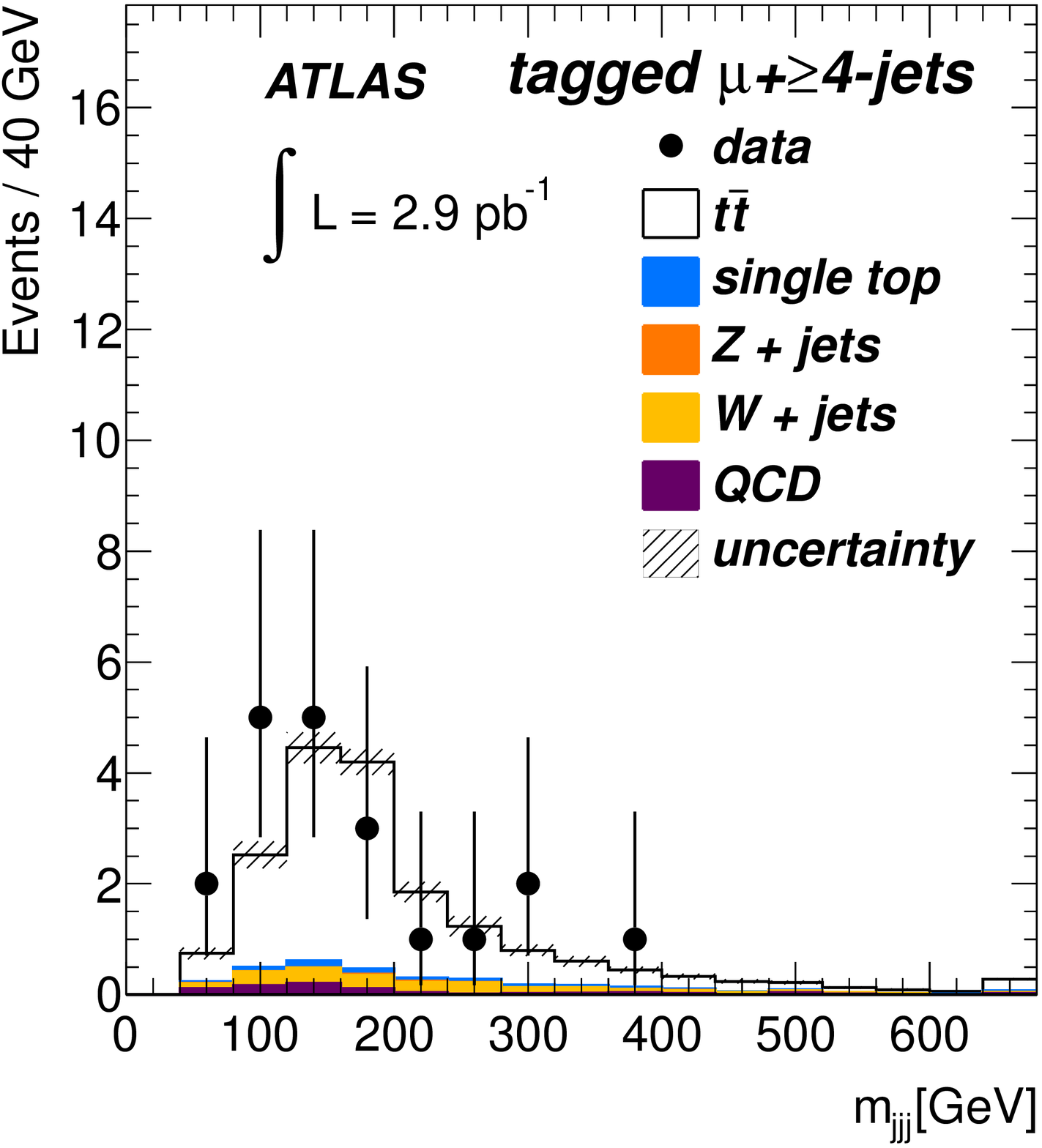} \\
(a) & (b) \\
\includegraphics[width=0.33\textwidth]{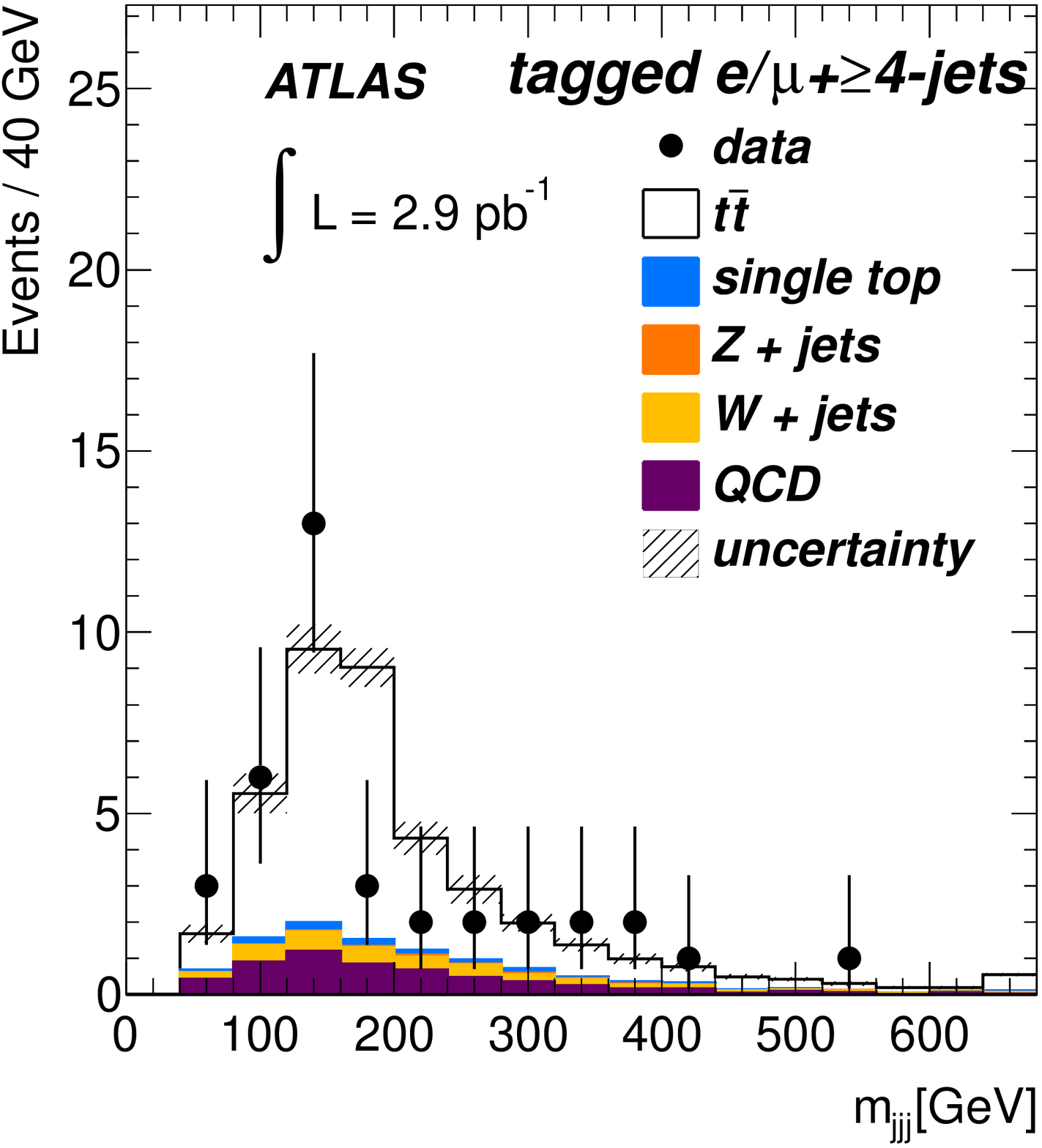} &
\includegraphics[width=0.33\textwidth]{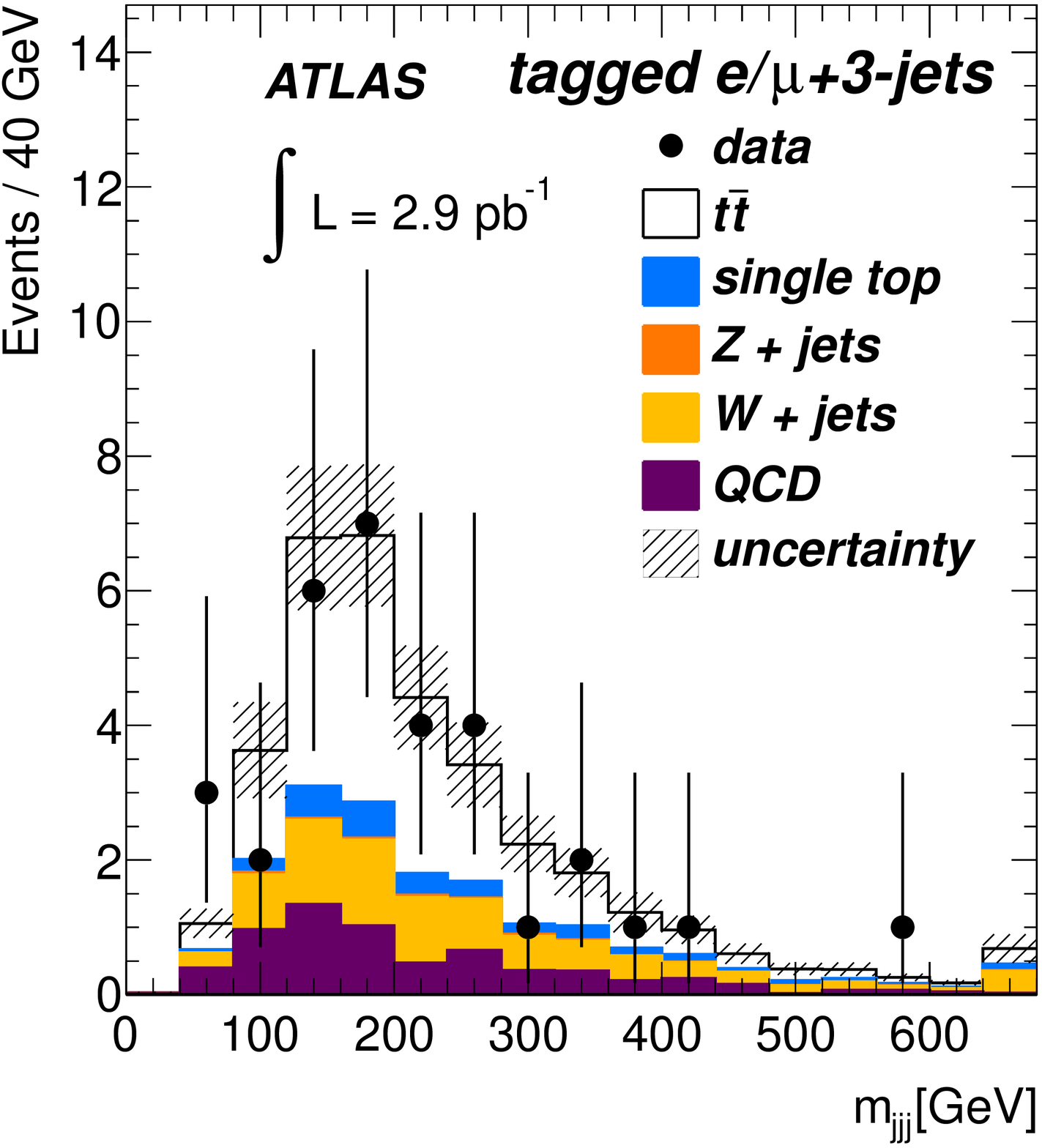} \\
 (c) & (d)\\
\end{tabular}
\caption{\small {Distributions of the invariant mass of the 3-jet combination having the highest $\pT$
for (a) the $\ge$4-jet tagged \eplus\ sample, (b) the $\ge$4-jet tagged
\muplus\ sample, (c) the $\ge$4-jet tagged samples combined and (d)
the combined 3-jet tagged sample. The data is compared to the sum of all
expected contributions. For the totals shown, simulation estimates
are used for all contributions except QCD multi-jet, where a
data-driven technique is used. The background
uncertainty on the total expectation is represented by the hatched
area.}}
\label{fig:Mjjj}
\end{center}
\end{figure}

While not used in the counting method, further information can be
gained from the use of kinematic event properties: in the $\ttbar$
candidate events, three of the reconstructed jets are expected to come
from a top quark which has decayed into hadrons.  Following
\cite{CSCbook}, the hadronic top quark candidate is empirically defined as
the combination of three jets (with $\pT>20\GeV$) having the highest
vector sum $\pT$. This algorithm does not make use of the $b$-tagging
information and selects the correct combination of the reconstructed
jets in about 25\,\% of cases.  The observed distributions of the invariant
mass ({\mjjj}) of the hadronic top quark candidates in the various
$\ge$4-jet samples, shown in
Figures~\ref{fig:Mjjj} (a) - \ref{fig:Mjjj} (c), demonstrate good
agreement between the data and the signal+background
expectation. Figure~\ref{fig:Mjjj}d highlights a substantial
contribution of \ttbar\ signal events in the 3-jet tagged sample and
demonstrates further information which is also not exploited by
the baseline counting method.

\subsubsection{Fit based cross-section measurement in the 3-jet and $\geq$4-jet samples}
\label{sec:fits}

A complementary approach to measuring the cross-section exploits the
data in both the 3-jet and $\geq$4-jet samples.  With the current data
sample, it gives an important cross-check of the counting method, as
it makes different physics assumptions for the signal and background
modelling.  This technique is expected to become more powerful once
more integrated luminosity has been collected.

In the first approach (A), the tagged 3-jet and $\geq$4-jet
samples are used. The \mjjj\ distribution for each sample is described
by the sum of four templates for {\ttbar}, $W$+jets, QCD multi-jet and
other backgrounds respectively.  This method fits simultaneously the
\ttbar\ and $W$+jets components, relying mostly on shape information.
The shapes of the templates for {\ttbar}, $W$+jets and smaller
backgrounds are taken from simulation. The template for the QCD multi-jet
background is taken from a data sample using a
modified lepton definition, which requires at least one of the
selection criteria listed in Section \ref{s:obj} to fail.  A
constraint,
similar to the $f^{\mathrm{corr}}_{2\rightarrow \ge4}$ 
correction factor discussed in Section \ref{s:wjets},
is introduced in the ratio of the $W$+jets yields in the
3-jet and $\geq$4-jet samples,
which reduces the uncertainty on the extracted signal yield.  Additionally,
the $W$+jets yields in the \eplus\ and
\muplus\ channels are related by their respective acceptances.

In the second approach (B), the tagged and zero-tag $\geq$4-jet samples
are used, with a template describing the sum of all backgrounds
in each of these two samples. The fraction of background events that are 
tagged in
the $\ge$4-jet bin is constrained in the fit to a prediction based on
the measured tagged fraction in the 3-jet sample and includes a
simulation-based correction for the expected difference between the
3-jet and $\ge$4-jet bins. The template for \ttbar\ and the relative
contributions to the different samples are taken from simulation,
while the template for the background is taken from a QCD multi-jet enhanced
sample in data. The assumed rate of \ttbar\ events in the 3-jet bin is
iteratively adjusted to the measured cross-section.

\subsubsection{Results}

The cross-sections obtained with the baseline counting method in the \eplus\
and \muplus\ channels are shown in Table \ref{t:sigxs}.  The fit
methods make different assumptions about the signal and background and
therefore serve as good cross-checks; their cross-sections are also
shown in Table \ref{t:sigxs} and are in good agreement with those
obtained from the baseline counting method.  Additionally, the estimate for the
$W$+jets background in $\ge$4-jet tagged sample as measured in fit A
is in agreement with the estimate quoted in Section
\ref{s:wjets}. 
\mbox{Table \ref{t:sigxs}} also shows the cross-section obtained with
the counting method for the \eplus\ and \muplus\ channels,
combined using the procedure described in Section~\ref{s:combination}. 
For the fit methods, the combined cross-sections
are obtained from a simultaneous fit to the electron and muon samples.

The systematic uncertainties of both fit-based methods are dominated
by acceptance-related systematic uncertainties. Compared to the
counting method, both fit-based techniques have a reduced sensitivity
to the QCD multi-jet background rate but have method specific systematics: the
ratio of tagged $W$+jets in the 3-jet and $\geq$4-jet bins and
shape-modelling uncertainties for fit A, and the modelling of the
$b$-tagged fraction for fit B. This trade-off results in a
comparable total uncertainty for both methods compared to the
counting method.

\begin{table}[hbt]
\centering
\begin{tabular}{|l|c|c|c|}\hline

  Method                 & \eplus                       & \muplus                      & $e$/$\mu$ +jets combined \\
\hline\hline
& & & \\[-1.5ex]
Counting $\sigma_{\ttbar}$ [pb]   & $105 \pm46$ $^{+45}_{-40}$   & $168 \pm49$ $^{+46}_{-38}$   & $142 \pm 34$ $^{+50}_{-31}$ \\[1.ex]
\hline\hline
& & & \\[-1.5ex]
Fitted $\sigma_{\ttbar}$(A) [pb]  & $98 \pm 58$ $^{+34}_{-28}$   & $167 \pm 68$  $^{+46}_{-39}$ & $130 \pm 44$ $^{ +38}_{ -30}$       \\[1.ex]
Fitted $\sigma_{\ttbar}$(B) [pb]  & $110 \pm 50 \pm 39  $    & $134 \pm 52 \pm 39 $  & $118 \pm 34 \pm 34 $       \\[1.ex]
\hline
\end{tabular}
\caption{\label{t:sigxs} Inclusive \ttbar\ cross-section measured in the 
single-lepton channel using
the counting method and the template shape fitting techniques (A and B).
The uncertainties represent respectively the statistical and systematic 
uncertainty including luminosity. The top row shows the counting-method results 
that are used for the combination presented in Section \ref{s:combination}.}
\end{table}

\section{Dilepton analysis}
\label{dilepton.section}

\newcommand{\DYZeeNJetsTwoJet}{0.25 $\pm$ 0.18}
\newcommand{\DYZmmNJetsTwoJet}{0.67 $\pm$ 0.38}
\newcommand{\ZtteeNJetsTwoJet}{0.07 $\pm$ 0.04}
\newcommand{\ZttemNJetsTwoJet}{0.13 $\pm$ 0.06}
\newcommand{\FakeWeeNJetsTwoJet}{0.16 $\pm$ 0.18}
\newcommand{\FakeWmmNJetsTwoJet}{-0.08 $\pm$ 0.07}
\newcommand{\FakeWemNJetsTwoJet}{0.47 $\pm$ 0.28}
\newcommand{\singletopeeNJetsTwoJet}{0.08 $\pm$ 0.02}
\newcommand{\singletopmmNJetsTwoJet}{0.07 $\pm$ 0.03}
\newcommand{\singletopemNJetsTwoJet}{0.22 $\pm$ 0.04}
\newcommand{\dibosoneeNJetsTwoJet}{0.04 $\pm$ 0.02}
\newcommand{\dibosonmmNJetsTwoJet}{0.07 $\pm$ 0.03}
\newcommand{\dibosonemNJetsTwoJet}{0.15 $\pm$ 0.05}
\newcommand{\ttbarmmNJetsTwoJet}{1.87 $\pm$ 0.26}
\newcommand{\ttbaremNJetsTwoJet}{3.85 $\pm$ 0.51}
\newcommand{\DataeeNJetsTwoJet}{2}
\newcommand{\DatammNJetsTwoJet}{3}
\newcommand{\DataemNJetsTwoJet}{4}

\subsection{Event selection}
\label{s:dilepton_selection}

The dilepton $\ttbar$ final state is characterized by two isolated
leptons with relatively high $p_T$, missing transverse energy
corresponding to the neutrinos from the $W$ leptonic decays, and two
$b$ quark jets.  The selection of events in the signal region for the
dilepton analysis consists of a series of kinematic requirements on
the reconstructed objects defined in Section~\ref{s:obj}:

\begin{itemize}

\item Exactly two oppositely-charged leptons ($ee$, $\mu\mu$ or
  $e\mu$) each satisfying $\pT>20\GeV$, where at least one must be
  associated to a leptonic high-level trigger object;

\item At least two jets with $\pT>20\GeV$ and with $|\eta|<2.5$ are
  required, but no $b$-tagging requirements are imposed;

\item To suppress backgrounds from $Z$+jets and QCD multi-jet events
  in the $ee$ channel, the missing transverse energy must satisfy
  $\MET>40\GeV$, and the invariant mass of the two leptons must differ by at
  least $5\GeV$ from the $Z$ boson mass, {\em i.e.}\/
  $|m_{ee}-m_Z|>5\GeV$. For the muon channel, the corresponding
  requirements are $\MET>30\GeV$ and $|m_{\mu\mu}-m_Z|>10\GeV$;

\item For the $e\mu$ channel, no \MET\ or
  $Z$ boson mass veto cuts are applied. However, the event \HT, defined as
  the scalar sum of the transverse energies of the two leptons and all
  selected jets, must satisfy $\HT>150\GeV$ to suppress
  backgrounds from $Z$+jets production;

\item To remove events with cosmic-ray muons, events with two
  identified muons with large, oppositely signed transverse impact
  parameters ($d_0 > 500~\mu m$) and consistent with being back-to-back
  in the $r-\phi$ plane are discarded.

\end{itemize}

\begin{figure}[htbp]
  \centering
  \subfigure[]{ \includegraphics[width=0.315\textwidth]{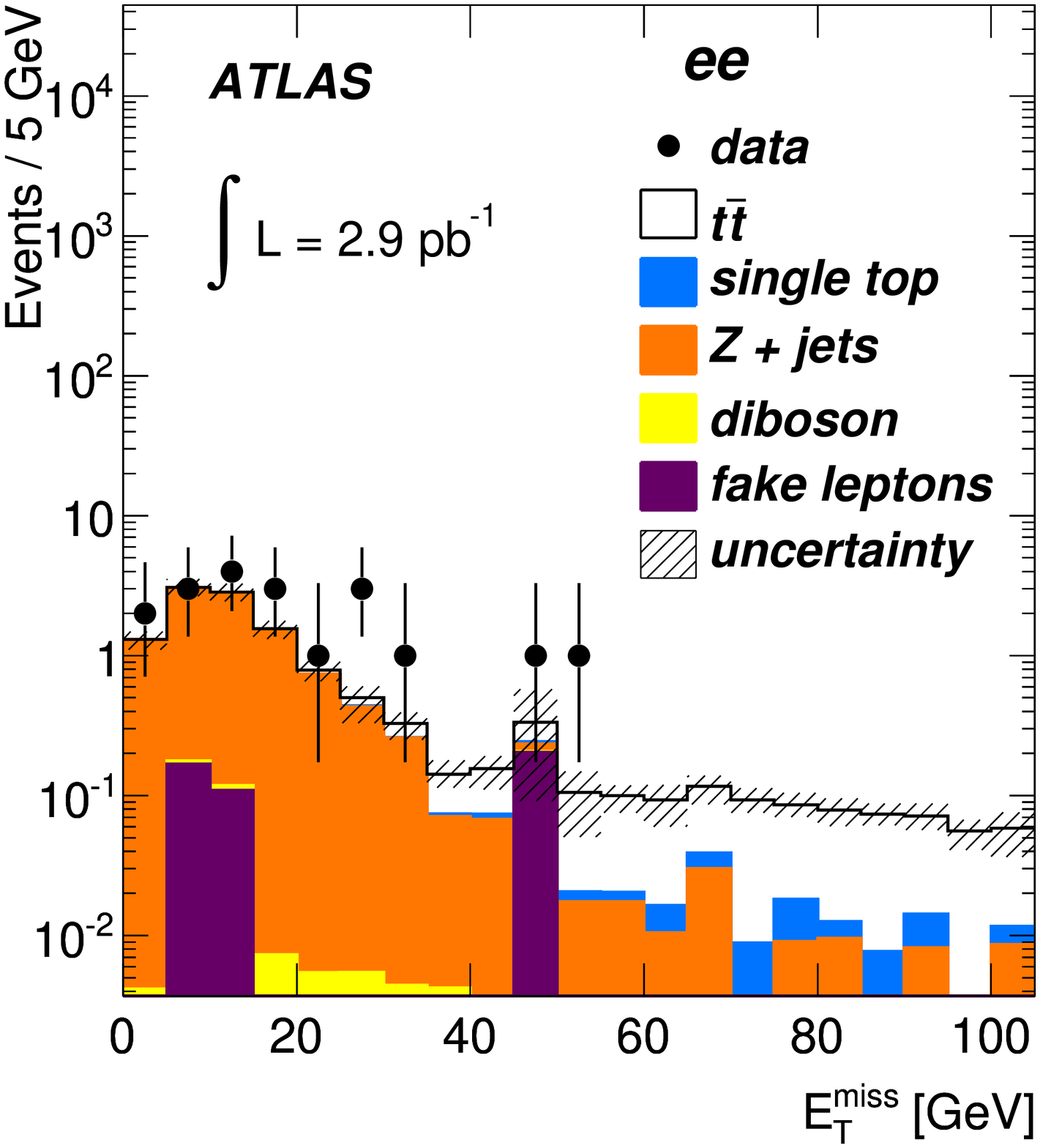} }
  \subfigure[]{ \includegraphics[width=0.315\textwidth]{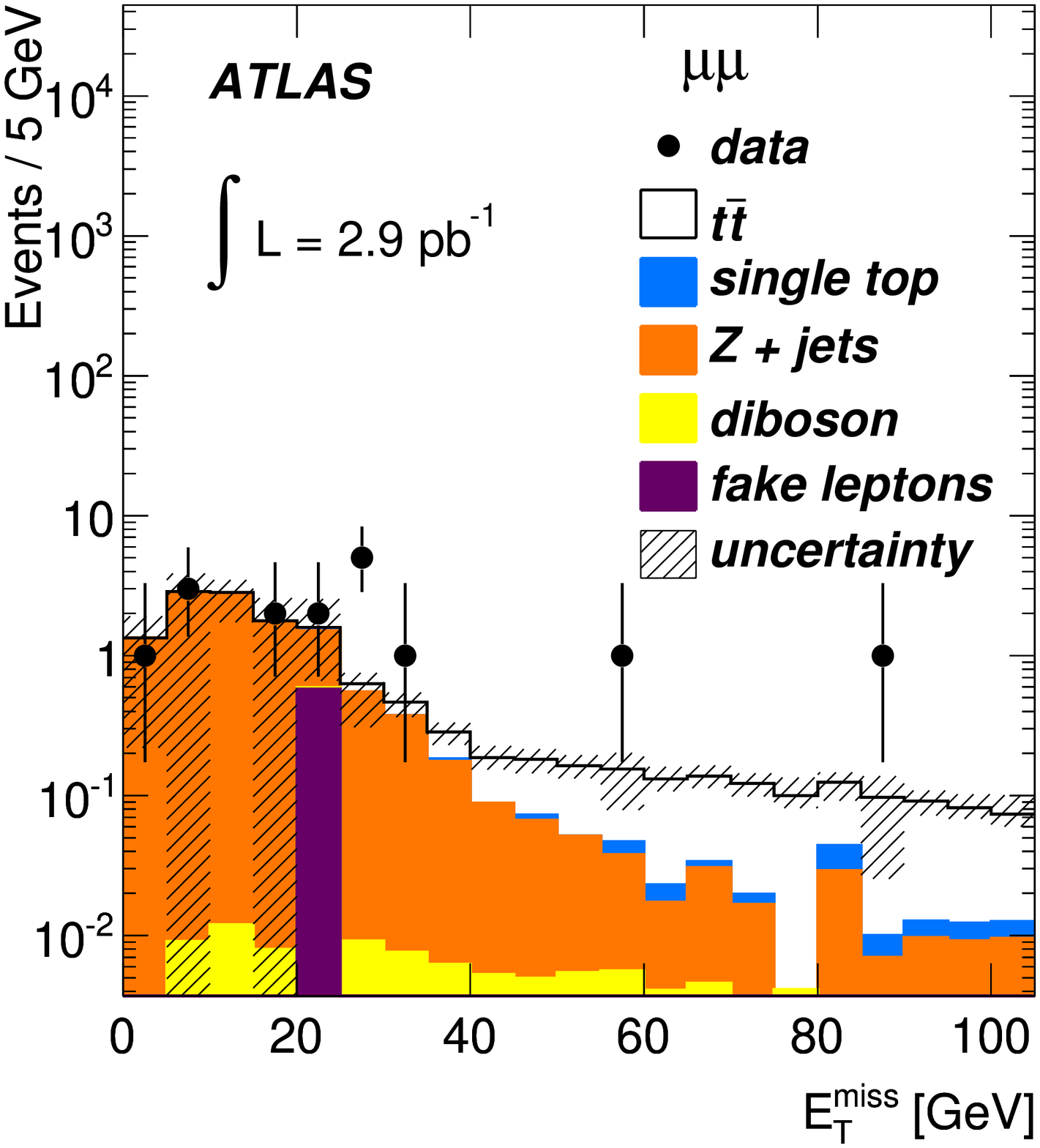} }
  \subfigure[]{ \includegraphics[width=0.315\textwidth]{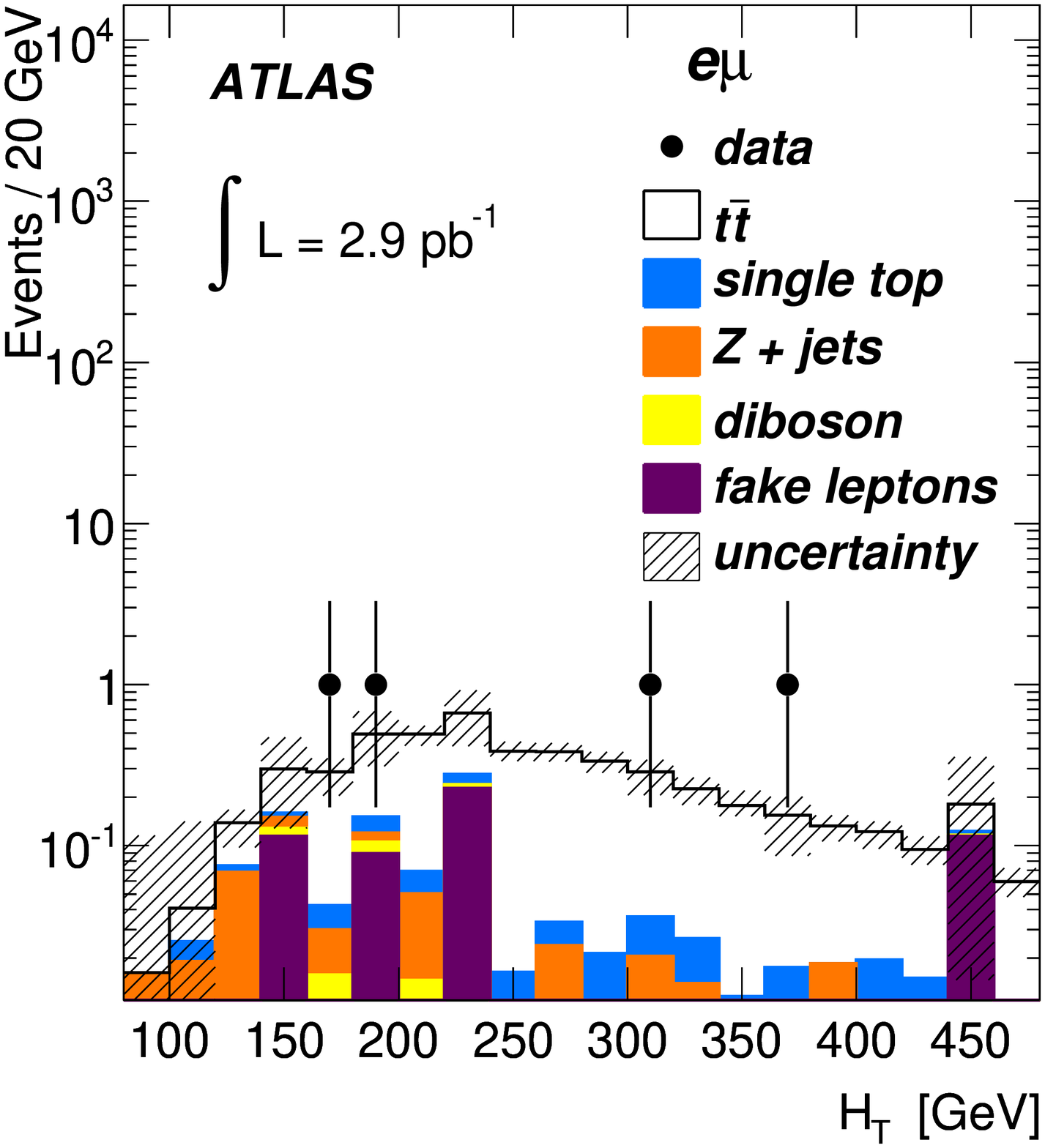} }
  \caption{
The $\met$ distribution in the signal region for (a) the $ee$ channel without the $\MET>40\GeV$ requirement,
(b) the $\mu\mu$ channel without the $\MET>30\GeV$ requirement, and (c)
the distribution of the $\HT$, defined as
the scalar sum of the transverse energies of the two leptons and all
selected jets, in the signal region
without the $\HT>150\GeV$ requirement.
}
  \label{f:ll_met_ht}
\end{figure}

\begin{figure}[htbp]
  \centering
  \subfigure[]{ \includegraphics[width=0.315\textwidth]{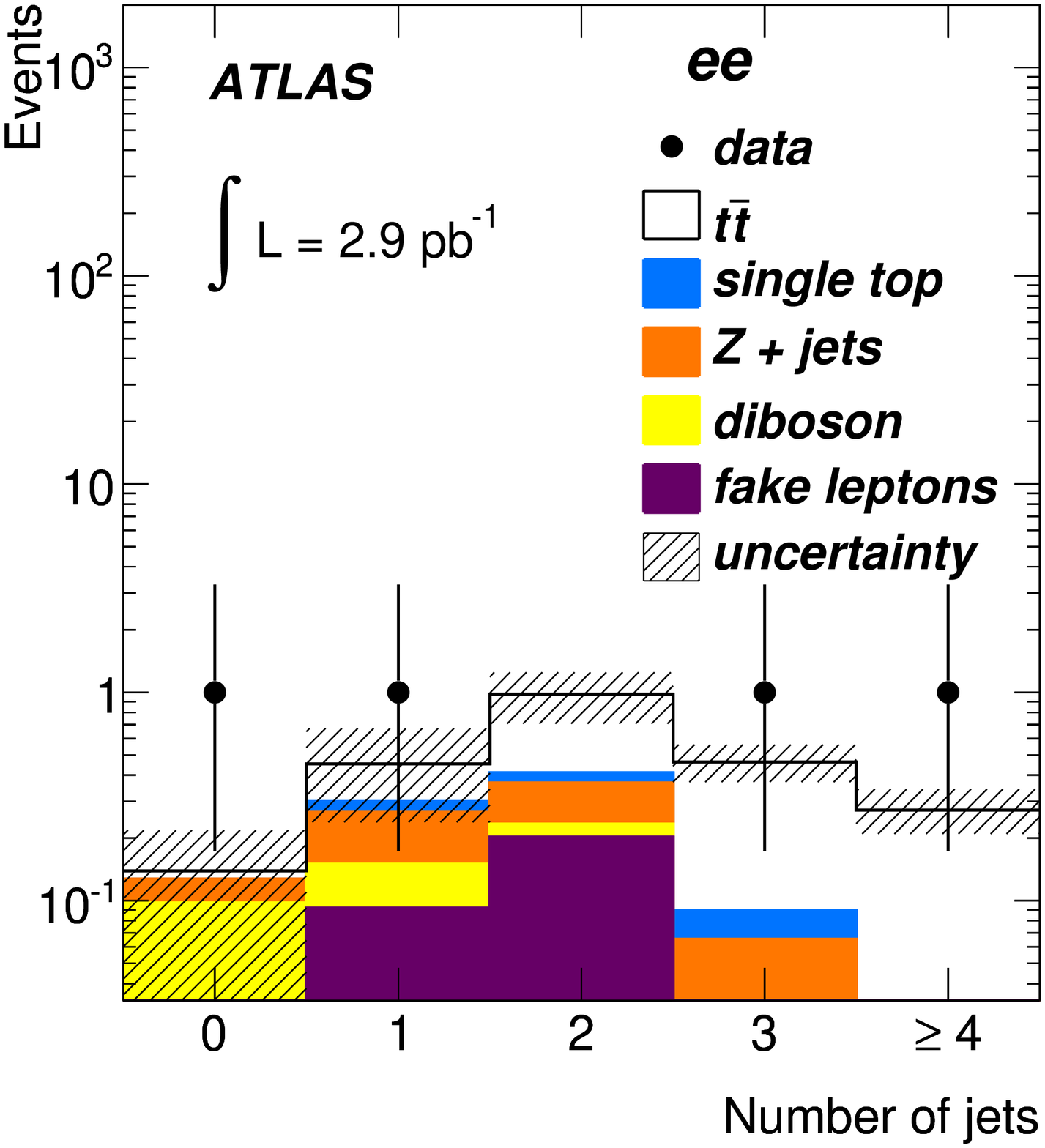} }
  \subfigure[]{ \includegraphics[width=0.315\textwidth]{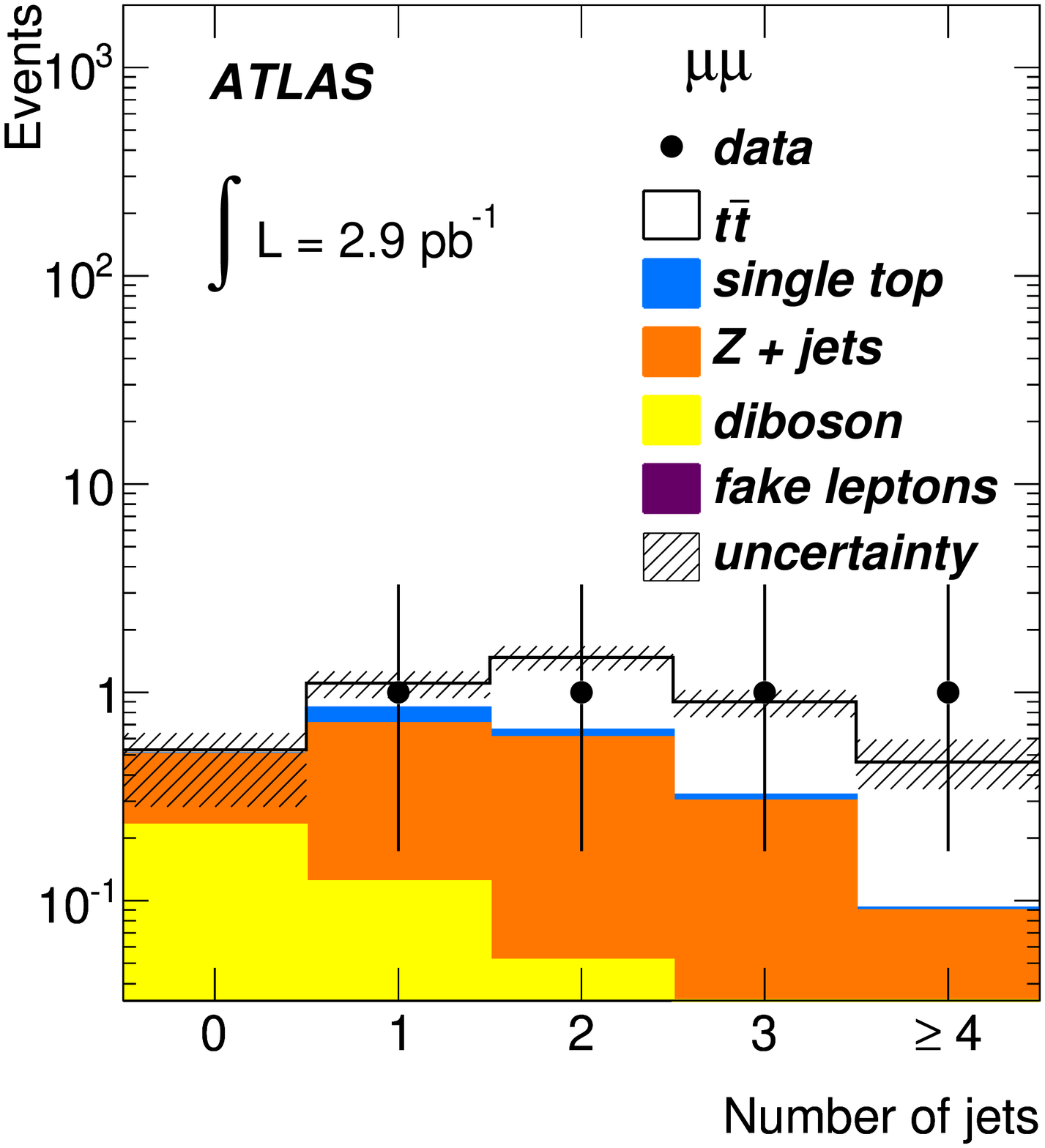} }
  \subfigure[]{ \includegraphics[width=0.315\textwidth]{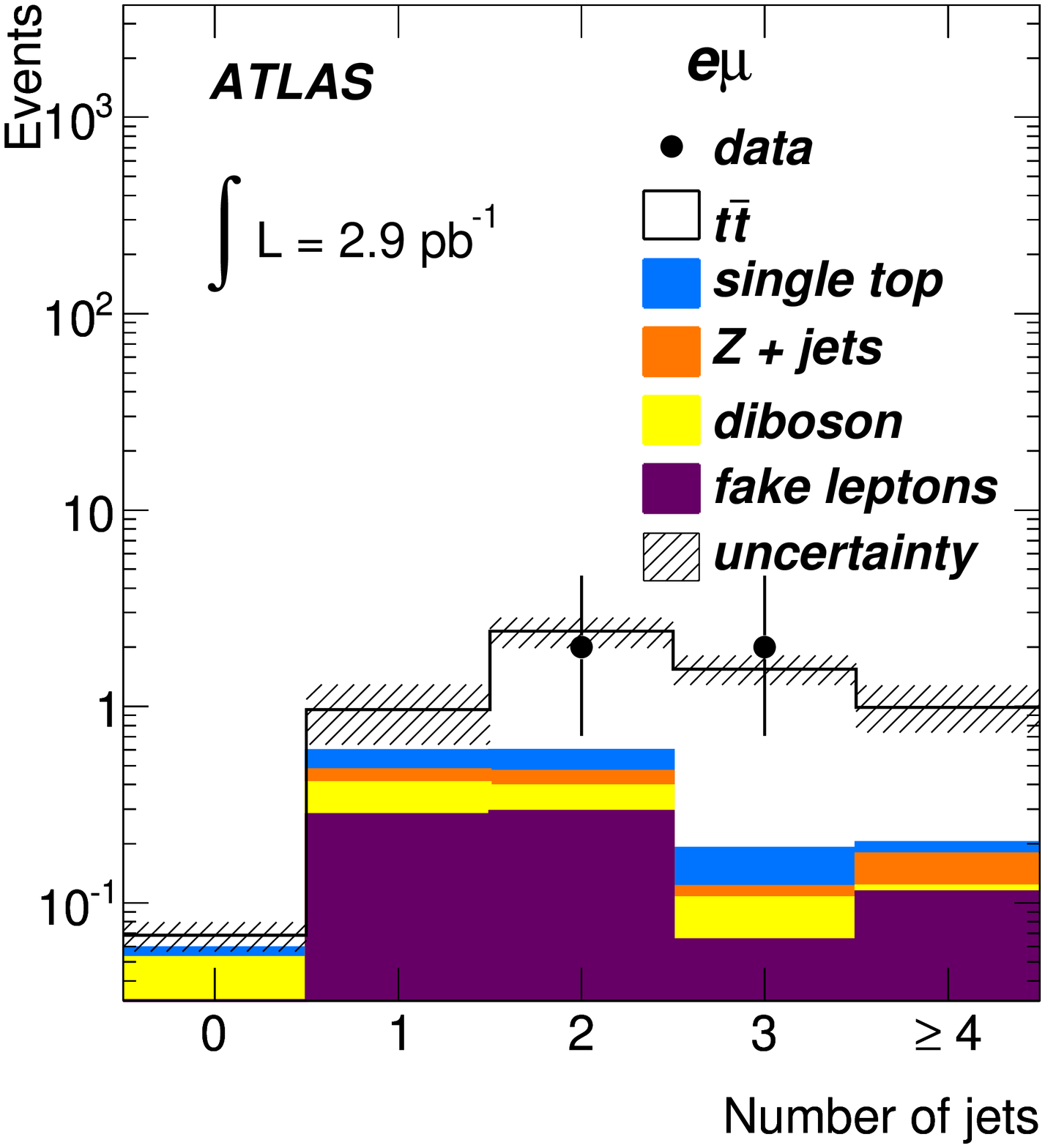} }
  \caption{ Jet multiplicities for the signal region omitting the $N_{jets}\ge 2$ requirement
 in (a) the $ee$ channel, (b) the $\mu\mu$ channel and (c) the $e\mu$ channel.}
  \label{f:ll_njets}
\end{figure}

\begin{figure}[htbp]
  \centering
  \subfigure[]{ \includegraphics[width=0.315\textwidth]{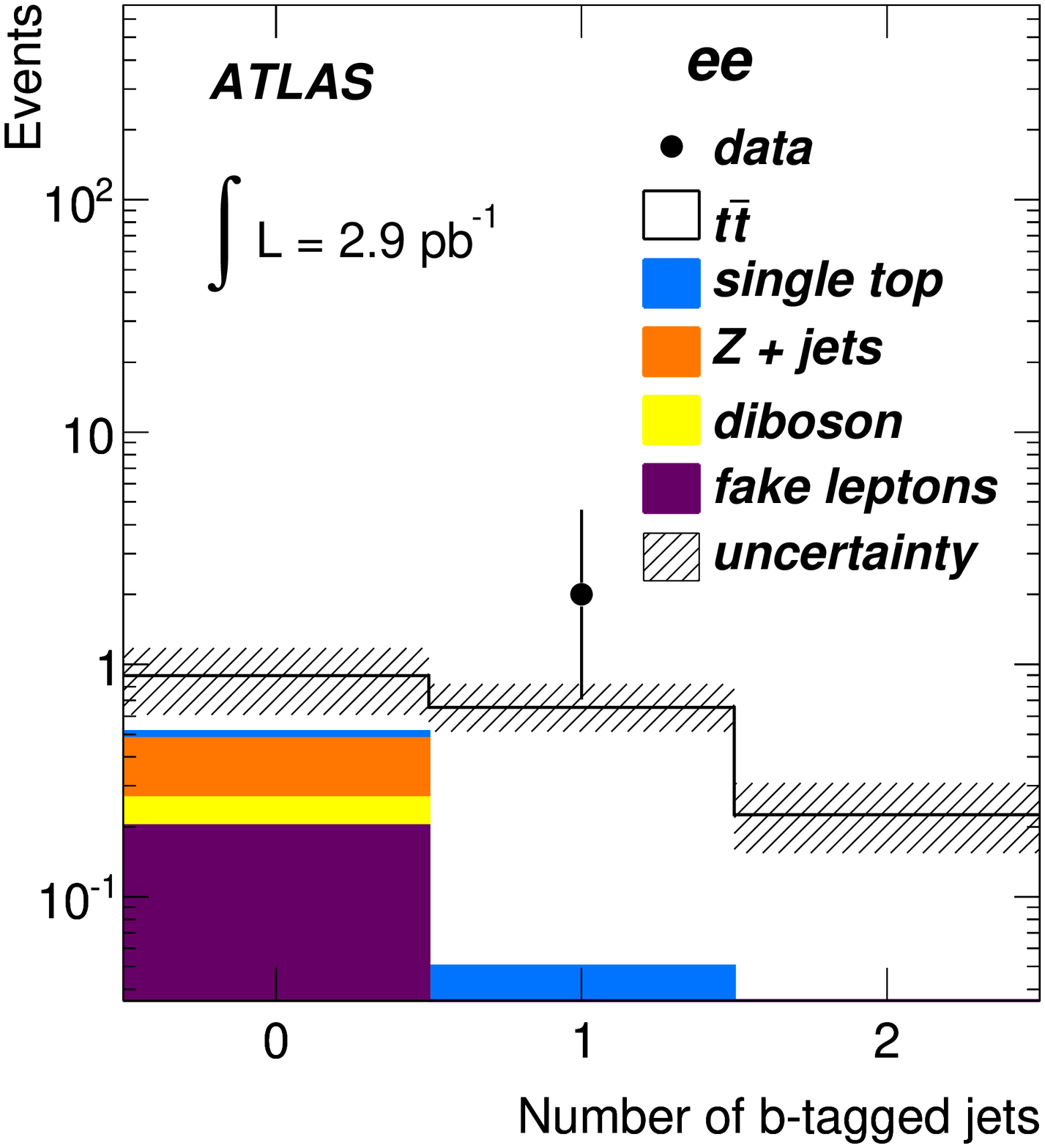} }
  \subfigure[]{ \includegraphics[width=0.315\textwidth]{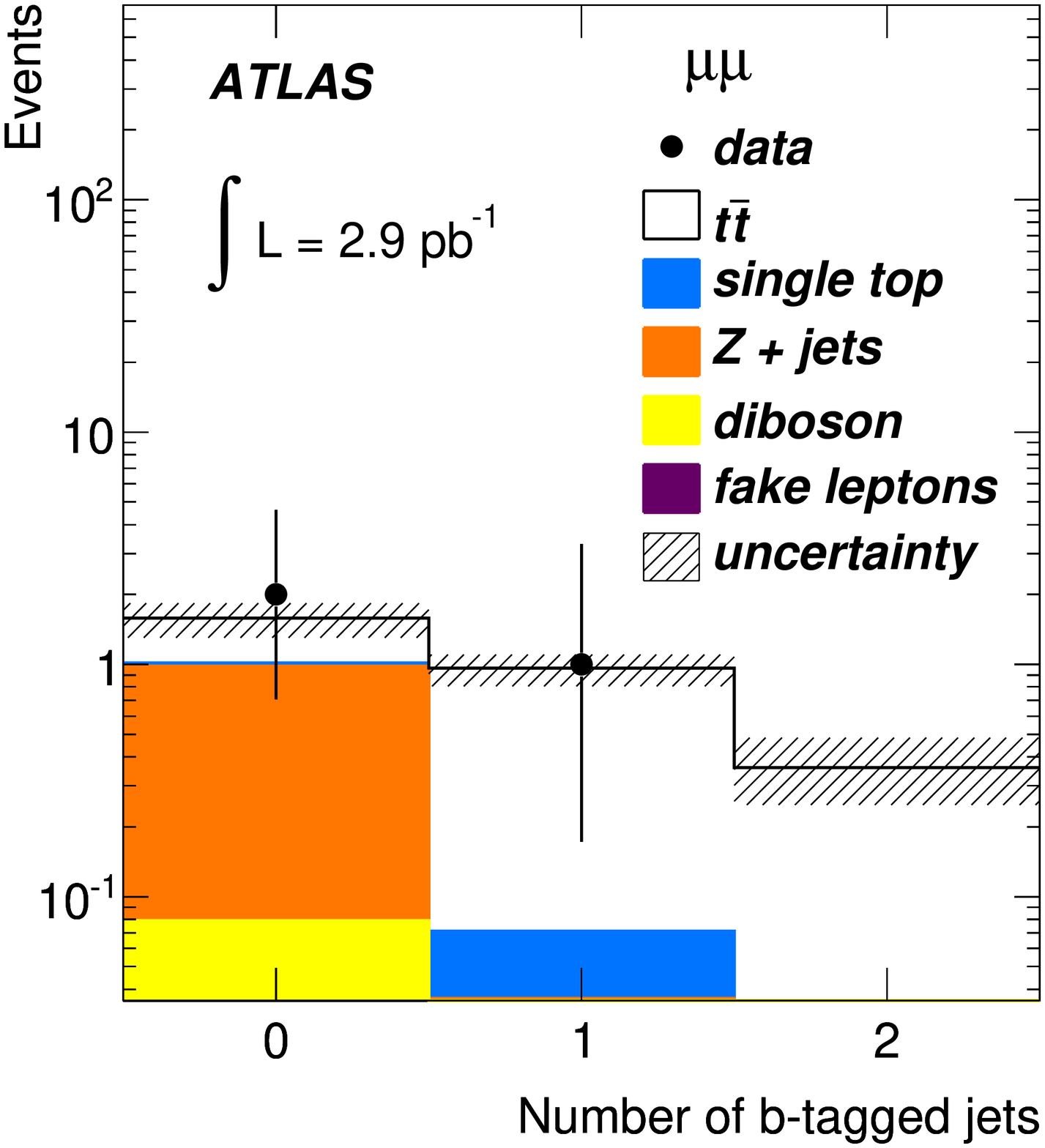} }
  \subfigure[]{ \includegraphics[width=0.315\textwidth]{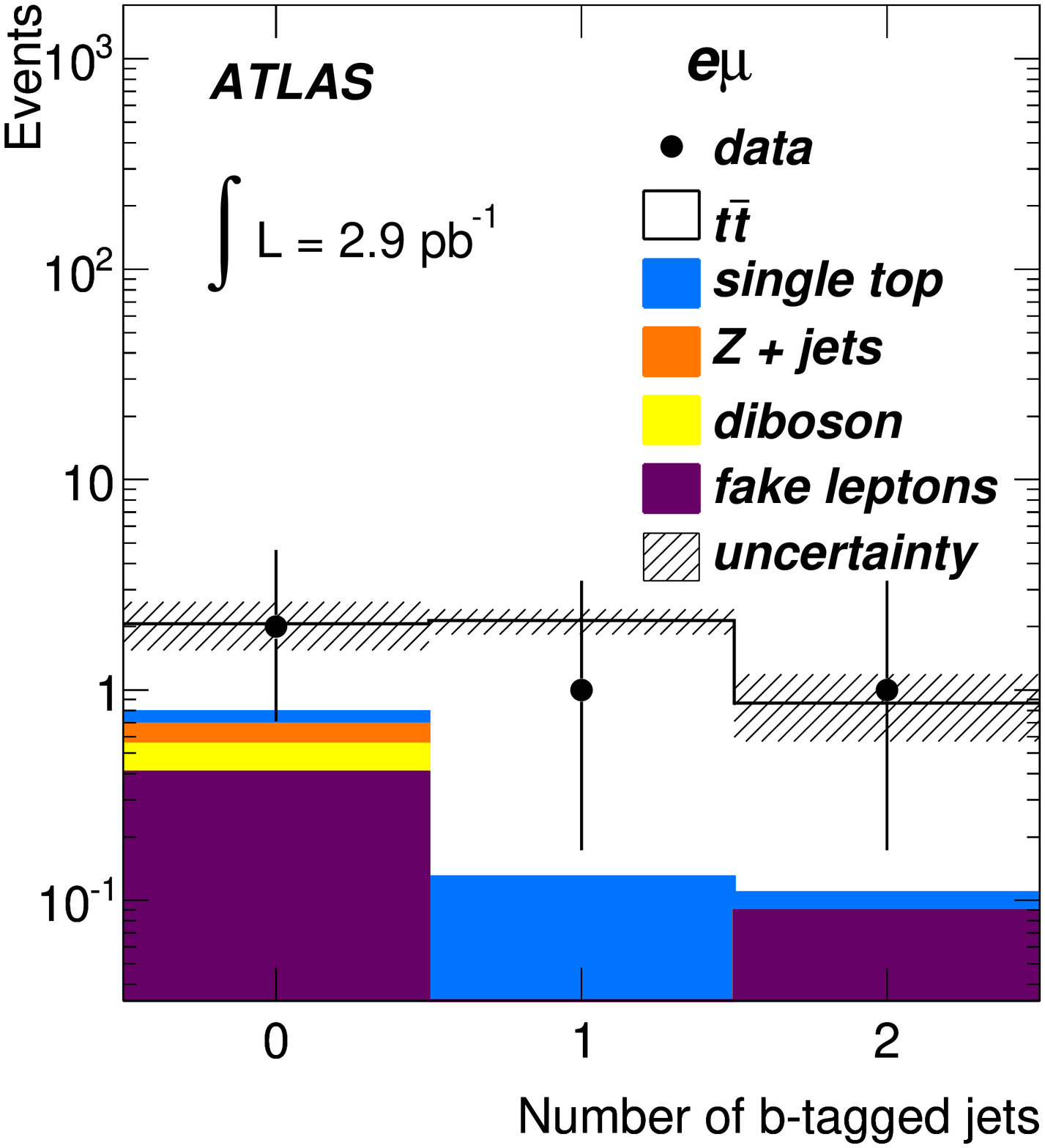} }
  \caption{The $b$-tagged jet multiplicities in the signal region for
(a) the $ee$ channel, (b) the $\mu\mu$ channel and (c) the $e\mu$ channel.
  }
  \label{f:ll_nbjets}
\end{figure}

\begin{figure}[htbp]
  \centering
  \subfigure[]{ \includegraphics[width=0.315\textwidth]{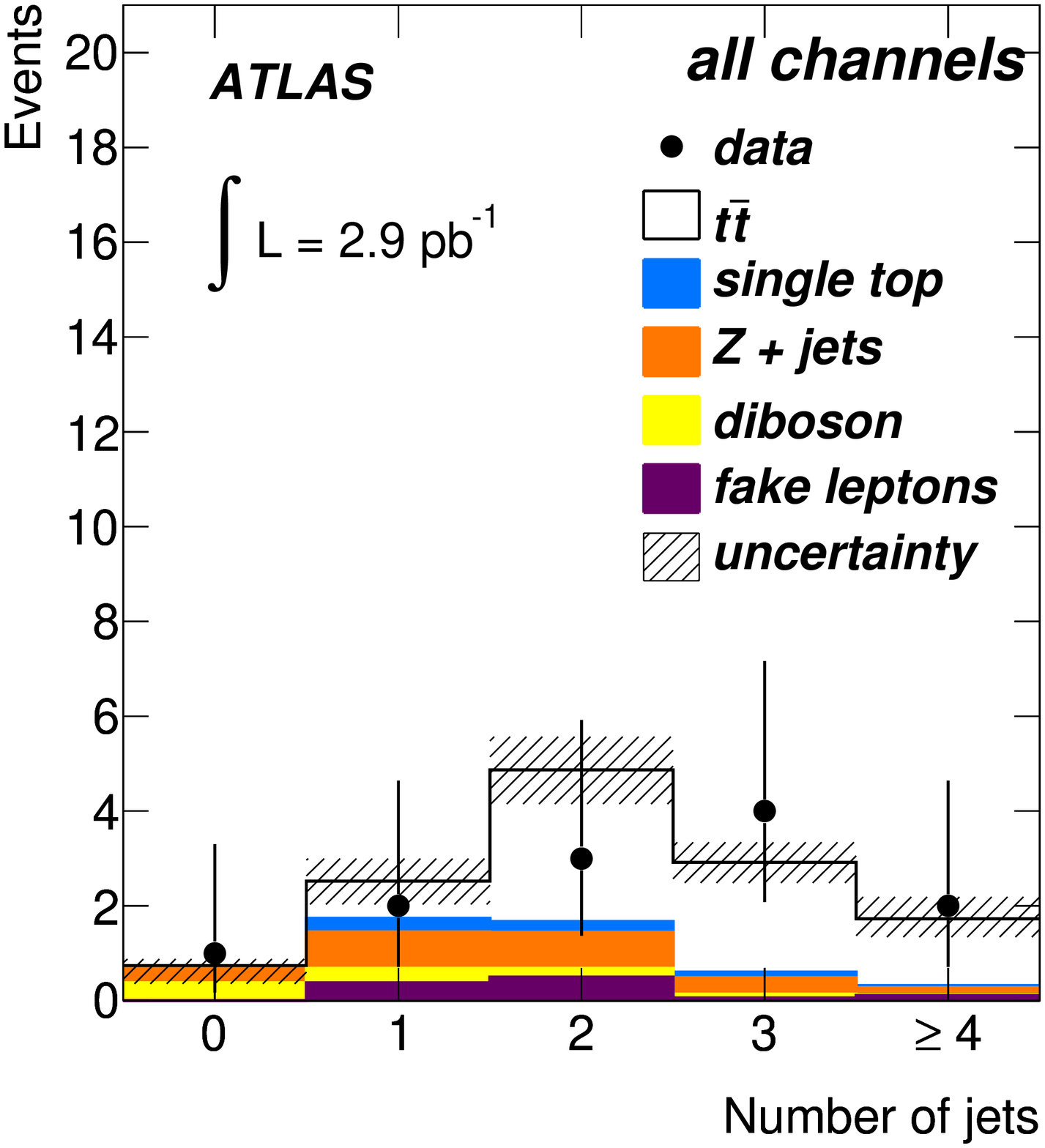} }
  \subfigure[]{ \includegraphics[width=0.315\textwidth]{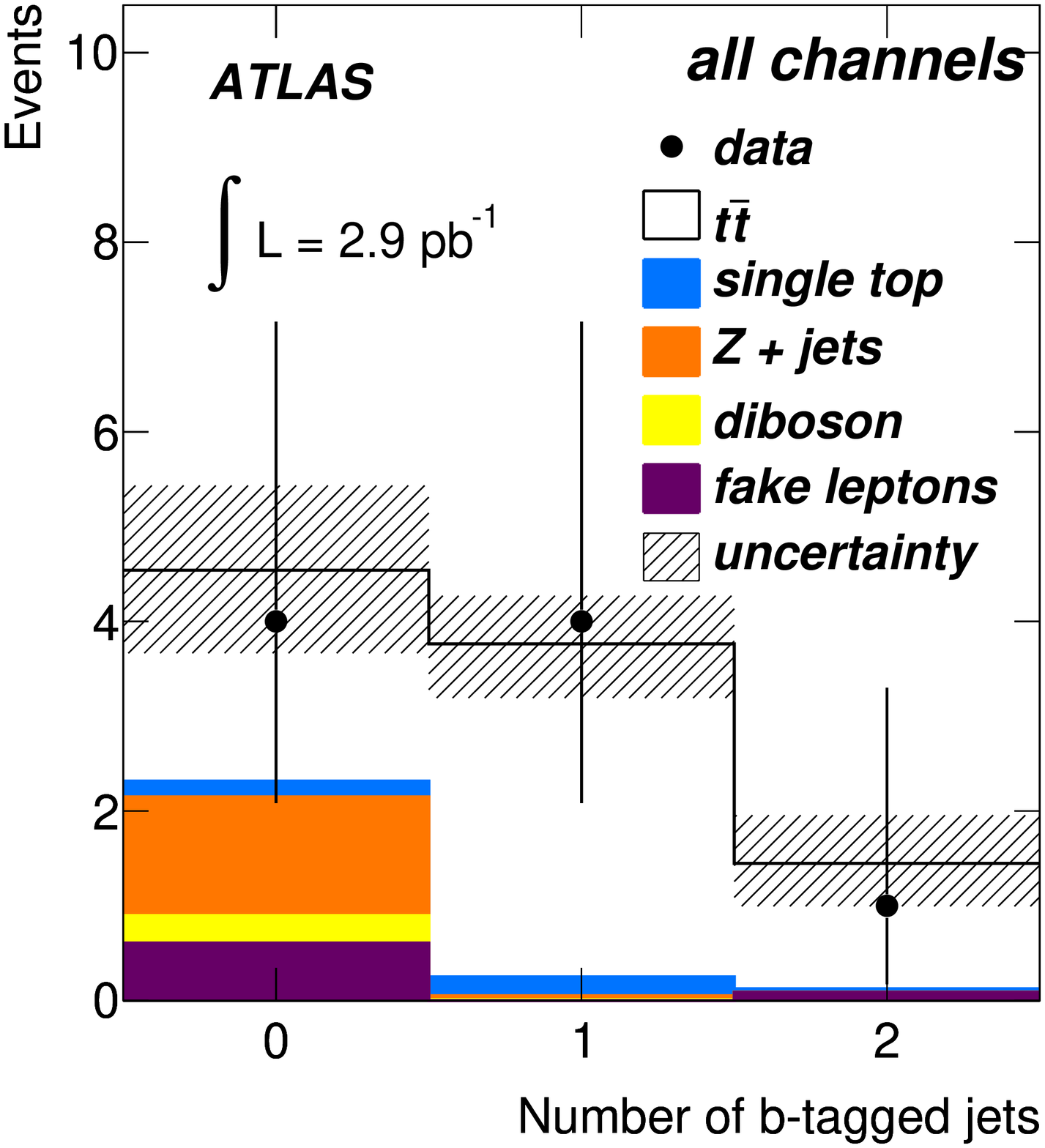} }
	\caption{(a) Jet multiplicity in the signal region
     without the $N_{jets}\ge 2$ requirement and (b)
     the $b$-tagged jet multiplicity in the signal region, both for the combined dilepton channels. } \label{f:ll_summed}
\end{figure}

The \MET{}, $Z$ boson mass window, and \HT\ cuts are derived from
a grid scan significance optimisation on simulated events which includes systematic
uncertainties.  The estimated $\ttbar$ acceptance, given a dilepton
event, in each of the dilepton channels are $14.8\pm 1.6\%$ ($ee$),
$23.3\pm 1.8\%$ ($\mu\mu$) and $24.8\pm 1.2\%$ ($e\mu$). The corresponding
acceptances including the \ttbar\ branching ratios are
$0.24\%$ ($ee$), $0.38\%$ ($\mu\mu$) and $0.81\%$ ($e\mu$). The final
numbers of expected and measured events in the signal region are shown
in Table~\ref{t:signal}. Figure~\ref{f:ll_met_ht} shows the predicted
and observed distributions of $\met$ for the $ee$ and $\mu\mu$
channels and of $\HT$ for the $e\mu$ channel. The predicted and
observed multiplicities of all jets and $b$-tagged jets are compared in
Figure~\ref{f:ll_njets} and Figure~\ref{f:ll_nbjets} for each channel
individually, and in Figure~\ref{f:ll_summed} for all channels
combined. Figure~\ref{f:ll_summed} (b) shows that a majority of the selected 
events have at least one $b$-tagged jet, consistent with the hypothesis that 
the excess of events over the estimated background originates from
\ttbar\ decay. In each of these plots the selection has been relaxed to omit 
the cut on the observable shown.

\begin{table}[!hbt]
\centering
\begin{tabular}{|c|c|c|c|} \hline
                                & $ee$                    & $\mu\mu$               & $e\mu$                    \\ \hline
$Z$+jets (DD)                   & \DYZeeNJetsTwoJet       & \DYZmmNJetsTwoJet      & -                          \\
$Z( \to \tau\tau)$+jets (MC)    & \ZtteeNJetsTwoJet       & $0.14 \pm 0.07$        & \ZttemNJetsTwoJet          \\
Non-$Z$ leptons (DD)               & \FakeWeeNJetsTwoJet     & \FakeWmmNJetsTwoJet    & \FakeWemNJetsTwoJet        \\
Single top (MC)                 & \singletopeeNJetsTwoJet & \singletopmmNJetsTwoJet& \singletopemNJetsTwoJet    \\
Dibosons (MC)	                & \dibosoneeNJetsTwoJet   & \dibosonmmNJetsTwoJet  & \dibosonemNJetsTwoJet      \\ \hline
Total (non \ttbar{})          & $0.60 \pm 0.27$         & $0.88 \pm 0.40$        & $0.97 \pm 0.30$            \\ \hline
\ttbar (MC)                 & $1.19 \pm 0.19$         & \ttbarmmNJetsTwoJet    & \ttbaremNJetsTwoJet        \\ \hline
Total expected                 & $1.79 \pm 0.38$         & $2.75 \pm 0.55$        & $4.82 \pm 0.65$            \\ \hline\hline
Observed                        & \DataeeNJetsTwoJet      & \DatammNJetsTwoJet     & \DataemNJetsTwoJet         \\ \hline
\end{tabular}
\caption
{ The full breakdown of the expected {\ttbar}-signal and background in the signal
  region compared to the observed event yields, for each of the dilepton channels (MC is
simulation based, DD is data driven). All systematic
  uncertainties are included.
}
\label{t:signal}
\end{table}

\subsection{Background determination strategy}

The expected dominant backgrounds in the dilepton channel are $Z$
boson production in association with jets, which can give rise to the
same final state as \ttbar\ signal, and $W$+jets. The latter 
can only contribute to the signal selection if the event
contains at least one fake lepton.

Both $Z$+jets background and backgrounds with fake leptons are
estimated from the data. The contributions from remaining electroweak
background processes, such as single-top, $WW$, $ZZ$ and $WZ$ boson
production are estimated from Monte-Carlo simulations.

\subsection{Non-$Z$ lepton backgrounds} 
\label{s:ll_fakes}

True $\ttbar$ dilepton events contain two leptons from $W$ boson
decays; the background comes predominantly from $W$+jets events
and single-lepton $\ttbar$ production with a fake lepton
and a real lepton, though there is a smaller contribution with two
fake leptons coming from QCD multi-jet production. As in the
single-lepton analysis, in the case of muons, the dominant fake-lepton
mechanism is a semi-leptonic decay of a heavy-flavour hadron, in which
a muon survives the isolation requirement. In the case of electrons,
the three mechanisms are heavy flavour decay, light flavour jets with
a leading $\pi^0$ overlapping with a charged particle, and conversion
of photons. Here `fake' is used to mean both non-prompt leptons and
$\pi^0$s, conversions etc misidentified as leptons taken together.

The `matrix method' introduced in Section~\ref{s:mujets_fakes} is
extended here to measure the fraction of the dilepton sample that
comes from fake leptons. A looser lepton selection is
defined, and then it is used to count the number of observed dilepton
events with zero, one or two tight (`T') leptons together with two,
one or zero loose (`L') leptons, respectively ($N_{LL}$, $N_{TL}$ and
$N_{LT}$, $N_{TT}$, respectively). Then two probabilities are defined,
$r$ ($f$), to be the probability that real (fake) leptons that pass
the loose identification criteria, will also pass the tight criteria.
Using $r$ and $f$, linear expressions are then obtained for the
observed yields as a function of the number or events with zero, one
and two real leptons together with two, one and zero fake leptons,
respectively ($N_{FF}$, $N_{FR}$ and $N_{RF}$, $N_{RR}$,
respectively).

The method
explicitly accounts for the presence of events with two fake leptons.
These linear expressions form a matrix that is inverted in order to
extract the real and fake content of the observed dilepton event
sample:

\begin{equation}
\label{eqn:mm_matrix}
\hspace{-0.5cm}
\begin{bmatrix}N_{TT}\\N_{TL}\\N_{LT}\\N_{LL}\end{bmatrix} = \begin{bmatrix}rr& rf& fr& ff\\ r(1-r)& r(1-f)& f(1-r)& f(1-f)\\ (1-r)r& (1-r)f& (1-f)r& (1-f)f\\ (1-r)(1-r)& (1-r)(1-f)& (1-f)(1-r)& (1-f)(1-f)\end{bmatrix}\begin{bmatrix}N_{RR}\\N_{RF}\\N_{FR}\\N_{FF}
\end{bmatrix}
\end{equation}

For muons, the loose selection is identical to the one described in
Section~\ref{s:mujets_fakes}.  For loose electrons, the $E/p$ cut and
isolation requirements are dropped, and the `medium' electron
identification criteria as defined in Ref.~\cite{wobspaper} is
replaced with the corresponding loose definition, with looser
calorimeter and tracking cuts.

The efficiency for a real loose lepton to pass the full tight
criteria, $r$, is measured in data in a sample of $Z\rightarrow
\ell\ell$ events as a function of jet multiplicity. The corresponding
efficiency for fake leptons, $f$, is measured in data in events with a
single loose lepton, which are dominated by QCD di-jet production.
Contributions from real leptons due to $W$+jets in the fake lepton
control region are subtracted using simulated data.

The dominant systematic uncertainty on the $W$+jets background, as
determined by the matrix method, comes from the possible difference in
the mixture of processes where the efficiency for fake leptons $f$ is
measured, di-jet events and, where it is applied, the signal
region. For electrons, a larger contribution is expected from heavy
flavour events in the signal region due to $t\bar{t}\rightarrow
\ell\nu b jjb$ events. This effect is accounted for by measuring the
dependence of the efficiency for fake leptons on the heavy-flavour
fraction and calculating a corrected efficiency for fake leptons based
on the expected heavy-flavour fraction in the signal region in
simulation studies. The fake estimate in the data includes
contributions from events with tight and loose leptons, whose
contributions have opposite signs. This can lead to some negative
background estimates in the case of small statistics, but always
consistent with zero. The results of the matrix method for the
non-$Z$ background are shown in Table \ref{t:fake} for 0, 1 and $\ge2$
jet bins.  The results for the signal region ($\ge 2$ jets) is also
reported in Table~\ref{t:signal}.

The most important cross-check comes from comparing the matrix method
with two additional methods. The first (the `weighting method') uses
fake candidates in the single lepton sample and a fake rate to build
an event weight for the fake lepton event. It uses a less restrictive
loose definition and so probes the extrapolation of the fake rate $f$
to the signal region. The method gives results consistent with the
matrix method, as shown in Table~\ref{t:fake}.  The second (the
`fitting method') makes no assumptions about the relative mixture of
fake-lepton mechanisms, but uses data-derived templates in variables which
can discriminate between real and fake leptons to fit for the
fake-lepton fraction in the signal region. For the signal region the
fitting method predicts $0.01^{+0.97}_{-0} \pm 0.01$ non-$W$ boson events
for the $ee$ channel, $0.01^{+0.29}_{-0} \pm 0.01$ for the $\mu\mu$
channel, and $0.13^{+0.42}_{-0.13} \pm 0.14$ for the $e\mu$ channel.
The estimate from the fitting method is based on data in the signal
region, whereas the other methods provide estimates for the signal
region based on measurement in control regions.

\begin{table}
\begin{center}
\begin{tabular}{|lc|rrr|}
\hline \rule{0pt}{2.0ex}
Method& $N_{jets}$ & \multicolumn{1}{c}{$ee$} & \multicolumn{1}{c}{$\mu\mu$} & \multicolumn{1}{c|}{$e\mu$} \\ [2pt]
 \hline \rule{0pt}{3.5ex}
\multirow{3}{*}{Matrix} & 0 & $-0.07 \pm 0.05 \pm 0.05$ & $-0.09 \pm 0.05 \pm 0.07$ & $0.00 \pm 0.01 \pm 0.01$ \\ [2pt]
& 1 & $0.09 \pm 0.14 \pm 0.07$ & $-0.03 \pm 0.03 \pm 0.04$ & $0.28 \pm 0.20 \pm 0.09$ \\ [2pt]
& $ge 2$ & $0.16 \pm 0.17 \pm 0.06$ & $-0.08 \pm 0.04 \pm 0.06$ & $0.47 \pm 0.26 \pm 0.11$ \\ [3.5pt]
 \hline \rule{0pt}{3.5ex}
\multirow{3}{*}{Weighting} &0& $0.03 \pm 0.03 \pm 0.02$     & $0.34 \pm 0.14 \pm 0.32$  & $0.00 \pm 0.04 \pm 0.04$ \\ [2pt]
			   &1& $0.06 \pm 0.04 \pm 0.06$     & $0.10 \pm 0.07 \pm 0.11$  & $0.08 \pm 0.06 \pm 0.06$ \\ [2pt]
			   & $ge 2$ & $0.10 \pm 0.06 \pm 0.08$     & $0.00 \pm 0.04 \pm 0.04$  & $0.10 \pm 0.05 \pm 0.09$ \\ [3.5pt]
\hline
\end{tabular}
\caption{
Overview of the estimated non-$Z$ background yields in the signal region
using two different data-driven methods with their statistical
and systematic uncertainties respectively. The matrix method is the
baseline method, the weighting method is used as a
cross-check. 
} 
\label{t:fake}
\end{center} 
\end{table}

\subsection{$Z$+jets background}
\label{s:dy}

Although the $\ttbar$ event selection is designed to reject $Z$+jets
events, a small fraction of events which populate the $\met$ tails and
dilepton invariant mass more than 5 GeV (for $ee$) or 10 GeV (for
$\mu\mu$) away from the $Z$ boson mass will enter the signal sample. These
events are difficult to model in simulations due to large
uncertainties on the non-Gaussian missing energy tails, the $Z$ boson
cross-section for higher jet multiplicities, and the lepton energy
resolution. The $Z$+jets events are expected to have significant
$\met$ tails, primarily originating from mis-measurements of the jet
energies.

The $Z$+jets background is estimated by extrapolating from a control
region orthogonal to the top quark signal region.  This control region is
defined using the cuts for the signal region, but with an inverted $Z$ boson
mass window (requiring $|m_{\ell\ell}-m_{Z}|<5$ GeV for $ee$ and
$|m_{\ell\ell}-m_{Z}|<10$ GeV for $\mu\mu$) and lowering the $\met$
requirement to $\met>20$ GeV. For $\met$ below the signal region, and
for $\met$ larger than 20 GeV, the $Z$ boson mass window is extended to
$|m_{\ell\ell}-M_{Z}|<15 \GeV$ to reduce systematic uncertainties from
the lepton energy scale and resolution. A scale factor from $Z$+jets simulation
is used to extrapolate from the observed yield in the control region
to the expected yield in the signal region. The small non-$Z$ boson background
in the control region is corrected using the Monte-Carlo expectation.

The yield estimates obtained with this procedure are shown in
Table~\ref{t:DY}, along with estimates of $Z$+jets background based on
simulation only. The comparison demonstrates that data-driven
normalisation using the control regions helps to reduce the effect of
the systematic uncertainties. The estimated yields from data are
higher than those from the Monte-Carlo prediction. This trend is also
observed in the control regions involving \met{} where jets are used
in the selection.

\begin{table}[htb]
\centering
\begin{tabular}{|l|c|c|} \hline
         & $ee$  & $\mu\mu$ \\ \hline
$Z$+jets (Monte-Carlo)     & $0.14 \pm 0.03 \pm 0.16$ & $0.56 \pm 0.06 \pm0.39$ \\
$Z$+jets (data-driven)     & $0.25 \pm 0.09 \pm 0.16$ & $0.67 \pm 0.22 \pm0.31$ \\ \hline
\end{tabular}
\caption{Yields and uncertainties for the estimates of the $Z$+jets background. The uncertainties
are statistical and systematic, respectively.
}
\label{t:DY}
\end{table}

Due to the very limited data statistics, simulation is used for the $Z
\rightarrow \tau\tau$ contribution instead of the data-driven method
used to estimate $Z\rightarrow ee$ and $Z \rightarrow \mu\mu$
contributions. The modelling of the $Z \rightarrow \tau\tau$ is
cross-checked in the $e\mu$ channel in the 0-jet bin, where five
events are observed in data versus a total expectation of 3.1 events,
with an expected $Z \rightarrow \tau\tau$ contribution of 2.4 events.
The largest systematic uncertainty comes from that on the integrated
luminosity. The estimated $Z$+jets backgrounds are summarised in
Table~\ref{t:signal}.

Data-driven backgrounds and simulated acceptances and efficiencies 
are validated in various control regions which are depleted
of \ttbar{} events.

\begin{figure}[htbp]
  \centering
  \subfigure[]{\includegraphics[width=0.315\textwidth]{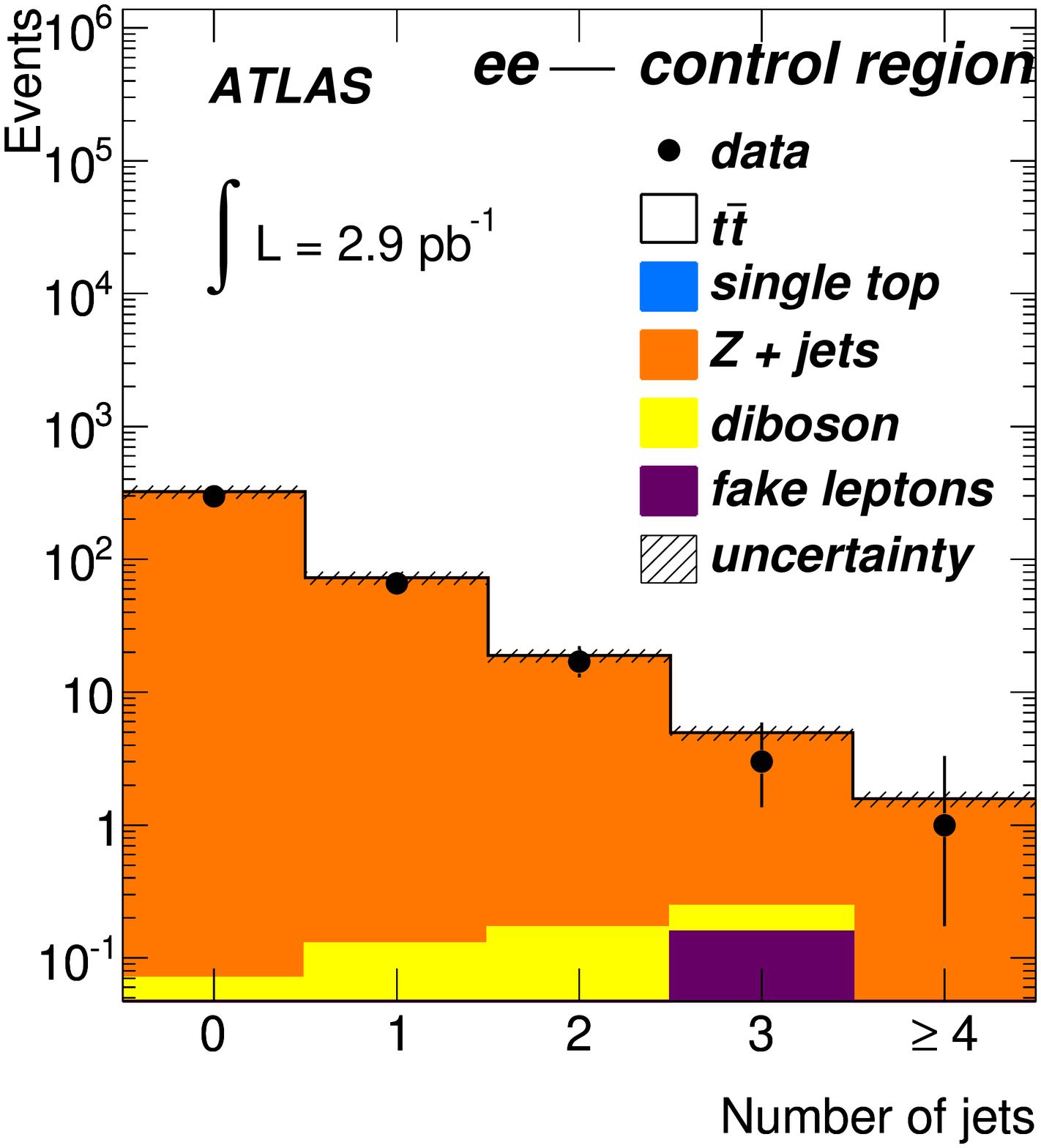}}
  \subfigure[]{\includegraphics[width=0.315\textwidth]{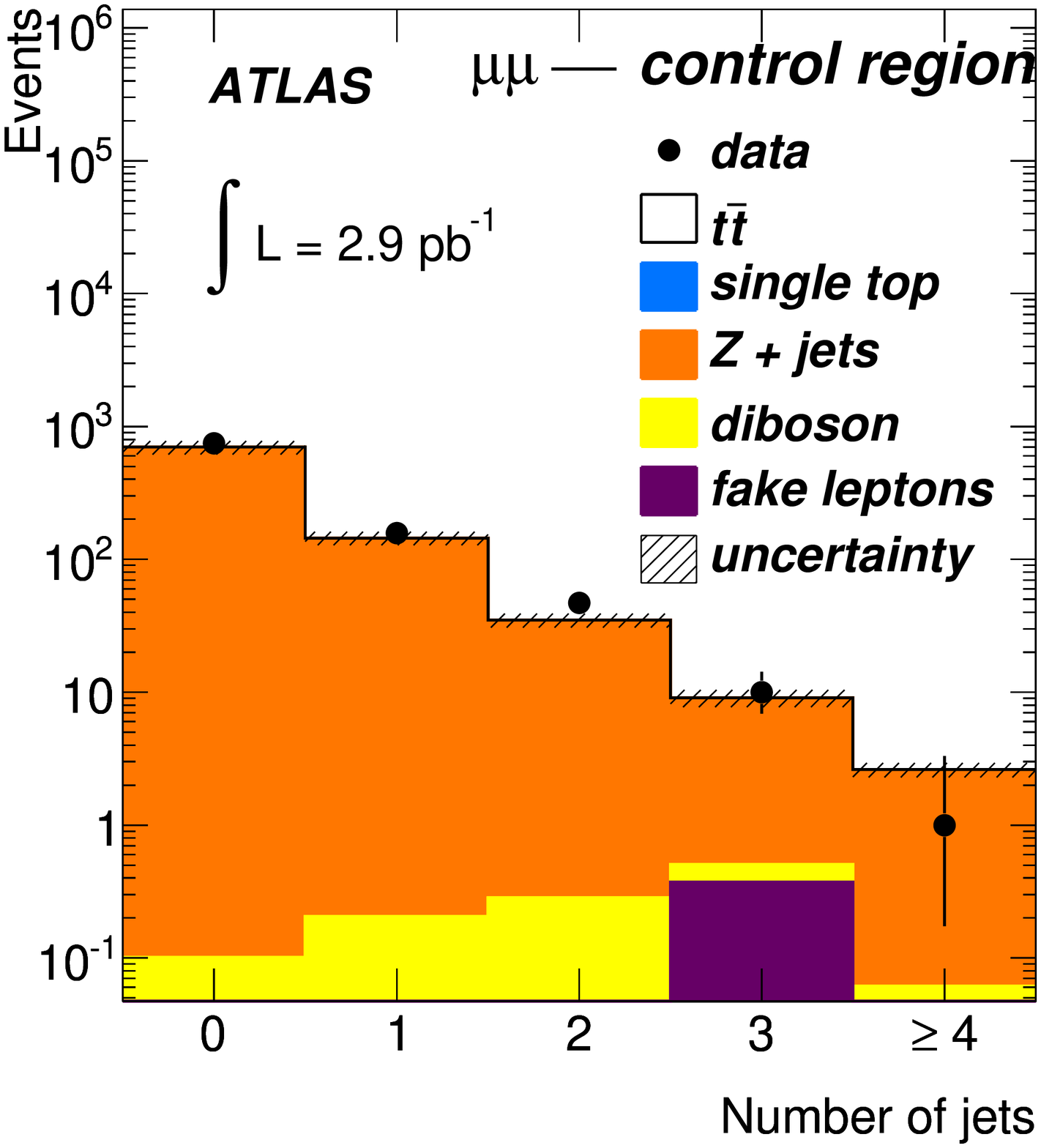}} \\
  \subfigure[]{\includegraphics[width=0.315\textwidth]{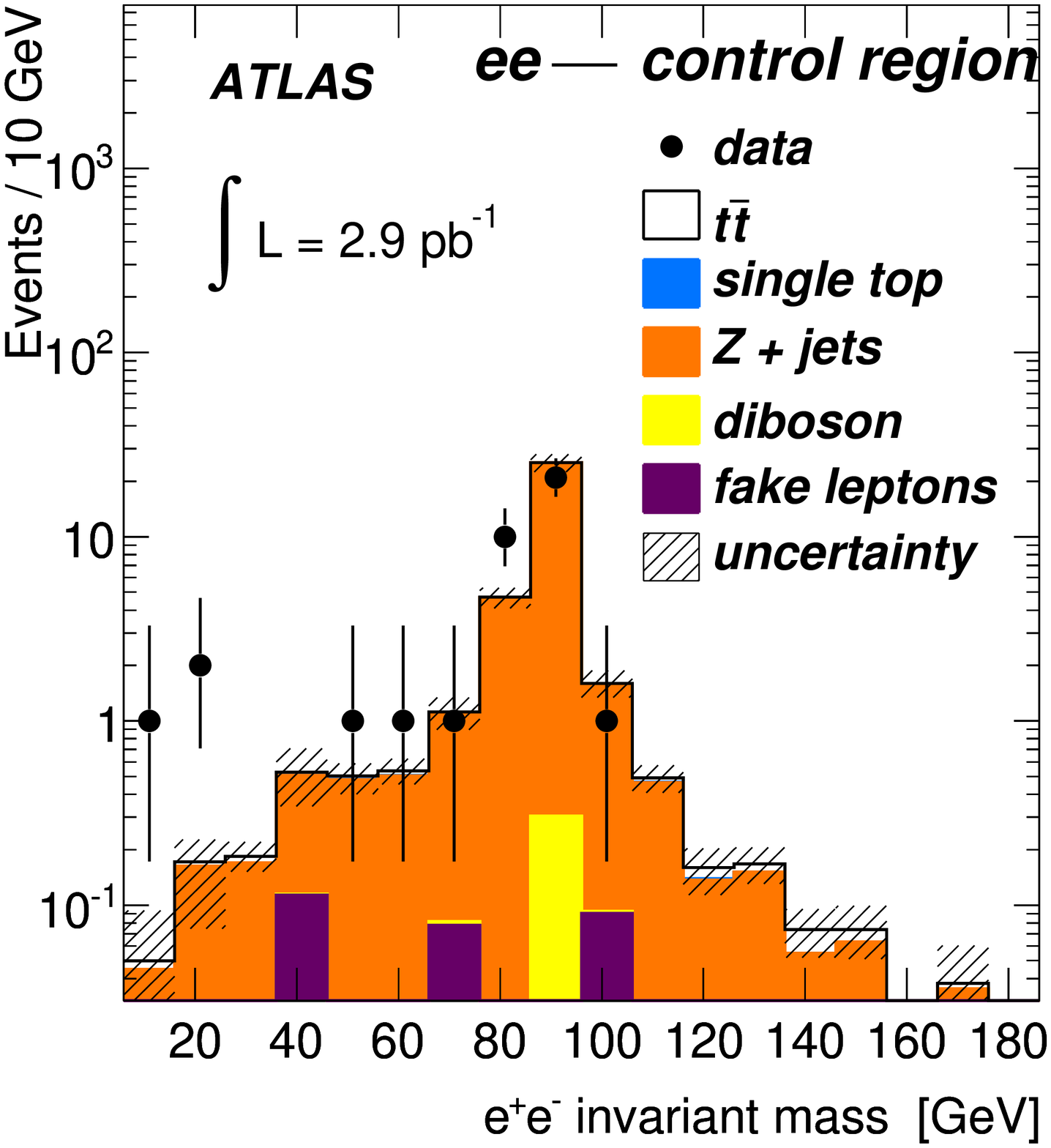}}
  \subfigure[]{\includegraphics[width=0.315\textwidth]{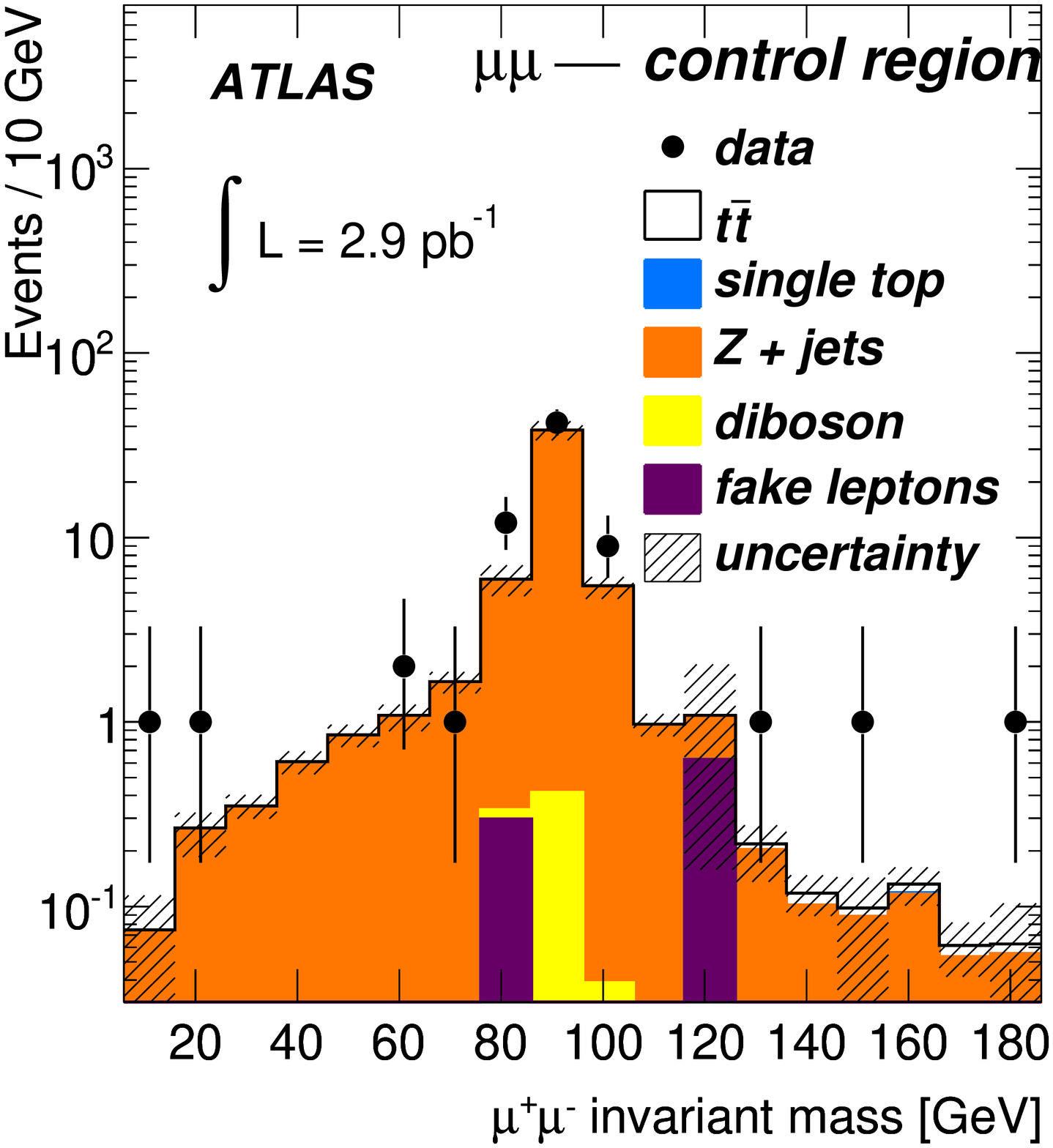}}
  \caption{Top row: Number of jets in events with the measured dilepton mass inside the $Z$ boson mass window
	for (a) the $ee$ channel and (b) the $\mu\mu$ channel. Bottom row: Invariant mass of opposite-signed
	lepton pairs in events with $\ge$2 jets in the low \met\ region for (c) the $ee$ channel and (d)
	the $\mu\mu$ channel.
  }
  \label{f:control}
\end{figure}

Figure~\ref{f:control} (a) and (b) show the jet multiplicity for
events where the dilepton mass lies inside the $Z$ boson peak and tests the initial state
radiation (ISR) modelling of jets for $Z$+jets processes. The dilepton mass
plots, Figure~\ref{f:control} (c) and (d), probe the lepton energy
scale and resolution.

The understanding of $\gamma\rightarrow e^+e^-$ conversions can be
tested by using same-sign events. Five same-sign events are observed
inside the $Z$ boson peak in the inclusive $ee$ channel and they are
compatible, within the limited statistics, with the conversions
modelled by the simulations.  No same-sign events have been observed in
the $\mu\mu$ or $e\mu$ channels.

\subsection{Cross-section determination in the dilepton channels}
\label{s:ll_xsec}

The cross-section is measured in each dilepton channel and translated
into an inclusive \ttbar{} cross-section using the $W \rightarrow
\ell\nu$ and $\tau \rightarrow \ell\nu\nu_\tau$ branching ratios.  The
cross-sections and uncertainties in the individual channels are
estimated using the likelihood method as will be described in
Section~\ref{s:combination}.  The cross-sections are summarised in
Table~\ref{t:dilxs}, and the breakdown of the individual sources of cross-section uncertainties are listed in
Table~\ref{t:sys_xs}. 
The dependence of the measured cross-section on the assumed top-quark mass
is small. A change of $\pm 1$ \GeV in the assumed mass results in a change
of $\mp 0.5$\% in the cross-section.

{
\renewcommand{\arraystretch}{1.4}
\setlength{\tabcolsep}{40pt}
\begin{table}[htb]
\centering
\begin{tabular}{|p{2cm}|c|} \hline
Channel  & $\sigma_{\ttbar}$ [pb] \\
\hline
$ee$     & 193 $^{ + 243}_{ - 152}$  $^{ + 84}_{ - 48}$  \\
$\mu\mu$ & 185 $^{ + 184}_{ - 124}$  $^{ + 56}_{ - 47}$   \\
$e\mu$   & 129 $^{ + 100}_{ - 72}$   $^{ + 32}_{ - 18}$  \\
\hline
Combined  &  $151 $ $^{+78}_{-62}$ $ ^{+37}_{-24}$   \\
\hline 
\end{tabular}
\caption { Measured cross-sections in each individual dilepton
  channel and in the combined fit. The uncertainties represent the statistical and
  combined systematic uncertainty, respectively. }
\label{t:dilxs}
\end{table}
}

\begin{table}[hbt]
\centering
\begin{tabular}{|l|c|c|c|}
\hline 
       &  \multicolumn{3}{c|}{Relative cross-section uncertainty [\%]} \\ \hline
Source & $ee$ &  $\mu\mu$ &  $e\mu$ \\ \hline
\hline
Statistical uncertainty      &   -79 / +126    &  -67 / +100     &  -56 / +77  \\\hline

{\em Object selection} & & &\\
Lepton reconstruction, identification, trigger  &   -2  / +11  & -4  / +3    &  -1  / +3    \\
Jet energy reconstruction &   -7  / +13  & -14 / +9    &  -3  / +5    \\
\hline
{\em Background rates} & & & \\ 
Fake leptons              &   -31 / +24  & -4  / +1    &  -15 / +8    \\
$Z$+jets                  &  -12  / +4   & -19 / +5    &  -2  / +1    \\
Monte-Carlo simulation statistics  &  -5   / +3   & -3  / +4    &  $\pm$ 2    \\
Theoretical cross-sections &  $\pm$ 3    & -5  / +4    &  $\pm$ 3    \\
\hline
{\em Signal simulation} & & & \\ 
Initial/final state radiation  &   -4  / +5   & -2  / +3    &  -2  / +3    \\
Parton distribution functions  &  -2   / +1   & -2  / +3    &  -2  / +3    \\
Parton shower and hadronisation &   -9  / +14  & -6  / +9    &  $\pm$ 3    \\
Next-to-leading order generator             &   -8  / +11  & -11 / +13   &  -3   / +4    \\
\hline
Integrated luminosity                &   -11 / +16  & -11 / +16   &  -12  / +14   \\
\hline
\hline
Total systematic uncertainty        &  -25 / +44    &  -25 / +30 &  -14 / +25  \\ \hline
Statistical + systematic uncertainty   &  -83 / +134   &  -72 / +104 &  -57 /  +81  \\\hline

\end{tabular}
\caption{
  Individual systematic uncertainties on the \ttbar{} cross-section
  in the dilepton channels. The combined uncertainties listed in the bottom two rows 
  include the luminosity uncertainty.
}
\label{t:sys_xs}
\end{table}

\section{Combination of the single lepton and the dilepton channels}
\label{s:combination}

The combined measurement of the $\ttbar$ production cross-section is
based on a likelihood fit in which the number of expected events is
modeled as

\begin{equation}
N^{exp}(\sigma_{\ttbar},\alpha_{j}) =
L \cdot \epsilon_{\ttbar}(\alpha_{j})\cdot \sigma_{\ttbar} + \sum_{bkg} L \cdot \epsilon_{bkg}(\alpha_{j}) \cdot \sigma_{bkg}(\alpha_{j}) + N_{DD}(\alpha_{j})
\end{equation}

\noindent where $L$ is the integrated luminosity, $\epsilon_{\ttbar}$ is the
signal acceptance, $\epsilon_{bkg}$, $\sigma_{bkg}$ are the
efficiency and cross-section for backgrounds as obtained from MC
simulation respectively, and $N_{DD}$ is the number of expected events
from data-driven estimates. The acceptance and background estimates
depend on sources of systematic uncertainty labelled as $\alpha_j$. The
likelihood for a single channel is defined as

\begin{equation}\label{Eq:likelihood}
\mathcal{L}(\sigma_{\ttbar}, L, \alpha_{j}) = \textrm{Poisson}
\left (N^{obs} \, | \, N^{exp}(\sigma_{\ttbar},\alpha_{j}) \right) \, \times \textrm{Gauss}(L_0 | L,
\delta_L)\,  \times \prod_{j\in \rm syst} \Gamma_j( \alpha_{j} ) \, .
\end{equation}

\noindent where $L_0$ is the integrated luminosity of the data sample
and $\delta_L = 11\% \cdot L_0$.
Sources of systematic uncertainties are grouped into subsets that are
uncorrelated to each other. However each group can have correlated
effects on multiple signal and background estimates.  The relationship
between the channels is enforced by identifying the $\alpha_j$ common
to different channels in the construction of  the combined likelihood
function. Ensembles of pseudo-data were generated and the resulting
estimate of the cross-section was confirmed to be unbiased. The method
is the same as the one used in~\cite{mcpub} and described in~\cite{likelihood}; 
however, in this case systematic uncertainties are modelled with gamma distributions,
which are more suitable priors for large systematics than truncated
Gaussians~\cite{gamma}. In the small systematic uncertainty limit, the
gamma distribution coincides with the conventional choice of a
Gaussian.

Table~\ref{t:sys_combxs} lists the cross-sections and signal
significance for the single-lepton, dilepton and the combined channels
with the corresponding statistical and systematic uncertainties
extracted from the likelihood fit. By combining all five channels, the
background-only hypothesis is excluded at a significance of
$\sigcomb\sigma$ obtained with the approximate method of
~\cite{likelihood}.  If Gaussian distributions are assumed for all
systematic uncertainties, a significance of $5.1\sigma$ is obtained. The
absence of bias in the fit is validated by
pseudo-experiments. Similarly, the traditional hybrid
Bayesian-frequentist approach in which the $\alpha_j$ are randomized
in an ensemble of pseudo-experiments finds a signal significance
consistent with the results from the likelihood method within 
$0.1\sigma$. The results also agree with those obtained from an
alternative method based on a purely Bayesian methodology.

\begin{table}[hbt]
\centering
\begin{tabular}{|l|c|c|}
\hline
                         &  Cross-section [pb]  & Signal significance [$\sigma$] \\
\hline
& & \\[-1.5ex]
Single lepton channels   &   $142 \pm 34 $ $^{+50}_{-31}$ &  4.0 \\[1ex]
Dilepton channels        &   $151 $ $^{+78}_{-62}$ $ ^{+37}_{-24}$ &  2.8 \\[0.9ex]
\hline
& & \\[-1.5ex]
All channels             &   $145 \pm 31$ $^{+42}_{-27}$ &  4.8 \\[0.9ex]
\hline
\end{tabular}
\caption{\label{t:sys_combxs} Summary of \ttbar\ cross-section and signal significance calculated by
combining the single lepton and dilepton channels individually and for all channels combined.}
\end{table}


\section{Summary}
\label{summary.section}

Measurements of the \ttbar\ production cross-section in the
single-lepton and dilepton channels using the ATLAS detector are
reported.  In a sample of \lumitot, \ncandlj\
\ttbar\ candidate events are observed in the single-lepton topology,
as well as \ncandll\ candidate events in the dilepton topology,
resulting in a measurement of the inclusive \ttbar\ cross-section of
\[
\sigmattbar= \xsecctot \xseccstat \xseccsyst\,~\rm pb \, .
\]

\singlefigure{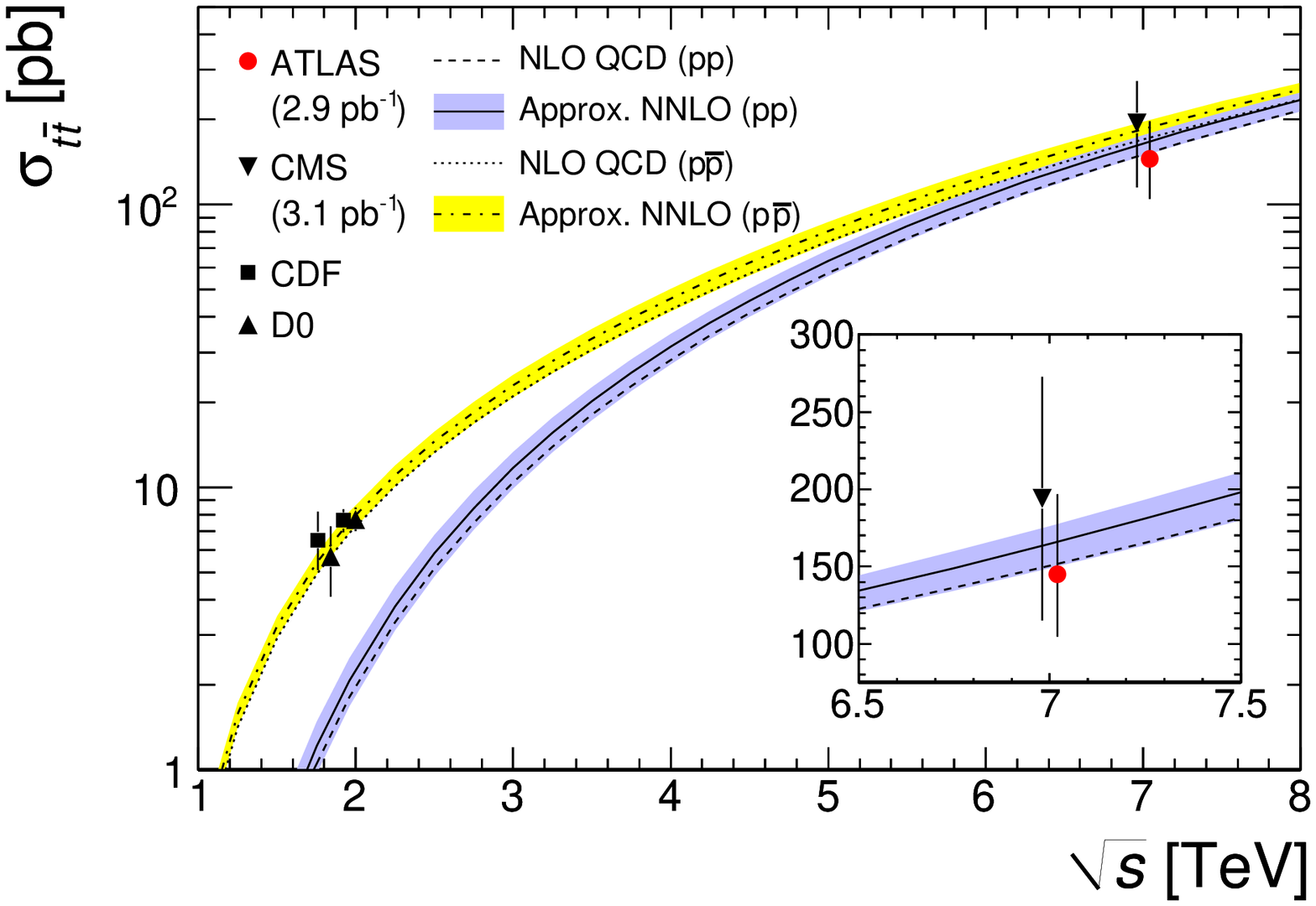}{0.7\textwidth}{f:xsec-vs-energy}%
{Top quark pair-production cross-section at hadron colliders as measured by CDF and D0 at Tevatron~\cite{cdfd0},
CMS~\cite{cmstopdileptons} and ATLAS (this measurement). The theoretical predictions for $pp$ and  $p\bar{p}$
collisions~\cite{other_refs} include the scale and PDF uncertainties, obtained using the HATHOR tool with
the CTEQ6.6 PDFs~\cite{hathor} and assume a top-quark mass of 172.5 \GeV.}

\noindent This is the first ATLAS Collaboration measurement making 
simultaneous use
of reconstructed electrons, muons, jets, $b$-tagged jets and missing
transverse energy, therefore exploiting the full capacity of the detector.
The combined measurement, consisting of the first measurement of the
\ttbar\ cross-section in the single-lepton channel at the LHC and a
measurement in the dilepton channel, is the most precise measurement to date
of the \ttbar\ cross-section at $\sqrt{s}=7$ TeV. 

The cross-sections measured in each of the five sub-channels are 
consistent with each other and kinematic properties of the selected 
events are consistent with SM $\ttbar$ production.
The measured \ttbar\ cross-section is in good agreement with the
measurement in the dilepton channel by CMS~\cite{cmstopdileptons},
as well as with NLO QCD predictions~\cite{other_refs} and the
approximate NNLO top quark cross-section calculation~\cite{hathor}.
Figure~\ref{f:xsec-vs-energy} shows the ATLAS and CMS measurements
together with previous Tevatron measurements~\cite{cdfd0}.

With the prospect of accumulation of larger data samples, the
statistical and systematic uncertainty on the \ttbar\ cross-section
will decrease and a precise measurement can challenge the SM
prediction based on QCD calculations and constrain the parton
distribution functions. Larger samples of \ttbar\ events
will also be instrumental in precision studies of the production, mass
and decay properties of top quarks, and be vital in new physics
searches in which SM \ttbar\ production is an important background.


\section{Acknowledgements}

We wish to thank CERN for the efficient commissioning and operation of
the LHC during this initial high-energy data-taking period as well as
the support staff from our institutions without whom ATLAS could not
be operated efficiently.

We acknowledge the support of ANPCyT, Argentina; YerPhI, Armenia; ARC,
Australia; BMWF, Austria; ANAS, Azerbaijan; SSTC, Belarus; CNPq and
FAPESP, Brazil; NSERC, NRC and CFI, Canada; CERN; CONICYT, Chile; CAS,
MOST and NSFC, China; COLCIENCIAS, Colombia; MEYS (MSMT), MPO and
CCRC, Czech Republic; DNRF, DNSRC and Lundbeck Foundation, Denmark;
ARTEMIS, European Union; IN2P3-CNRS, CEA-DSM/IRFU, France; GNAS,
Georgia; BMBF, DFG, HGF, MPG and AvH Foundation, Germany; GSRT,
Greece; ISF, MINERVA, GIF, DIP and Benoziyo Center, Israel; INFN,
Italy; MEXT and JSPS, Japan; CNRST, Morocco; FOM and NWO, Netherlands;
RCN, Norway; MNiSW, Poland; GRICES and FCT, Portugal; MERYS (MECTS),
Romania; MES of Russia and ROSATOM, Russian Federation; JINR; MSTD,
Serbia; MSSR, Slovakia; ARRS and MVZT, Slovenia; DST/NRF, South
Africa; MICINN, Spain; SRC and Wallenberg Foundation, Sweden; SER,
SNSF and Cantons of Bern and Geneva, Switzerland; NSC, Taiwan; TAEK,
Turkey; STFC, the Royal Society and Leverhulme Trust, United Kingdom;
DOE and NSF, United States of America.

The crucial computing support from all WLCG partners is acknowledged
gratefully, in particular from CERN and the ATLAS Tier-1 facilities at
TRIUMF (Canada), NDGF (Denmark, Norway, Sweden), CC-IN2P3 (France),
KIT/GridKA (Germany), INFN-CNAF (Italy), NL-T1 (Netherlands), PIC
(Spain), ASGC (Taiwan), RAL (UK) and BNL (USA) and in the Tier-2
facilities worldwide.


\begin{flushleft}
{\Large The ATLAS Collaboration}

\bigskip

G.~Aad$^{\rm 48}$,
B.~Abbott$^{\rm 111}$,
J.~Abdallah$^{\rm 11}$,
A.A.~Abdelalim$^{\rm 49}$,
A.~Abdesselam$^{\rm 118}$,
O.~Abdinov$^{\rm 10}$,
B.~Abi$^{\rm 112}$,
M.~Abolins$^{\rm 88}$,
H.~Abramowicz$^{\rm 153}$,
H.~Abreu$^{\rm 115}$
% ...
\bigskip

% ...
$^{10}$ Institute of Physics, Azerbaijan Academy of Sciences, H. Javid Avenue 33, AZ 143 Baku, Azerbaijan\\
$^{11}$ Institut de F\'isica d'Altes Energies, IFAE, Edifici Cn, Universitat Aut\`onoma  de Barcelona,  ES - 08193 Bellaterra (Barcelona), Spain\\
% ...
$^{48}$ Albert-Ludwigs-Universit\"{a}t, Fakult\"{a}t f\"{u}r Mathematik und Physik, Hermann-Herder Str. 3, D - 79104 Freiburg i.Br., Germany\\
$^{49}$ Universit\'e de Gen\`eve, Section de Physique, 24 rue Ernest Ansermet, CH - 1211 Geneve 4, Switzerland\\
% ...
$^{88}$ Michigan State University, Department of Physics and Astronomy, High Energy Physics Group, East Lansing, MI 48824-2320, United States of America\\
% ...
$^{111}$ University of Oklahoma, Homer L. Dodge Department of Physics and Astronomy, 440 West Brooks, Room 100, Norman, OK 73019-0225, United States of America\\
$^{112}$ Oklahoma State University, Department of Physics, 145 Physical Sciences Building, Stillwater, OK 74078-3072, United States of America\\
% ...
$^{115}$ LAL, Univ. Paris-Sud, IN2P3/CNRS, Orsay, France\\
% ...
$^{118}$ Oxford University, Department of Physics, Denys Wilkinson Building, Keble Road, Oxford OX1 3RH, United Kingdom\\
% ...
$^{153}$ Tel Aviv University, Raymond and Beverly Sackler School of Physics and Astronomy, Ramat Aviv, IL - Tel Aviv 69978, Israel\\
% ...
\end{flushleft}

\end{document}